\documentclass[11pt]{article}

\usepackage{xcolor}
\renewcommand{\textcolor}[2]{#2}
\renewcommand{\color}[1]{}

\usepackage{algorithm} 
\usepackage{algpseudocode} 
\usepackage{amsmath}
\usepackage{amsfonts}
\usepackage{amssymb}
\usepackage[english]{babel}
\usepackage{booktabs}
\usepackage{caption}
\usepackage{doi}
\usepackage{enumitem}
\usepackage{float}
\usepackage{graphicx}
\usepackage{hyperref}
\usepackage[capitalise]{cleveref}
\usepackage[utf8]{inputenc}
\usepackage{mathptmx}
\usepackage{microtype}
\usepackage{siunitx}
\usepackage{stfloats}
\usepackage{tabularx}
\usepackage{textcomp}
\usepackage{xspace}
\usepackage{xcolor}
\usepackage{mathtools}
\usepackage{subfigure}
\usepackage{tikz}

\usepackage[backend=biber, style=numeric]{biblatex}
\usepackage{csquotes}
\bibliography{main}

\usepackage{amsthm}
\newtheorem{remark}{Remark}

\usepackage{./template/arxiv}

\usepackage{./input/ao-math-std}
\usepackage{./input/ao-math-graphs}
\usepackage{./input/ao-math-fields}

\usepackage{ulem}
\newcommand\jh{\textcolor{blue}}

\newcommand\orcauth[2]{\href{https://orcid.org/#1}
  {\includegraphics[height=0.7em]{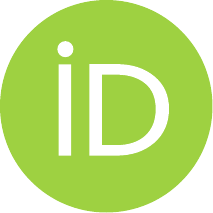}\enspace#2}}

\DeclareSIUnit\mmHg{mmHg}

\title{A Multiscale-Multiphysics Framework for Modeling Organ-scale Liver Regrowth}
\date{}

\author{
    \orcauth{0000-0002-8762-6158}{Adnan Ebrahem}\\
    \texttt{adnan.ebrahem@tu-darmstadt.de} \\
\And
    \orcauth{0009-0002-1016-2698}{Jannes Hohl} \\
	\texttt{jannes.hohl@tu-darmstadt.de} \\
\And
    \orcauth{0000-0003-0276-6029}{Etienne Jessen} \\
	\texttt{etienne.jessen@tu-darmstadt.de} \\
\And
     \orcauth{0000-0002-1153-146X}{Marco F.P. ten Eikelder} \\
	\texttt{marco.eikelder@tu-darmstadt.de} \\
\And
    \orcauth{0000-0002-9068-6311}{Dominik Schillinger} \\ 
    \texttt{dominik.schillinger@tu-darmstadt.de} \\ \\
	Institute for Mechanics, Computational Mechanics Group\\
	Technical University of Darmstadt, 64287 Darmstadt, Germany\\
}

\begin{document}
\maketitle

\begin{abstract}
%The human liver is able to grow back after a portion has been removed. %The human liver is able to regenerate after partial surgical resection, driven by rapid cell division at the microscale. % tissue growth associated with liver regeneration significantly impacts the liver's meso- and macroscale perfusion capability, which liver functionality critically depends on. 
We present a framework for modeling liver regrowth on the organ scale that is based on three components: % associated with physics at multiple spatial scales: 
(1) a multiscale perfusion model that combines synthetic vascular tree generation with a multi-compartment homogenized flow model, including a homogenization procedure to obtain effective parameters; (2) a poroelastic finite growth model that acts on all compartments and the synthetic vascular tree structure; (3) an evolution equation for the local volumetric growth factor, driven by the homogenized flow rate into the microcirculation as a measure of local hyperperfusion and well-suited for calibration with available data. We apply our modeling framework to a prototypical benchmark and a full-scale patient-specific liver, for which we assume a common surgical cut. Our simulation results demonstrate that our model represents hyperperfusion as a consequence of partial resection and accounts for its reduction towards a homeostatic perfusion state, exhibiting overall regrowth dynamics that correspond well with clinical observations. In addition, our results show that our model also captures local hypoperfusion in the vicinity of orphan vessels, a key requirement for %potential applications, e.g., 
the prediction of ischemia or the preoperative identification of suitable cut patterns.
%In addition, we motivate an empirical driving force that initiates compensatory growth of liver tissue until a physiological blood flow rate is reached at each point of the liver domain, and show how it is calibrated by available liver data. Considering a patient-specific liver, we demonstrate that our multiscale-multiphysics model is able to correctly predict the typical perfusion outcome associated with a common surgical cut pattern.
\end{abstract}

\keywords{Liver regeneration \and Tissue regrowth \and Partial resection \and Synthetic vascular trees \and Multi-compartment homogenized flow \and Poroelastic finite growth \and Hyperperfusion \and Hypoperfusion}

\newpage
{
  \hypersetup{linkcolor=black}
  \tableofcontents
}

\newpage

\section{Introduction}
\label{sec:Introduction}

%OPENING: for drawing reader's attention to the research topic.
Liver cancer is the seventh most commonly diagnosed cancer and the third leading cause of cancer mortality worldwide \cite{RefCancerStatistics}. The removal of cancerous portions of the liver via surgical intervention is known as partial liver resection or partial hepatectomy. The liver is able to regenerate back to full size, where up to 70\% of the original liver mass can be removed \cite{RefMichalopoulos}. %Liver regeneration - and the functionality of the liver remnant - depend on a multitude of factors, such as the current volume, the size of the in- and outflow areas, and the impairment of the micro- and macrocirculation due to the individual cut pattern. 
The tissue growth associated with liver regeneration severely affects the perfusion capability of the liver, which liver functionality critically depends on \cite{RefDebbautDiss}. %The patient-specific assessment of the postoperative success of partial liver resection prior to the operation remains a challenge for surgeons.

Computational liver models that predict regeneration, regrowth and perfusion for a patient-specific cut liver have the potential to support surgeons in assessing and planning partial liver resections. % and the  are potentially useful and have clinical relevance.
%BACKGROUND INFORMATION: Provide a brief overview of the broader research area or field in which the study is situated. Include relevant background information, key concepts, and historical context to help readers understand the significance of your research.
Various models based on continuum mechanics have been proposed to model the growth of biological tissues \cite{RefRodriguez,RefReviewGrowth,RefLubarda,RefMenzel,RefKuhl2014,RefGarikpati,RefJonesChapman,cyron2017growth,ambrosi2019growth}, including applications to the heart \cite{RefKroon,RefGoektepe2010,RefGoektepe}, brain \cite{RefBrain1,RefBrain2}, tumor tissue \cite{RefAmbrosi}, arterial wall growth \cite{RefKuhl2007,sankaran2013efficient}, cardiovascular modeling \cite{schwarz2023fluid} and vein graft growth \cite{ramachandra2017gradual}. A major focus lies on specifying individual growth laws and identifying factors that drive growth for each specific application. These models can simulate the volumetric growth of tissue stimulated by a mechanical driving force without the necessity of resolving the underlying processes at the cell level \cite{RefChrist}. To the best of our knowledge, there is currently no continuum mechanics model that addresses liver regrowth after partial hepatectomy.

% Dominik: What others did.
%{\color{red} Adnan: please distribute the references better.}
Modeling of liver growth is strongly linked to modeling perfusion, and therefore requires an appropriate patient-specific representation of the hierarchical liver vasculature. Given the limited resolution of in-vivo imaging, reconstruction from imaging data is generally only possible for the few largest vessels in the hierarchy. Discrete representations of hierarchical vascular trees can be synthetically generated by computer-based methods built on energy minimization principles \cite{Schreiner1,Schreiner2,RefSchwen,guy20193d}. %such as constrained constructive optimization (CCO) \cite{Schreiner1,Schreiner2,RefSchwen} built on Murray's minimization principles \cite{RefMurray}.
Several continuum models have been developed to simulate liver perfusion. %\cite{RefLiverPerfusion,RefLiverPerfusion1,RefLiverPerfusion2,RefDebbaut3,RefDebbaut4,RefRicken,RefRicken2,RefLambers,RefStoter} 
Many of them focus on the microcirculation in idealized liver lobules \cite{RefLiverPerfusion1,RefLiverPerfusion2,RefDebbaut3}, and they have also been extended to simulate the relation between liver function and perfusion \cite{RefRicken,RefLambers}. %In \cite{siddiqui2024reduced}, an efficient reduced-order model was introduced for simulating blood perfusion in liver lobules, using Darcy’s equation.
Some models describe perfusion in the whole organ, coupling continuum flow models with synthetically generated vascular trees \cite{RefLiverPerfusion,RefAdnan} or experimentally determined trees \cite{RefDebbaut4,RefStoter}. To account for the vascular features over many spatial scales, one approach is to partition the vasculature into spatially co-existing compartments and homogenize them based on Darcy's law \cite{RefHyde}. This concept has been applied to model the perfusion in the heart \cite{RefHeartPerfusion,RefReducedDarcy} and in the liver \cite{RefLiverPerfusion}.

Only a few models exist that represent liver regeneration on the organ scale \cite{RefChrist}. They are primarily phenomenological, % and have been investigated in the context of postoperative liver failure. They are 
mostly based on ordinary differential equations and focus on the temporal evolution of the regeneration process, integrating risk factors for postoperative liver failure \cite{RefYamamoto,RefShestopaloff,RefPeriwal,RefCook}. Models of angiongenic and remodeling processes have been proposed in the context of the continuum theory of porous media for two-dimensional liver lobules \cite{RefRicken2}. At the cell scale, discrete models, e.g.\ based on cellular automata, have been developed to represent the proliferation of liver cells that occur during regeneration \cite{RefCellularAutomata1,RefCellularAutomata2}.
Additionally, agent-based models have been widely used to simulate cell behavior and tissue mechanics, utilizing both lattice-based (e.g., cellular automata) and off-lattice approaches (e.g., deformable cell models) \cite{drasdo2007role}, as well as hybrid discrete-continuum models \cite{osborne2010hybrid,ghallab2019influence}. A model that integrates cell-scale dynamics within a lobule-scale framework is published in \cite{hohme2007mathematical}. In another study \cite{drasdo2005single}, a single-cell-based biophysical model has been developed to investigate the spatio-temporal growth dynamics of two-dimensional tumor monolayers and three-dimensional tumor spheroids. In \cite{hoehme2010prediction}, a mathematical model has been developed to predict the mechanism of hepatocyte-sinusoid alignment during liver regeneration. A separate study \cite{konig2012quantifying} introduced a model that focuses on the overall kinetic processes of hepatic glucose metabolism. Another approach at the cell scale is based on Monte Carlo simulations, which offer a stochastic framework for modeling the dynamics of tissue-cell populations and investigating complex processes such as pattern formation and tissue repair \cite{radszuweit2009comparing,drasdo1995monte}. A comprehensive whole-organ model, however, that connects liver regrowth and the associated change in perfusion capability across the relevant scales is still lacking. %In particular, such a model requires the combination of tissue growth and liver perfusion, as liver growth is driven by and determines the perfusion capability. 

In this work, we will develop such a multiscale-multiphysics framework that integrates three models associated with different physics at multiple spatial scales to simulate the effect of liver tissue regrowth on the perfusion capability of a full-scale liver. 
Our article is organized as follows: in Section \ref{sec:Physiology}, we briefly review the multiscale mechanisms of hepatic vasculature and perfusion, the perfusion-related factors that drive liver regrowth, and derive basic concepts for their modeling. 
In Section \ref{sec:MultiCompartmentPerfusion}, we present a multiscale perfusion model by incorporating a discrete vascular tree approach that represents blood supply and drainage at the organ scale and a multi-compartment homogenized flow model that represents perfusion at the lower levels of the hierarchical tree network and the liver microcirculation, based on our prior work in \cite{RefEtienne,RefEtienneMurray,Jessen3}. 
In Section \ref{sec:Poroelasticity_growth}, we develop our framework by successively combining the multiscale perfusion model with an isotropic growth model of a poroelastic medium that represents hyperplasia of liver tissue at the microcirculation. We also discuss appropriate coupling mechanisms and boundary conditions, and motivate the driving force that initiates compensatory growth of liver tissue. We briefly illustrate the characteristics of each modeling step via a two-dimensional test problem. In Section \ref{sec:Numerical_examples_Liver}, we demonstrate the capabilities of our computational framework for a full-scale patient-specific liver example by correctly predicting the typical perfusion outcome associated with a common surgical cut pattern. Section \ref{sec:Conclusion} closes with
a summary and an outlook.

% Refer to poroelasticity here Within biomechanics,  for the lung \cite{RefLungTree} or for the heart \cite{Refheart1} have been developed.
%\section{Physiology}
%\section{Modeling liver regrowth}
\section{Physiological mechanisms of liver regrowth across different scales}
\label{sec:Physiology}

We start with a brief description of the multiscale anatomy of the liver, with a particular focus on its vasculature that determines liver perfusion. We then summarize the perfusion-related mechanisms at the microcirculation scale that drive regrowth. %We focus on the mechanisms from the sinusoid scale upwards, which are the ones relevant for the current modeling framework. 
As a starting point for our contribution, we motivate basic concepts how these physiological mechanisms at the microcirculation scale can be transferred into a computational modeling framework that represents liver regrowth at the organ scale.  %We close this section by providing a computational modeling framework for liver regrowth. 

\subsection{Multiscale vasculature and liver perfusion}
\label{sec:2.1}

\begin{figure}[ht]
    %\centering
    %\includegraphics[width=165mm]{Images/Liver_anatomy_new1.pdf}
 \includegraphics[width=1.0\textwidth]{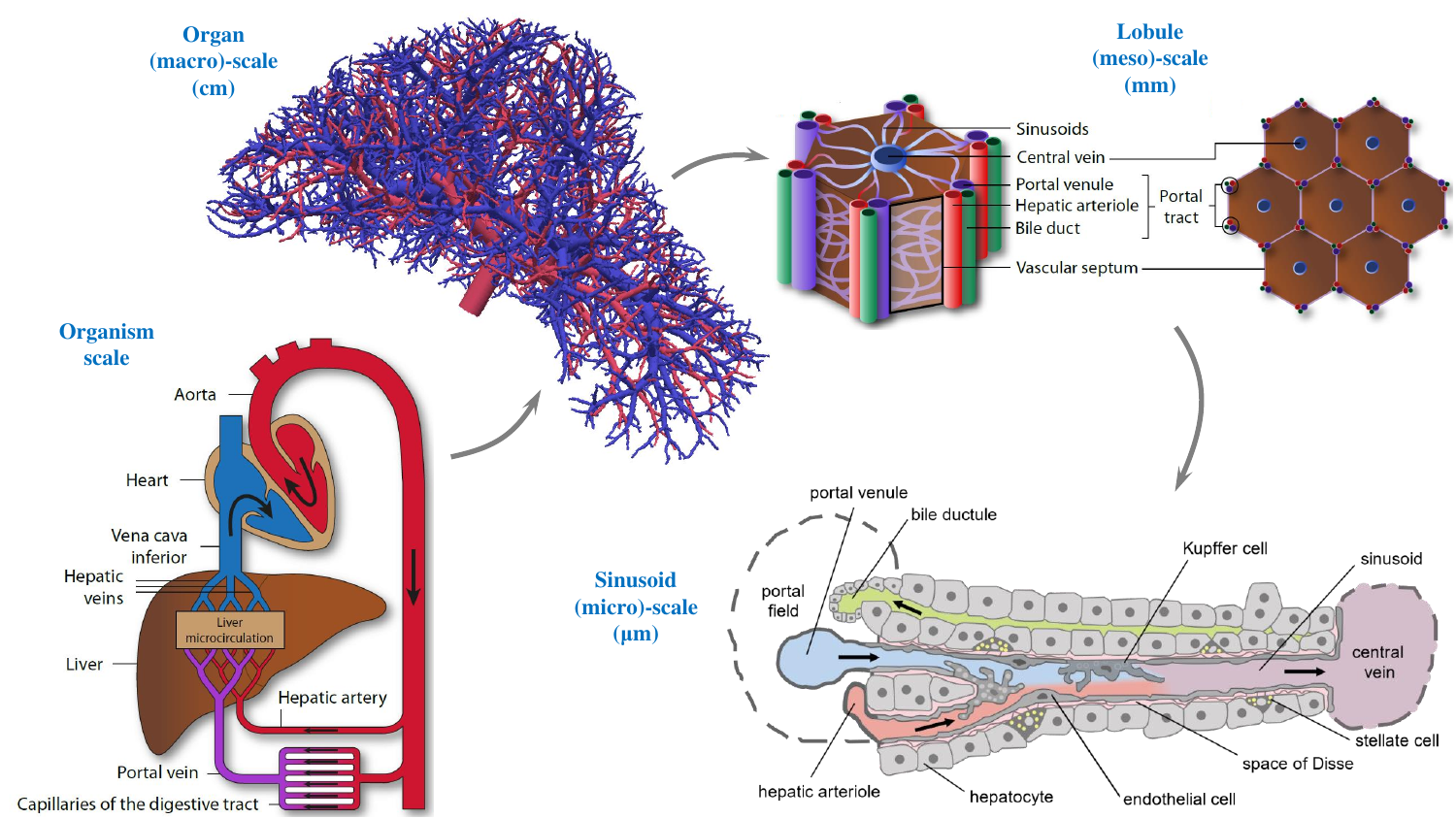}
    \caption{Multiscale liver vasculature (organism- and lobule-scale pictures from \cite{RefDebbautDiss}, sinusoid-scale picture from \cite{RefSinusoid}).}
    \label{fig:liveranatomy}
\end{figure}

The liver is primarily known as the main site of metabolization and detoxification of xenobiotics in the human body. However, the liver serves a multitude of further functions, such as combating infections; storing iron, vitamins and other essential chemicals; manufacturing, breaking down and regulating numerous hormones; or producing enzymes and proteins which are responsible for most chemical reactions in the body. All liver functions rely on its vasculature that enables the perfusion of the complete liver with blood \cite{lorente2020liver}. Figure \ref{fig:liveranatomy} illustrates the multiscale nature of liver vasculature. 

At the organ (or macro-) scale, the hepatic artery and the portal and hepatic veins with diameters of up to 1 cm provide blood supply and drainage. The hepatic artery originates from the heart and carries oxygen-rich blood, and the portal vein stems from the digestive tract and carries nutrient-rich blood. The hepatic vein re-inserts the blood back into the vena cava and the cardiovascular system. At the lobule (or meso-) scale, the blood is driven through sinusoids, small capillaries with a diameter of approximately 10 $\mu$m. The organ-scale vessels and the sinusoid microcirculation are connected by a hierarchical vascular tree that consists of up to 20 levels of bifurcations \cite{debbaut2014analyzing}. The sinusoids are uniformly distributed throughout the entire liver volume, forming a three-dimensional network around rows of hepatocytes that are responsible for the metabolic liver functions. The sinusoids are arranged in lobules with a characteristic size of 1.5 mm, forming the fundamental building blocks of the liver. Each lobule is classically idealized as a prismatic volume of hexagonal cross section with a supplying triad made of a hepatic artery, a portal vein and a bile duct at each of the six hexagon corners, and a draining central vein along the axis of the lobule.

At the sinusoid (or micro-) scale, the liver is composed of specialized cells and structures that work together to maintain its functions. Figure \ref{fig:liveranatomy} also illustrates the main functional units at the microscale. 

\subsection{Liver regeneration and driving mechanisms at the microscale}
\label{sec:2.2}
 
\begin{figure}[t]
\centering
\includegraphics[width=1.0\textwidth]{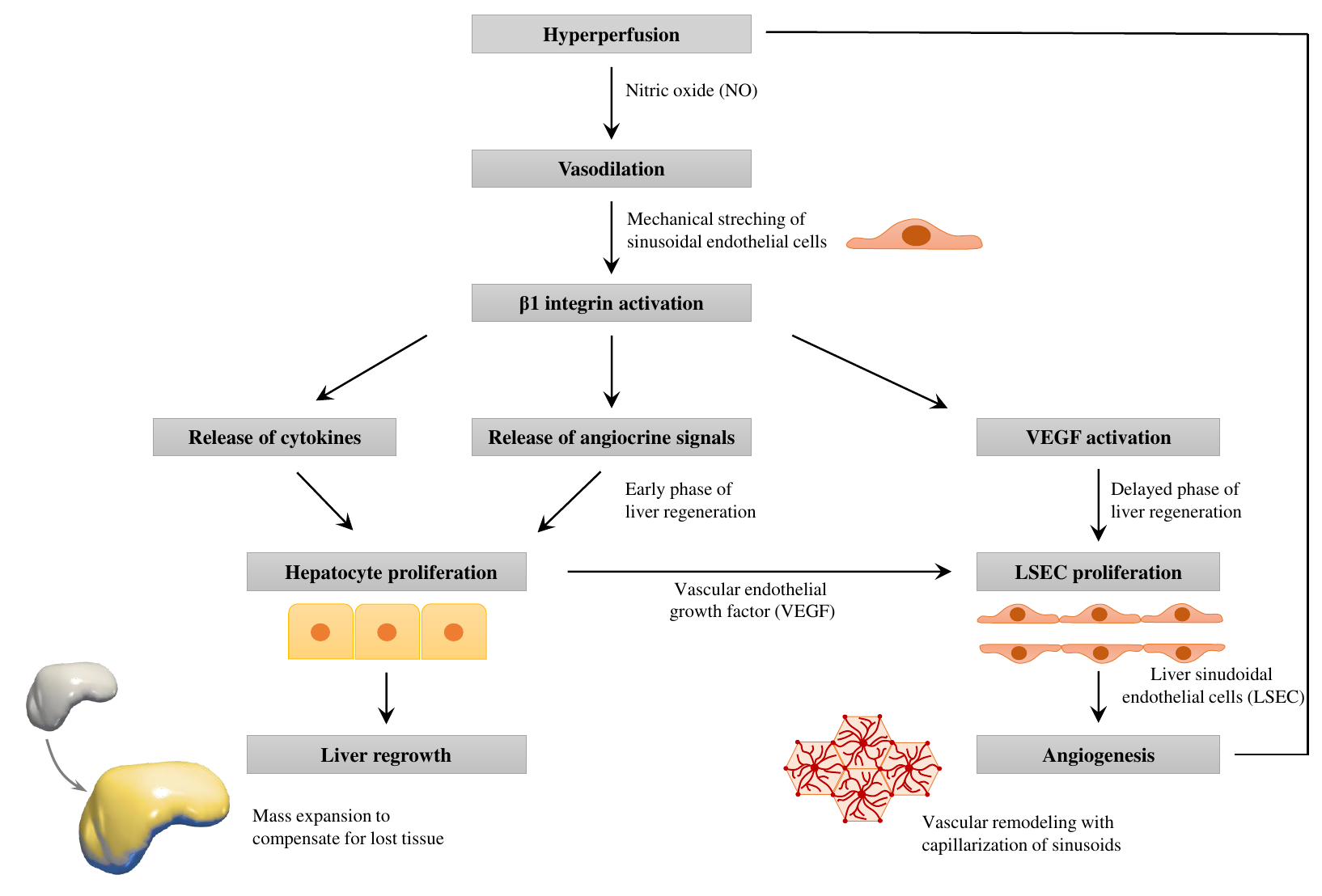}
    \caption{Liver size regulation after surgical resection (picture adapted from \cite{grosse2021role}).}
    \label{fig:drivingfactor}
\end{figure} 

The liver possesses the remarkable ability to regenerate itself after partial resection \cite{RefMichalopoulos,RefMichalopoulos2}. %It can reach its original size within a few weeks. 
Since blood from the portal vein cannot easily bypass the liver to return to the heart, almost the same volume of blood needs to flow through the smaller liver remnant after partial resection (on average, about 1.25 liters of blood per minute). As a consequence, massive hemodynamic changes occur immediately after partial resection. To enable the same blood flow through a reduced cross section, the portal pressure needs to increase (portal hypertension). In contrast to the portal vein, blood from the hepatic artery can continue to flow through the aorta to other parts of the gastrointestinal tract and legs. Therefore, the hepatic artery contracts, when the portal pressure increases (hepatic arterial buffer response).
As a result of hypertension, a larger difference in pressure occurs between the portal venules on the one hand and the central veins on the other hand. Therefore, the flow increases in the remaining vasculature, leading to a larger blood velocity and therefore increased shear stresses within the sinusoids. This state is called \textit{hyperperfusion}.

The restoration of liver mass primarily occurs through the rapid cell division of hepatocytes (proliferation) \cite{RefLiverSize}.
Hepatocytes are the predominant cells in liver tissue, comprising about 80\% of the liver mass. As illustrated in Fig.~\ref{fig:liveranatomy}, they are arranged in plates or cords, radiating outward from the central veins within lobules. Hepatocytes perform numerous essential tasks, including metabolism of carbohydrates, lipids, and proteins, and detoxification of drugs and toxins.
There is concensus that \textit{hyperperfusion} is the central stimulus that triggers liver regeneration. Within the biomedical literature, there are different models that connect hyperperfusion and hypatocyte proliferation at the cell level, see e.g.\ \cite{abshagen2012critical}. In this work, we focus on the recent model by Lorenz et al. \cite{lorenz2018mechanosensing,grosse2021role}, which is illustrated in Fig. \ref{fig:drivingfactor} and that we briefly outline in the following.

Immediately after partial resection, the increase in blood flow through the remaining sinusoids and the associated increase in shear stress at the vessel walls stimulates sinusoidal endothelial cells to release vasodilators such as nitric oxide (NO). Sinusoidal endothelial cells line the liver sinusoids and form a barrier between the blood and hepatocytes, see Fig.~\ref{fig:liveranatomy}. The dilation of the sinusoids enables a larger flow of blood at a given pressure level and therefore limits portal hypertension, blood velocity and wall shear stress. Vasodilation requires the mechanical stretching of the endothelial cell layer, resulting in the release of growth-promoting angiocrine signals and mechanoresponsive $\beta$1 integrin. Angiocrine signals constitute the main stimulus for hepatocyte proliferation, and hence are essential for liver regrowth. Integrin activation results in the reorganization of the components of the extracellular matrix, triggering the release of cytokines. Both integrin and cytokines contribute to the release of further growth factors to support hepatocyte proliferation. % that are stored in the ECM or localized at the plasma membrane

The proliferation of hepatocytes increases the volume of the remaining lobules (hyperplasia), where the hepatocyte plates are up to twice as thick as they initially are \cite{RefMichalopoulos}. Thus, the remaining liver expands in mass to compensate for the lost tissue (compensatory growth), but does not restore its original shape and the excised parts do not grow back \cite{RefNelson}. To guarantee liver function, the added hepatocyte cells need to be supplied with blood in the same way as before, which requires the formation of new vessels (angiogenesis), primarily at the level of the microcirculation.  
Angiogenesis is primarily based on the proliferation of sinusoidal endothelial cells. It is driven by vascular endothelial growth factor (VEGF) released by proliferating hepatocytes and the mechanically stretched endothelial cells themselves. The resulting vascular expansion process leads to the formation of an extended vascular network of sinusoids (lobular remodeling).

The rate of regeneration depends on several factors, including the extent of the resection, the general physical condition and health of the patient, and any preexisting liver diseases \cite{RefLiverSize,RefFurchtgott,RefKoniaris}. In general, hepatocyte proliferation starts immediately and heavily after partial liver resection (early phase of liver regeneration). Angiogenesis and lobular remodeling starts after a few days (delayed phase of liver regeneration). For instance, following a two-third removal, the normal liver weight is restored within 8 to 15 days in humans, followed by several weeks of slow lobular remodeling \cite{RefMichalopoulos}. 

Once the lost tissue mass and the vascular network is restored, hyperperfusion ceases, as enhanced blood flow and pressure is not required in the hepatic vasculature to guarantee the required volume of blood to pass through the liver. Therefore, all cell proliferation processes stop and liver regeneration is complete.

\begin{figure}[t!]
    %\centering
    %\includegraphics[width=160mm]{Images/Modeling_framework_new.pdf}
    %\includegraphics[width=170mm]{Images/Modeling_framework_new3.pdf}
    \includegraphics[width=\textwidth]{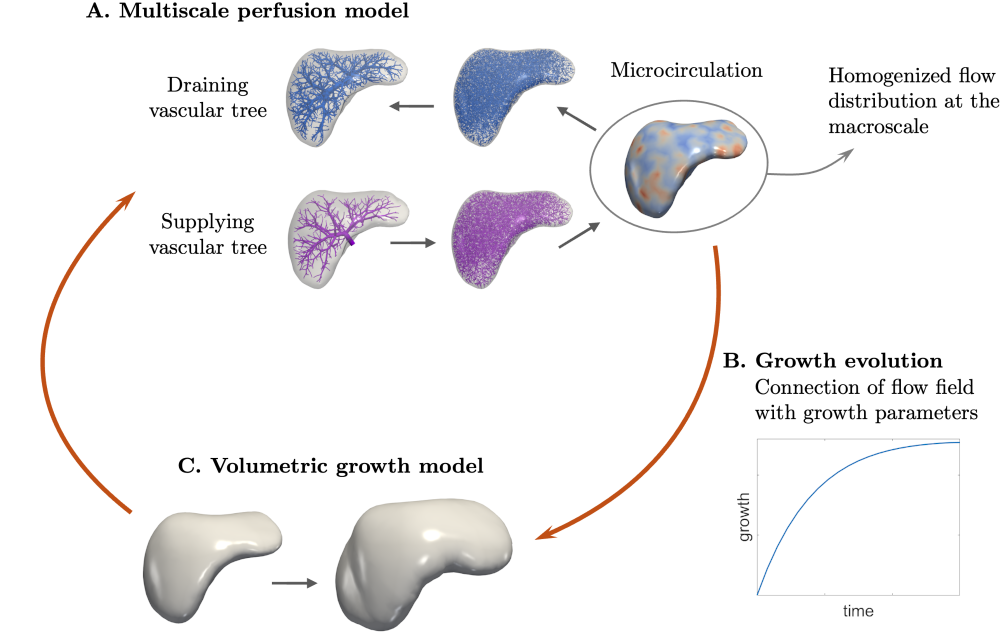}
    \caption{Concept for modeling liver regrowth.}
    \label{fig:Modelingframework}
\end{figure}

\subsection{Fundamental concepts for modeling hyperperfusion-driven liver regrowth}

Building upon our understanding of the multiscale driving mechanisms behind liver regrowth that we outlined above, we develop the following fundamental modeling concept, illustrated in Fig. \ref{fig:Modelingframework}: %The concept of heightened shear stresses in the sinusoids following resection underscores the need for simulating liver perfusion. Therefore, our modeling framework of liver regrowth consists of various compartments at different scales and is driven by a multiscale perfusion model: 

\begin{enumerate}[label=\Alph*.]

% Understand perfusion of the liver, separate in distribution, hierarchical vasculature, mesoscale flow model at the lobule scale
\item \textbf{Multiscale perfusion model:} 
As the key stimulus for liver regrowth is hyperperfusion in the microcirculation, understanding the current state of blood perfusion at the lobule scale is the key prerequisite for assessing liver regrowth. We therefore require a detailed perfusion model, based on suitable scale-bridging concepts, which is able to separate the overall liver perfusion into macro- and mesoscale blood distribution and collection, and microscale blood flow through the functional units of the liver at the lobule scale. We can then use the corresponding model representation of microscale blood flow to obtain a measure of hyperperfusion in a particular region of the liver. %the associated wall shear stress in the microcirculation, which is the primary trigger for hepatocyte proliferation.

\item \textbf{Growth evolution model:} Once a multiscale perfusion model is established that can assess hyperperfusion at the lobule scale at each point of the liver domain, we can set up a (phenomenological) growth evolution equation that relates an increased level of blood flow in the microcirculation to a growth rate of homogenized liver tissue, which represents the effect of hepatocyte proliferation. An important requirement is that the growth evolution equation is based on parameters that can be calibrated by available experimental data.
%The next step consists of connecting the flow field of the lobule scale with growth parameters of a volumetric growth model and, consequently, incorporating them into the growth evolution equation.

\item \textbf{Organ-scale growth model:} Once the growth evolution model for homogenized liver tissue is known, it can be integrated into an organ-scale growth model, for which well-established concepts exist in continuum poroelasticity. It represents macroscale growth, locally driven by the growth evolution equation, which depends on the local state of hyperperfusion in the microcirculation. We assume that lobular remodeling in the microcirculation takes effect immediately, such that macroscale growth of liver tissue also implies an increase in volume of the vasculature. In turn, growth of the liver tissue enlarges the microcirculatory domain of the perfusion model, such that hyperperfusion can be regulated at each point of the liver.

\end{enumerate}

In the following, we will motivate and describe in detail the technical aspects regarding our multiscale perfusion model as well as regarding our growth evolution and organ-scale growth models.

% A further important aspect is the patient-specific calibration of the model. In the current case, imaging data macroscale liver domain, entry points of the large vessels on macroscale vasculature 
% Goal of the model is to understand, starting from a resected liver geometry, whether the 
%A further aspect is patient-specific 
% Say something about patient-specific nature.

%In the following two sections, we will present the 

% Say something about which section we talk about what,

%The modeling objectives comprise several key aspects: understanding the evolution of fluid pressure within liver tissue and the distribution of blood supply, identifying potentially sensitive areas or regions prone to flow or pressure accumulation, predicting tissue growth, and assessing the eventual size and shape of the regenerated liver. Based on expected surgical tissue cuts, mentioned goals may aid in predicting surgical outcome and provide additional guidance for surgeons before performing a liver resection.

%\section{Coupling liver perfusion to the poroelastic-growth model}
\section{A multi-compartment perfusion model}
%\section{Multi-compartment poroelastic-growth model}
\label{sec:MultiCompartmentPerfusion}
%Increased perfusion, increased shear stress (assumptions), requires simulation of liver perfusion.
%Patient-specific modeling of surgical liver resection and liver tissue regrowth necessitates the adequate modeling of liver's hierarchical vasculature. For this purpose, 
In this section, we describe our multiscale perfusion model that corresponds to part A of our modeling framework illustrated in Fig. \ref{fig:Modelingframework}. It consists of two fundamental components: (a) a synthetic discrete model that represents the upper levels of the hierarchical vasculature for blood distribution and collection, and (b) a homogenized flow model that is divided into multiple compartments to individually represents the lower levels of the hierarchical vasculature for blood distribution and collection, and the microcirculation. % at the lobule scale. %The multicompartment concept ensures that the microscale blood flow 

%. Due to multiscale nature of the liver's vessel anatomy, we will then briefly present an approach where the vascular network is partitioned into a coupled multi-compartment model with several simplified Darcy-type flow models that account for the different spatial scales \cite{RefHeartPerfusion,RefReducedDarcy,RefHyde,RefLiverPerfusion}. \jh{Finally, the liver regrowth model is closed by introducing the poroelastic, finite growth model. For this, we provide the kinematics, balance laws, and constitutive equations. The idea is to extend the multi-compartment perfusion model by a poroelastic-growth model at compartment 2 where the interaction between tissue deformation and fluid flow is coupled to simulate the macroscopic volumetric growth determined by lobule enlargement at a lower scale. Additionally, we present our evolution equation of liver growth.}

\subsection{Discrete non-intersecting vascular trees for blood supply and drainage}
\label{sec:VascularTrees}
Our multiscale perfusion model necessitates a detailed description of the vascular trees inside the liver. 
Since in-vivo imaging methods are limited in resolution, we generate these trees synthetically, {\color{red}where it is also possible to incorporate patient-specific vessel data up to the resolution available.}
Our generation framework is based on our previous work \cite{RefEtienne,RefEtienneMurray}, which we recently extended to generate multiple non-intersecting trees inside the same perfusion domain \cite{Jessen3}.
%In the following, we review and summarize the mathematical formulation and algorithmic framework.

\subsubsection{Mathematical formulation}
A discrete vascular tree is represented as a directed graph $\Tree = \left(\Nodes, \Arcs\right)$ with nodes $u \in \Nodes$ and segments $a \in \Arcs$.
Each segment $a = uv$ of $\Arcs$ is defined by the geometric locations of its proximal and distal node $x_u$ and $x_v$, its length $\ell_a = ||x_u - x_v||$, volumetric flow $\hat
{Q}_a$ and radius $r_a$. 
Thus, a $\emph{vessel}$ is simplified to a rigid and straight cylindrical tube.
A tree has a single \emph{root} node $0$ and (multiple) \emph{leaves} $v \in \Leaves$, which are the distal nodes of terminal segments.
We assume blood to be incompressible, homogeneous and a Newtonian fluid in the laminar regime.
The pressure drop over a segment is described by Poiseuille's law with
\begin{equation}
  \Delta p_a = R_a \hat
{Q}_a \quad\forall a \in \Arcs,
\end{equation}
where the hydrodynamic resistance $R_a$ of each segment $a$ is 
\begin{equation}
  R_a = \frac{8 \eta}{\pi} \frac{\ell_a}{r_a^4} \quad\forall a \in \Arcs.
\end{equation}
We set the dynamic blood viscosity $\eta$ to a (constant) value of $\SI{3.6}{cP}$ and further assume a homogeneous flow distribution of our (given) root flow $\hat{Q}_\text{perf}$ to all $N$ leaves, leading to the terminal flow $\hat{Q}_\text{term} = \hat{Q}_\text{perf}/N$.
At intermediate branching nodes, the flow can be computed using Kirchhoff's law with
\begin{equation}\label{eq:kirchhoff-flow}
\hat{Q}_{uv} = \sum_{vw \in \Arcs}  \hat{Q}_{vw} \quad\forall v \in \Nodes \setminus \left({0} \cup \Leaves\right).
\end{equation}
We note that more complex viscosity laws and flow distributions to take into account non-Newtonian behavior of the blood such as the F\aa{}hr\ae{}us Lindqvist effect can be also incorporated into the framework \cite{RefEtienneMurray}.

The tree can be based on a combination of different goal functions and constraints \cite{RefEtienneMurray}.
Here, we choose to minimize the total power of the tree, which consists of the power to maintain blood inside the vessels $P_\text{vol}$ and the (viscous) power to move blood through vessels $P_\text{vis}$. 
The cost function for the vascular tree thus becomes
\begin{equation}\label{eq:tree-power}
    f_\Tree = P_\text{vol} + P_\text{vis} = \sum_{a \in \Arcs} m_b\pi\ell_a r_a^2 +\frac{8 \eta}{\pi} \frac{\ell_a}{r_a^4}\hat{Q}_a^2,
\end{equation}
where $m_b$ is the metabolic demand factor of blood, which we set to \SI{0.6}{\uW\per\cubic\mm} \cite{liu2007vascular}.
Since we precompute all flow values using \eqref{eq:kirchhoff-flow}, we can rewrite \eqref{eq:tree-power} to 
\begin{align}\label{eq:weighted-length-problem}
    f_\Tree &= \sum_{a \in \Arcs} w_a\ell_a,
\end{align}
where weight $w_a$ is defined at each segment $a$ with
\begin{align}
    w_a = m_b\pi r_a^2 +\frac{8 \eta}{\pi r_a^4}\hat{Q}_a^2.
\end{align}
Our formulation does not include global constraints between nodes and each summand in \eqref{eq:weighted-length-problem} is decoupled. 
As shown in \cite{Jessen3}, the (optimal) radius of each segment $a$ can then be independently computed with
\begin{align}\label{eq:radius-murray}
    r_a = \sqrt[6]{\frac{16\eta}{m_b\pi^2}} \sqrt[3]{\hat{Q}_a}.
\end{align}
The problem of finding the optimal geometry of a tree now only consists of finding the optimal nodal positions $x$ and corresponding lengths $\ell$.

We achieve the optimality requirement for the geometry of a single tree by optimizing the global geometry (position of all branch nodes) using a nonlinear optimization problem (NLP) \cite{bertsekas1997nonlinear,boyd2004convex}.
We include the nodal positions $x$ and the lengths $\ell$ of all segments inside the vector of optimization variables $y = (x, \ell)$.
With physical lower bounds $\ell^-$ and numerical upper bounds $\ell^+$, the best geometry is found in 
\begin{equation}
    Y = \mathbb{R}^{3|\Nodes|} \times [\ell^-, \ell^+]^\Arcs.
\end{equation}
%and our NLP reads:
%\begin{align}\label{eq:nlp-murray}
%  \min_{y \in Y} \quad
%  & \sum_{a \in \Arcs} w_a\ell_a,\\
%  \text{s.t.}\quad
%  \label{eq:nlp-fix-x}
%  &0 = x_u - \bar{x}_u, & u &\in \Nodes_0 \cup \Leaves,\\
%  \label{eq:nlp-length}
%  &0 = \ell_{uv}^2 - ||x_u - x_v||^2, & uv &\in \Arcs.
%\end{align}
%where \eqref{eq:nlp-fix-x} fixes the position of terminal nodes and \eqref{eq:nlp-length} ensures consistency between nodal positions and segment lengths.

For the liver, however, we require the combination of several trees at the same time. It is generally accepted that the hepatic artery and portal vein can be combined to a single supplying tree, as their vessels are mostly aligned with each other after a few generations, see \cite{RefDebbaut3}.
Therefore, our objective is to generate one supplying tree (hepatic artery and portal vein) and one draining tree (hepatic vein), which do not intersect and are optimal both in topology and geometry in regards to \eqref{eq:weighted-length-problem} .
To ensure that our supplying tree $\Tree^1$ and draining tree $\Tree^2$ do not intersect, we need to introduce coupling constraints between both trees \cite{Jessen3}. %, this involves two main steps.
We first introduce a set of (virtual) connections $\Arcs^{12}$ between neighboring nodes of both trees, defined with
\begin{align}
    \Arcs^{12} = \lbrace (v_1,v_2) |\ \ v_1 \in \Nodes^1,\ v_2 \in \Nodes^2,\ ||x_{v_1} - x_{v_2}|| < 1.25 (r_{{u_1v_1}}+ r_{u_2v_2}) \rbrace.
\end{align}
We then introduce a set of \emph{excursion nodes} $\Excursions$, which includes all nodes with exactly one proximal and one distal node.
These nodes are added to each set of intersecting vessels \cite{Jessen3}.
Consequently, the set of optimization variables $y^{12} = (y^1, y^2)$ now consists of both trees and the best geometry is found in
\begin{align}
    Y^{12} = Y^1 \times Y^2 \times \Arcs^{12}.
\end{align}
Our extended NLP then reads
\begin{align}\label{eq:nlp-coupled}
  \min_{y^{12} \in Y^{12}} \quad
  & \sum_{i = 1}^2  \sum_{a_i \in \Arcs^i} \ell_{a_i} w_{a_i}\\
  \stq
  \label{eq:nlp-murray_1}
  &0 = x_{u_1} - \_x_{u_1}, & u_1 &\in \Nodes_0^1 \cup \Leaves^1\\
  &0 = x_{u_2} - \_x_{u_2}, & u_2 &\in \Nodes_0^2 \cup \Leaves^2\\
  &0 = \ell_{u_1v_1}^2 - \norm{x_{u_1} - x_{v_1}}^2, & u_1v_1 &\in \Arcs^1\\
  \label{eq:nlp-murray_2}
  &0 = \ell_{u_2v_2}^2 - \norm{x_{u_2} - x_{v_2}}^2, & u_2v_2 &\in \Arcs^2\\
  \label{eq:nlp-coupling_1}
  &0 = \ell_{v_1v_2}^2 - \norm{x_{v_1} - x_{v_2}}^2, & v_1v_2 &\in \Arcs^{12}\\
  \label{eq:nlp-coupling_2}
  &\ell_{v_1v_2} > (r_{{u_1}{v_1}} + r_{{u_2}{v_2}}) + \epsilon, & v_1v_2 &\in \Arcs^{12}\\
  \label{eq:nlp-excursion_1}
  &\ell_{u_1v_1} > \_\ell_{u_1v_1}, & v_1 &\in \Excursions^1\\
  &\ell_{v_1w_1} > \_\ell_{v_1w_1}, & v_1 &\in \Excursions^1\\
  &\ell_{u_2v_2} > \_\ell_{u_2v_2}, & v_2 &\in \Excursions^2\\
  \label{eq:nlp-excursion_2}
  &\ell_{v_2w_2} > \_\ell_{v_2w_2}, & v_2 &\in \Excursions^2.
\end{align}
Here, \eqref{eq:nlp-coupling_1} and \eqref{eq:nlp-coupling_2} ensure that the distance between two nodes (of tree 1 and 2) is at least their vessel radii plus a threshold $\epsilon$.
By using \eqref{eq:nlp-excursion_1}-\eqref{eq:nlp-excursion_2}, we prohibit excursion nodes to move along the path between their proximal and distal nodes.
After the extended NLP is solved, both trees are checked for intersections.
New excursion nodes are subsequently created at newly identified intersections.
The process of adding excursions and solving the extended NLP is repeated until no further intersections are found. We note that it is straightforward to extend this formulation and procedure to more than two trees \cite{Jessen3}.

\begin{figure}[t!]
    \centering
    \includegraphics[trim=0 0 0 0, clip=true,width=0.726\textwidth]{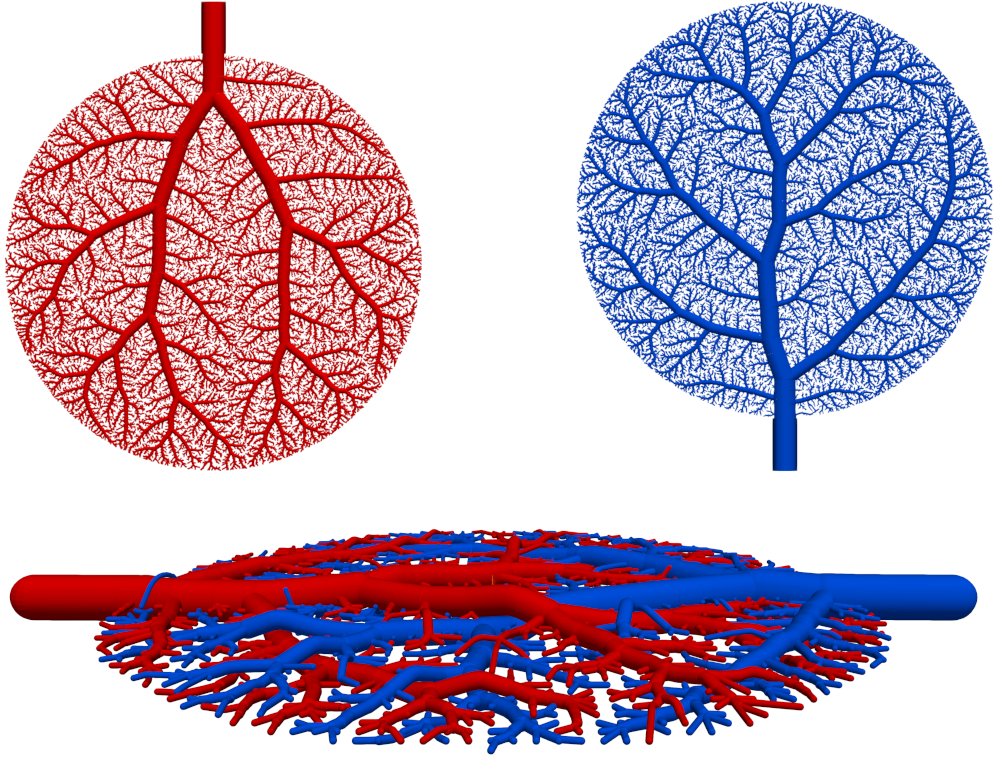}
    \caption{Model problem: supplying (in red) and draining (in blue) vascular trees.}
    \label{fig:tree_2D}
\end{figure}

\subsubsection{Algorithmic framework and computer implementation}
We use the framework described in \cite{RefEtienne, Jessen3} to generate each tree with a (locally) optimal topology.
First, we sample $N$ terminal nodes $\bar{x}$ for each tree inside the (non-convex) perfusion volume (liver) and connect them to the manually set root positions.
From these initial (fan) shapes, new topologies are explored by swapping segments. 
Each \emph{swap} changes the parent nodes between a sampled pair of nodes.
The new topology is accepted heuristically based on a simulated annealing (SA) approach \cite{RefEtienne, Jessen3}. % with probability
%\begin{align}
%    P = \exp{\left(\frac{-\Delta f_\Tree}{T}\right)}.
%\end{align}
%$\Delta f_\Tree$ is the change in cost induced by the swap and $T$ is the SA temperature, which is decreased ('cooled down') after each iteration.
After a fixed number of swaps, the global geometry is optimized and intersections are resolved as explained.
Once the trees are generated, we fix all nodal positions, so that we can retrieve the length $\ell_a$ for each segment $a$. For both trees, we prescribe a constant pressure drop $\Delta p$ between all terminal nodes and the root node, so that we can compute the radius $r_a$, the volumetric flow $\hat{Q}_a$ and further parameters such as the mean blood velocity $\bar{v}_a = \frac{\hat{Q}_a}{\pi r_a^2}$.

For illustration purposes, we consider a circular domain with a radius of 10 mm. We choose the root positions of the supplying and draining trees at two opposite top and bottom points of the circle, generate corresponding groups of (randomly distributed) terminal points within the circle, and choose a pressure difference at the roots and a root flow rate of $\hat{Q}_\text{perf}$. Table \ref{tab_vesseltrees2D} lists key parameters. % that characterize the tree structure. 
Figure \ref{fig:tree_2D} illustrates the resulting supplying and draining trees. We observe that %although both sets of terminal points are located within the planar circular domain, 
the two trees are not intersecting each other, using the third dimension to move their vessel segments around each other. 

{\color{red}We carried out first steps to validate our synthetically generated tress structures via a comprehensive comparison with experimentally characterized vessel networks for a human liver, with very good results \cite{RefEtienne,RefEtienneMurray}.}

\begin{table}[t]
%\centering
\caption{Parameters of the discrete vascular structure generated on a circular domain.}
\label{tab_vesseltrees2D}
\begin{center}
\begin{tabular}{l l l l l l l l} 
 \hline
  & $N_\text{seg}$ & $N_\text{term}$ & $p_\text{root}$ [$\frac{\text{kg}}{\text{mm} \, \text{s}^2}$] & $\hat{Q}_\text{perf}$ [$\frac{\text{mm}^3}{\text{s}}$] & $m_b$ [$\frac{\mu \text{W}}{\text{mm}^3}$] & $\eta$ [$\frac{\text{kg}}{\text{mm}\,\text{s}}$]\\ [0.5ex] 
 \hline
 Supplying tree & 43,981 & 21,991 & 0.4 & 80.0 & 0.6 & $3.6 \times 10^{-6}$ \\ 
 %\hline
 %\hline
 Draining tree & 43,857 & 21,929 & 0.0  & 80.0 & 0.6 & $3.6 \times 10^{-6}$ \\ [1ex] 
 \hline
\end{tabular}
\end{center}
\end{table}

%The two-dimensional simulation is not meant to represent the liver tissue itself; rather, its objective is to examine the model's behavior and sensitivity to model parameters. All important information regarding the network flow model of the vessel trees such as number of vessels $N_\text{vessel}$, number of terminal vessels $N_\text{term}$, the root pressure $p_\text{root}$ and the root flow $\hat{Q}_\text{perf}$ are summarized in 

%\subsection{Multi-compartment approach}
\subsection{Homogenization of blood flow through hierarchical vascular networks}
\label{sec:MultiCompartment}

According to Figure \ref{fig:liveranatomy} and the associated discussion in Section \ref{sec:Physiology}, the characteristic size of vessel diameters across the hierarchy of the supplying and draining trees ranges from several millimeters for the macroscale arteries and veins to 100 micrometers for the smallest arterioles and venules \cite{RefDebbautDiss}. Direct flow modeling on the vascular tree geometry provided by the discrete model presented above is possible, see e.g.\ the technologies reviewed in \cite{formaggia2010cardiovascular}, but requires a significant computing effort, especially within an iterative solution procedure. To balance resolution accuracy and computational effort, %we follow a different approach that still takes into account the hierarchically complex structure of the tissue and vessels. To this end, 
we use the lower hierarchies of the discrete vascular trees {\color{red}as the basis for parameter estimation of} a porous medium, where flow can be homogenized in a continuum sense. To enable us to separately represent homogenized flow in different parts of the vascular tree, we compartmentalize the vasculature into spatially co-existing compartments and associate only those vessels to each compartment that belong to the selected (supplying or draining) tree and the selected range of spatial scales \cite{RefHeartPerfusion,RefReducedDarcy,RefHyde,RefLiverPerfusion}. 

%Each compartment is associated with a particular range of length scales and characterized by an intracompartment permeability and intercompartment coupling term for mass exchange. Essential parameters are determined for each compartment using averaging procedures. The interaction between the different compartments is considered via the pressure-dependent mass exchange and is applied in an averaged sense. 

\subsubsection{Multi-compartment homogenized flow equations}
To set up the model, we divide the discrete vessel structure (or parts of it) into $N$ compartments, assigning to each compartment $i = 1,...,N$ its own (positive definite) permeability tensor $\mathbf{K}_i$, pore pressure $p_i$, source term $\theta_i$, and homogenized flow velocity $\mathbf{w}_i$. We consider the system of first-order differential equations that is governed by Darcy's law and the continuity equation
\begin{subequations}\label{eqn:full_multi_compartment} 
    \begin{align}
    \mathbf{w}_i + \mathbf{K}_i \nabla p_i &= \mathbf{0}   \; \; \;\; \, \text{in} \; \Omega, \label{eqn:full_multi_compartment1} \\
    \nabla \cdot \mathbf{w}_i + q_{i} &= \theta_i \; \;\; \, \text{in} \; \Omega,
    \label{eqn:full_multi_compartment2}
\end{align}
\end{subequations}
in each compartment $i$, in an open bounded region $\Omega \subset \mathcal{R}^d$, with space dimension $d$ and impermeable outer boundary $\Gamma$. The quantities $q_{i}$ denote the pressure-dependent intercompartmental flow rate density between compartment $i$ and all other compartments and are given by:
\begin{align}
    q_{i} = \sum_{k=1}^N \beta_{i,k}(p_i-p_k),
\end{align}
where $\beta_{i,k}\geq 0$ denotes the perfusion coefficient for coupling compartments $i$ and $k$ for $i \ne k$. As such, fluid exchange is absent in pressure equilibrium. We assume $\beta_{i,k} = \beta_{k,i}$ which is consistent with the mass balance $\sum_i q_i = 0$. Due to the assumption of an incompressible fluid, we use the terms mass flow and volumetric flow interchangeably.

Substituting \eqref{eqn:full_multi_compartment1} into \eqref{eqn:full_multi_compartment2} converts the first-order system into an equivalent system of one second-order differential equation per compartment:
\begin{align}
    -\nabla \cdot (\mathbf{K}_i \nabla p_i) + q_{i} &= \theta_i \; \;\; \, \text{in} \; \Omega.
    \label{eqn:reduced_multi_compartment1}
\end{align}
Here, the pressure $p_i$ is the sole unknown variable, and %This reduced formulation permits evaluating the compartment pressures $p_i, i = 1, ..., N$ and subsequently determining t
the compartment velocities $\mathbf{w}_i, i=1, ..., N$ follow from Darcy's law in \ref{eqn:full_multi_compartment1}. We supplement the model with a Neumann boundary condition to account for the impermeability: 
\begin{align}\label{eqn: bnd condition}
    -(\mathbf{K}_i \nabla p_i) \cdot \mathbf{n} = 0 \; \;\; \, \text{on} \; \Gamma_N = \Gamma.
\end{align}

\subsubsection{{\color{red}Estimation of} permeability tensors and perfusion coefficients}\label{sec:Modelparameters} %Calibrating permeability tensors from the vessel network
In the next step, we describe the computation of the model parameters directly from the vessel network, particularly the permeability tensors $\mathbf{K}_i$ for each compartment and the intercompartmental perfusion coefficients $\beta_\text{i,k}$, using averaging procedures \cite{RefHyde,RefHuyghe,RefHuyghe2,RefHuygheSchreiner}.  
In the following, the index set $\mathbb{K}_{i}(\mathbf{x})$ contains all indices of the vessels of compartment $i$ that are located at least partially within an averaging volume (AV) of spatial position $\mathbf{x}$. Furthermore, the index set  $\mathbb{I}_{ik}$ contains all indices of the vessels within the AV that belong to compartment $i$ but share a node with one or more vessels of compartment $k$. 
% As a consequence of the hierarchical structure of the compartments without skipped compartments, it follows that $\mathbb{I}_{i,k}  = \emptyset$ if $|i-k|\ne 1$ and $\mathbb{I}_{i,k} \subseteq \mathbb{K}_{i}(\mathbf{x}) $ if $|i-k|= 1$.

\begin{figure}[ht]
    \centering
    \includegraphics[width = 0.86\linewidth]{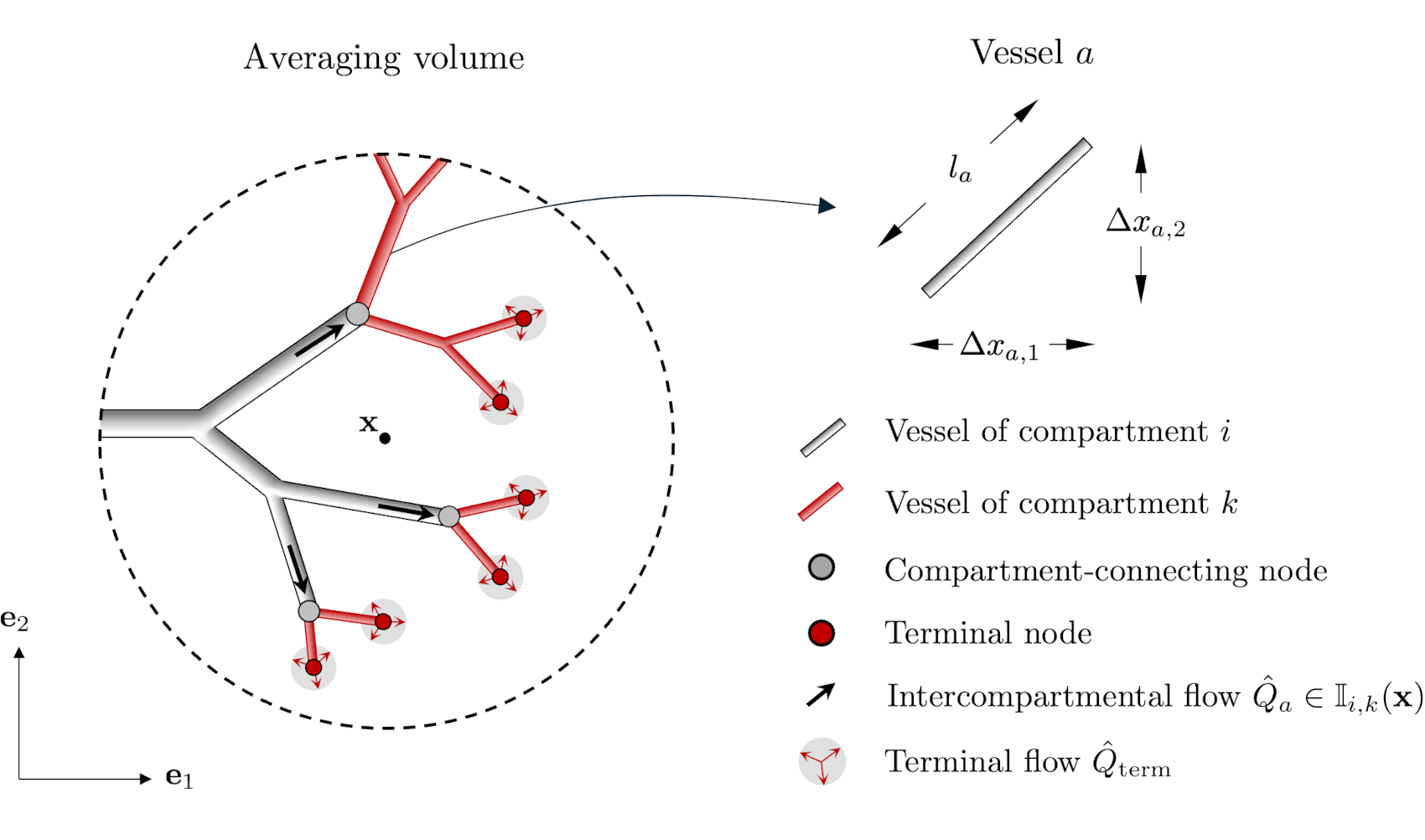}
    \caption{Averaging volume at spatial position $\mathbf{x}$.}
    \label{fig:AveragingVolume}
\end{figure}

The components of the permeability tensor at spatial position $\mathbf{x}$ for a network of straight rigid tubes subject to Poiseuille's law can be computed according to
\begin{align}
    K_{IJ}(\mathbf{x}) = \frac{\pi}{8\eta V_{\text{AV}}}\sum_{a\in\mathbb{K}_{i}(\mathbf{x})} \frac{(r_a)^4\Delta x_{a,I} \Delta x_{a,J}}{l_a}, \qquad I,J = 1,2,3
\end{align}
where $V_{\text{AV}}$ denotes the volume of the AV, $\eta$ the dynamic viscosity, $r_a$ the vessel radius, $l_a$ the vessel length, and $\Delta x_{a,i}$ and $\Delta x_{a,j}$ the components of the spatial vessel segment vector of vessel $a$ (see Fig. \ref{fig:AveragingVolume}). Note that only the segment of the vessel that actually intersects the AV is considered for $\Delta x_{a}$ and $l_a$.

Unlike the permeability tensor, which in combination with a given (and constant) dynamic viscosity can be determined solely based on geometry, the perfusion coefficients require flow quantities \cite{RefHyde}. The synthetically generated vascular trees assume Poiseuille's flow, providing the flow in each vessel segment and the pressure at each node. Using the same AV as above, we obtain the perfusion coefficient via
\begin{equation}\label{eq:perfusion_coefficient}
    \beta_{i,k}=
    \begin{cases}
      \frac{\overline{q}_{i,k}(\mathbf{x})}{|\overline{p}_i(\mathbf{x})-\overline{p}_k(\mathbf{x})|} & \text{if} \;\; |\overline{p}_i(\mathbf{x})-\overline{p}_k(\mathbf{x})|\neq 0 ,\\
      0 & \text{else},
    \end{cases}
\end{equation}
where $\overline{q}_{i,k}(\mathbf{x})$ is the bulk-volume-average for the intercompartmental flow rate density and can be obtained by division with the volume of the AV
\begin{align}
    \overline{q}_{i,k}(\mathbf{x}) = \frac{1}{V_\text{AV}}\sum_{a\in\mathbb{I}_{i,k}(\mathbf{x})}\hat{Q}_a,
\end{align}
where $\overline{p}_i(\mathbf{x})$ is the real-volume-average of the pressure in the compartment $i$. For straight vessel segments with constant diameter, $\overline{p}_i$ can generally be determined by
\begin{align}
    \overline{p}_i(\mathbf{x}) = \frac{\sum_{a\in \mathbb{K}_i(\mathbf{x})} p_a V_a}{\sum_{a\in \mathbb{K}_i(\mathbf{x})} V_a},
\end{align}
where $V_a = \pi d_a^2 l_a / 4$ is the volume of the vessel $a$, and $p_a$ represents the average pressure within the vessel. Since the pressure in the vessel segment varies linearly according to Poiseuille's flow, the average pressure can be determined simply as
%\begin{align}
    $p_a = \left(p_u + p_v\right) / 2$ from the known pressures
%\end{align}
$p_u$ and $p_v$ at the proximal and distal node, respectively.
We note that we consider all vessel segments whose end nodes are within the AV of the spatial point $\mathbf{x}$ in their entire length, % To circumvent the need to determine the mean pressure of a segment section, we assume that this is true 
even if the starting point of the vessel segments lies outside this volume.
For computing the perfusion coefficients coupling the lowest hierarchy of the vascular trees with the microcirculation we use the constant and prescribed reference pressure $\overline{p}_{\text{\tiny{micro}}} = p_{\text{root}} - \Delta p$.

\begin{remark}
    The case $|\overline{p}_i(\mathbf{x})-\overline{p}_k(\mathbf{x})|= 0$ in \eqref{eq:perfusion_coefficient} is not relevant for a hierarchically structured network as the pressure decreases continuously from the root segment to the terminal nodes within the supplying tree, as well as from the terminal nodes back to the root segment within the draining tree.
\end{remark} 

%\subsubsection{Resolved discrete vessels vs.\ homogenized compartments}
\subsubsection{Compartmentalization strategy}
Homogenization relies on the separation of scales, see e.g.\ \cite{hill1963elastic,hornung2012homogenization,fish2013practical}.
 %the existence of a representative averaging volume that is statistically homogeneous from a macroscopic point of view, is not unique in its choice and is selected based on the assumption of a periodic microstructure
%Such an averaging volume exists if the following separation of scales is satisfied:
%\begin{align}
%    r \ll l \ll L,
%    \label{eqn:scaleseparation}
%\end{align}
%where $r$, $l$, and $L$ denote the characteristic lengths of the microstructure of the porous medium, the averaging volume, and the macroscopic process under consideration, respectively.
Scale separation cannot be guaranteed for all vessels of the vascular tree structure. 
We assume that vessel segments with radii below or equal to an appropriate threshold of $r_{\textit{thresh}}$ are suitable for homogenization. Therefore, we categorize the vessels $a \in \mathbb{A}$ of the vascular tree $\mathbb{T}(\mathbb{V},\mathbb{A})$ in two groups: the lower hierarchies $\mathbb{A}_{\textit{lower}}$, suitable for homogenization, and the upper hierarchies $\mathbb{A}_{\textit{upper}}$, unsuitable for homogenization and which therefore must remain as a network of resolved vessels. We define
\begin{align}
    \mathbb{A}_{\textit{upper}} &= \{ a\in\mathbb{A}:\, r_a > r_{\textit{thresh}} \}, \quad \mathbb{A}_{\textit{upper}}\subseteq \mathbb{A},\\
    \mathbb{A}_{\textit{lower}} &= \{ a\in\mathbb{A}:\, r_a \le r_{\textit{thresh}} \} ,\quad \mathbb{A}_{\textit{lower}} \subseteq \mathbb{A},
\end{align}
%We choose a threshold with respect to the vessel radius, which we denote as $r_{\textit{thresh}}$. 
% all vessels with radii above the threshold need to remain in a vessel network. 
%For the current model problem, we choose a threshold of $r_{\textit{thresh}} = 0.1 \text{mm}$. 
where we can also define the respective subsets of nodes
\begin{align}
    \mathbb{V}_{\textit{upper}} &= \{ u,v\in a=uv:\, a \in \mathbb{A}_{\textit{upper}}\},\quad \mathbb{V}_{\textit{upper}}\subseteq \mathbb{V}, \\
    \mathbb{V}_{\textit{lower}} &= \{ u,v\in a=uv:\, a \in \mathbb{A}_{\textit{lower}}\},\quad \mathbb{V}_{\textit{lower}}\subseteq \mathbb{V}.
\end{align}
With that, we introduce the sets
\begin{align}
    \mathbb{V}_{\textit{conn}} &= \mathbb{V}_{\textit{upper}} \cap \mathbb{V}_{\textit{lower}},\\
    \mathbb{A}_{\textit{conn}} &= \{ a = uv \in \mathbb{A}_{\text{\tiny{lower}}}:\, u \in \mathbb{V}_{\textit{conn}}\}, 
\end{align}
where $\mathbb{V}_{\textit{conn}}$ denotes the set of nodes connecting the upper with the lower hierarchies and $\mathbb{A}_{\textit{conn}}$ denotes the set of vessels of the lower hierarchies with proximal node in $\mathbb{V}_{\textit{conn}}$. %connecting to the upper hierarchies.

\begin{figure}[t!]
    \centering
    \includegraphics[width=\textwidth]{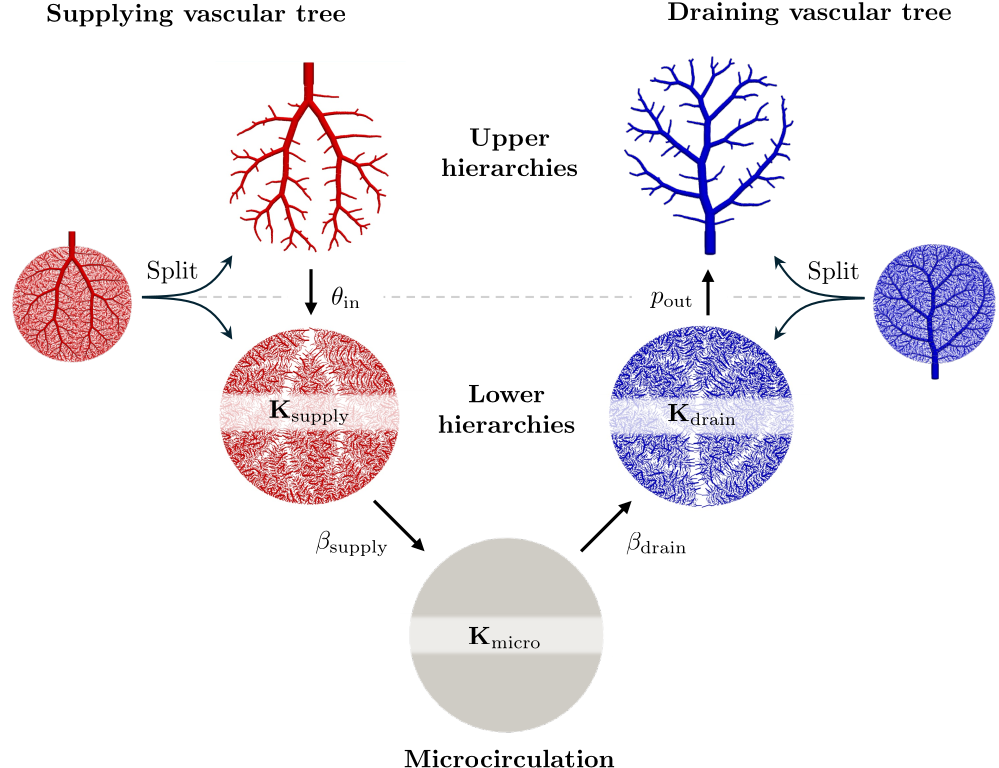}
    \caption{Compartmentalization of vascular tree structure.}
    \label{fig:Multicompartment}
\end{figure}

Note that $\mathbb{A}_{\textit{lower}}$ and $\mathbb{V}_{\textit{lower}}$ can be further divided into several compartments. In the scope of the present study, we will focus on the following simple compartmentalization strategy, illustrated in Fig. \ref{fig:Multicompartment} for the model problem of the circular domain: One compartment for each of the lower hierarchies of the respective vascular tree (compartment supply and compartment drainage) and the microcirculatory compartment: labeled \textit{supply}, \textit{micro} and \textit{drain}.

Due to the symmetry of the perfusion coefficient and the absence of intercompartmental flow between compartment supply and drainage, i.e. $\beta_{1,3} = 0$, two perfusion coefficients remain and we write $\beta_{1,2} = \beta_{\text{\tiny{supply}}}$ and $\beta_{2,3} = \beta_{\text{\tiny{drain}}}$. The permeability tensors of compartments supply and drainage as well as the perfusion coefficients $\beta_{\text{\tiny{supply}}}$ and $\beta_{\text{\tiny{drain}}}$ are determined using the averaging procedure described in Section \ref{sec:Modelparameters}. 

%When modeling perfusion at the lobular (meso-)scale, see Fig. \ref{fig:liveranatomy},  %anisotropy tends to average out across multiple lobules, yielding an effectively isotropic behaviour. This homogenization is a necessary simplification for mesoscopic modeling. 
%It is important to note that %that the resulting microcirculatory velocities predicted by the model are not physiologically accurate. Instead, the 
For the microcirculation, we use the isotropic permeability $K_{\text{\tiny{micro}}}$.
The compartment microcirculation of our model and its permeability represents the resistance of the sinusoid network at the microscale, see the illustration in Fig. \ref{fig:liveranatomy}. It cannot represent the flow patterns in the lobular structures at the mesoscale, as for instance the model in \cite{ricken2010remodeling}. In this sense, our model can be interpreted as the averaged flow redistribution across the microcirculation, which occurs through the network of the smallest-scale venules and arterioles.  %, which deviates from physiological reality, as mesoscopic perfusion redistribution primarily occurs within the vascular tree compartments.}

% \begin{subequations}
% \begin{align}
%     -\nabla \cdot (\mathbf{K}_{\text{\tiny{in}}} \nabla p_{\text{\tiny{in}}}) + \beta_{\text{\tiny{in}}} (p_{\text{\tiny{in}}} - p_{\text{\tiny{micro}}}) &= \theta_{\text{in}} \; \;\; \, \text{in} \; \Omega, \label{eqn:reduced_multi_compartment1a}\\
%     -\nabla \cdot (\mathbf{K}_{\text{\tiny{micro}}} \nabla p_{\text{\tiny{micro}}}) + \beta_{\text{\tiny{in}}} (p_{\text{\tiny{micro}}} - p_{\text{\tiny{in}}}) + \beta_{\text{\tiny{out}}} (p_{\text{\tiny{micro}}} - p_{\text{\tiny{out}}}) &= 0 \; \;\; \, \text{in} \; \Omega, \label{eqn:reduced_multi_compartment1b}\\
%     -\nabla \cdot (\mathbf{K}_{\text{\tiny{out}}} \nabla p_{\text{\tiny{out}}}) + \beta_{\text{\tiny{out}}} (p_{\text{\tiny{out}}} - p_{\text{\tiny{micro}}}) &= - \beta_{\text{out}} (p_{\text{\tiny{out}}} - p_{\text{ref}}) \; \;\; \, \text{in} \; \Omega, \label{eqn:reduced_multi_compartment1c}
% \end{align}
% \end{subequations}

The respective resolved vascular tree structure (upper hierarchies) and the lower hierarchy compartments can be connected by deriving suitable inflow and outflow conditions at the the connecting nodes $\mathbb{V}_{\textit{conn}}$. This can be achieved via suitably {\color{red}estimated} source and sink terms $\theta_{\text{\tiny{in}}}$ and $\theta_{\text{\tiny{out}}}$ in the continuity equation \eqref{eqn:full_multi_compartment2} \cite{RefAdnan}. For the inflow into compartment supply, we compute for each node $u\in\mathbb{V}_{\textit{conn}}$ the net flow $\hat{Q}_{u,\text{in}}$ that enters from the resolved part of the supplying tree into compartment supply:
\begin{align}
   \hat{Q}_{u,\text{in}} = \sum\limits_{\substack{a=uv\\ a\in\mathbb{A}_{\textit{conn}} }}  \hat{Q}_{a} \,.
\end{align}

 %As we assume a strictly hierarchical vascular network, where no part of the hierarchy is skipped, only the uppermost compartment $i=1$ is supplied with fluid in the region of inflow $\Omega_\text{inflow} \subseteq \Omega$ with $\Omega_\text{inflow}\ne \emptyset$:
\begin{align}
    \theta_{\text{\tiny{in}}}(\mathbf{x}) = \sum_{u\in\mathbb{V}_{\textit{conn}}} \frac{\hat{Q}_{u,\text{in}}}{(2\pi\sigma^{2})^{\tfrac{d}{2}}}\exp{\left(-\frac{1}{2}\frac{\left\lVert \mathbf{x}-\mathbf{x}_u \right\rVert^2 }{\sigma^{2}}\right)},  
\end{align}
where we spatially distribute the net flow $\hat{Q}_{u,\text{in}}$ at each node $u\in\mathbb{V}_{\textit{conn}}$ in a symmetric way in the form of a weighted multivariate Gaussian distribution \cite{RefAdnan}. We note that $d=\dim(\Omega)$ is the spatial dimension of the problem and $\sigma^2$ is the variance of the radially symmetric distribution that controls the effective spread of the source distribution. In the sense of homogenization, we choose a sufficiently small variance with respect to the radius of the AV such that the distributed inflow $\hat{Q}_{u,\text{in}}$ from the supplying vascular tree lies well within the range of the AV. 

\begin{remark}
Due to mass conservation, the following relation should hold:
\begin{align}
   \int_\Omega \theta_{\text{\tiny{in}}}(\mathbf{x})\,\mathrm{d}V = \hat{Q}_{\text{perf}} \, ,
\end{align}
where $\hat{Q}_\text{perf}$ is the inflow at the root of the supplying vascular tree. Due to the Gaussian distribution, which has unbounded support, this relation will in general not be exactly satisfied. In this paper, we assume that the corresponding mass error is sufficiently small for our application. As an alternative, one could also scale the Gaussian distribution so that its integration over the finite domain $\Omega$ yields one.  
\end{remark}

For the outflow, a pressure-dependent sink term $\theta_{\text{\tiny{out}}}$ is imposed in the compartment drainage. Similar to the intercompartmental mass exchange, $\theta_{\text{\tiny{out}}}$ is introduced as a pressure-dependent outflow
\begin{align}
    \theta_{\text{\tiny{out}}} = -\beta_\text{\tiny{out}}(p_{\text{\tiny{drain}}}-p_\text{\tiny{out}}) \; \;\; \, \text{in} \; \Omega_\text{\tiny{outflow}},
\end{align}
where $\Omega_\text{\tiny{outflow}} \subseteq \Omega$ with $\Omega_\text{\tiny{outflow}}\ne \emptyset$ denotes the region of outflow. 
The outflow perfusion coefficient $\beta_{\text{\tiny{out}}}$ is determined similarly to the intercompartmental perfusion coefficient using \eqref{eq:perfusion_coefficient}. It relates the outflow $\theta_{\text{\tiny{out}}}$ to the pressure difference between $p_{\text{\tiny{drain}}}$ and a prescribed outflow pressure $p_{\text{\tiny{out}}}$. The reference pressure is obtained by linearly interpolating the pressure at nodes shared by the upper and lower hierarchies.

Finally, the system of equations \eqref{eqn:reduced_multi_compartment1} for our compartmentalization strategy reads
\begin{subequations}
\begin{align}
    -\nabla \cdot (\mathbf{K}_{\text{\tiny{supply}}} \nabla p_{\text{\tiny{supply}}}) + q_{\text{\tiny{supply}}} &= \theta_{\text{in}} \quad \; \;\text{in} \; \Omega, \label{eqn:reduced_multi_compartment1a}\\
    -\nabla \cdot (\mathbf{K}_{\text{\tiny{micro}}} \nabla p_{\text{\tiny{micro}}}) + q_{\text{\tiny{micro}}}  &= 0 \qquad \, \text{in} \;\Omega, \label{eqn:reduced_multi_compartment1b}\\
    -\nabla \cdot (\mathbf{K}_{\text{\tiny{drain}}} \nabla p_{\text{\tiny{drain}}}) + q_{\text{\tiny{drain}}} &= \theta_{\text{out}} \quad \, \text{in} \; \Omega, \label{eqn:reduced_multi_compartment1c}
\end{align}
\end{subequations}
with the intercompartmental flow rate densities
\begin{subequations}
\begin{align}
    q_{\text{\tiny{supply}}} &= \beta_{\text{\tiny{supply}}} (p_{\text{\tiny{supply}}} - p_{\text{\tiny{micro}}}), \label{eq:45a}\\
    q_{\text{\tiny{micro}}} &= \beta_{\text{\tiny{supply}}} (p_{\text{\tiny{micro}}} - p_{\text{\tiny{supply}}}) + \beta_{\text{\tiny{drain}}} (p_{\text{\tiny{micro}}} - p_{\text{\tiny{drain}}}), \label{eq:45b}\\
    q_{\text{\tiny{drain}}}&= \beta_{\text{\tiny{drain}}} (p_{\text{\tiny{drain}}} - p_{\text{\tiny{micro}}}).
    \label{eq:45c}
    %\theta_{\text{\tiny{out}}} &= \beta_{\text{\tiny{out}}} (p_{\text{out}} - p_{\text{\tiny{drain}}}).
\end{align}
\end{subequations}

\subsubsection{{\color{red}Averaging volume size and vessel radius threshold}}

{\color{red}
On the one hand, the AV size must be chosen sufficiently large to ensure that scale separation holds and homogenization can be applied. On the other hand, the AV size must not be too large such that macroscale variations in the material parameters are not lost by averaging. Homogenization therefore requires a sufficiently large depth of the tree structure.

In this paper, the characteristic length of the macro-scale (e.g., the smallest liver diameter), the characteristic length of the AV (in terms of its radius), and the radii of the largest fine-scale vessels to be homogenized (specified via the vessel radius threshold $r_{thresh}$) are chosen at least one order of magnitude apart. We emphasize that the chosen AV size and separation threshold $r_{thresh}$ do not satisfy the classical requirements for scale separation in homogenization. 

In the scope of the current study, however, the separation by one order of magnitude for both is deemed acceptable to achieve a computationally feasible model. We carried out sensitivity studies for different AV sizes, reported in} \cite{Jannes}, {\color{red}which confirm the validity of these choices.}

\subsection{Prototypical model problem in 2D}\label{sec:Numerical_example_disk}

We further illustrate the concepts of our multi-compartment perfusion model via the model problem of a 2D circular disk with a radius of $10\,\text{mm}$. {\color{red}We start by discretizing the circular domain into a mesh of 45,955 standard six-noded triangular elements with quadratic basis functions.} We use the supplying and draining tree structures illustrated by Fig. \ref{fig:tree_2D} and Table \ref{tab_vesseltrees2D} and apply our compartmentalization strategy illustrated in Fig. \ref{fig:Multicompartment}.
To this end, we assign a circular AV to each mesh element, located at the center. The resulting permeability tensors and perfusion coefficients from homogenization are assigned to the corresponding element, where it is assumed constant in that element. Following our strategy to ensure scale separation of at least one order of magnuítude, the AV radius is set to $1\,\text{mm}$, and the maximum radius of the vessels to be homogenized is chosen to be $0.1\,\text{mm}$, hence $r_{thresh}=0.1\,\text{mm}$.
%Following the exposition in \cite{RefHyde}, 
% We specify the dimension of each AV by choosing its radius one order of magnitude smaller than the characteristic length of the domain, i.e. $1\,\text{mm}$.  We choose the threshold for separation to $r_{thresh}=0.1\,\text{mm}$, one order of magnitude smaller than the AV size. 
% \begin{remark}
%     With the present choices of the AV size and the separation criterion $r_{thresh}$, the classical requirements of scale separation for homogenization are not fully met. However, in terms of an efficient computation with the given trees, the scale separation with one order of magnitude each is considered sufficient.
% \end{remark}

 % for efficient computation with the available tree structures.
%\end{remark}

\begin{figure}[ht]
\centering
\subfigure[AV radius of 1 mm \label{a2}]{
 \begin{tikzpicture}
  %\node[] (pic) at (-3,-1.75) {\includegraphics[width=10mm, angle = 0]{"./Images/axes".png}};
   \node[] (pic) at (0,0) {\includegraphics[width=55mm, angle = 0]{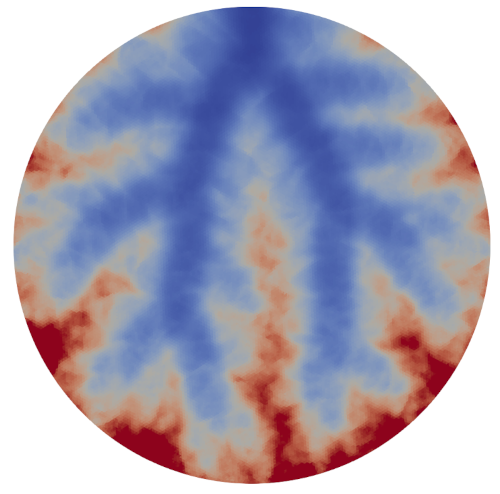}};
   \node[] (pic) at (3.5,0) {\includegraphics[height=35.0mm]{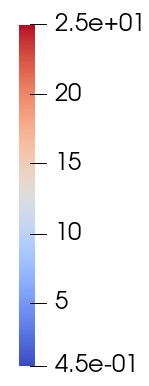}};
\end{tikzpicture}}
\hspace{1cm}
\subfigure[AV radius of 5 mm \label{b2}]{
 \begin{tikzpicture}
   \node[] (pic) at (0,0) {\includegraphics[width=55mm, angle = 0]{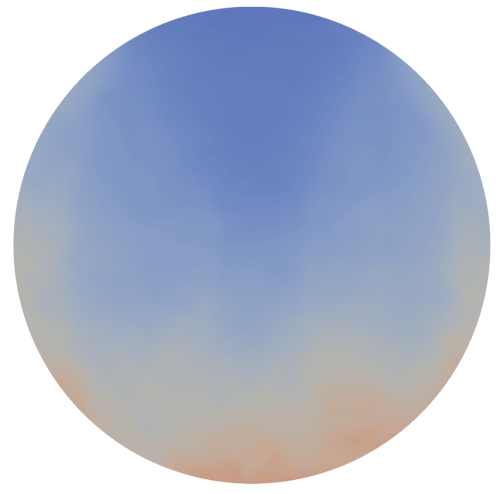}};
   \node[] (pic) at (3.5,0) {\includegraphics[height=35.0mm]{Images/beta_12_colorbar_review.png}};
\end{tikzpicture}}
  \caption{Perfusion coefficient $\beta_{\text{\tiny{supply}}}$ [$\text{mm}~\text{s}~\text{kg}^{-1}$] for coupling the compartments supply and microcirculation.}
\label{fig:beta12_RVE1_5}
\end{figure}

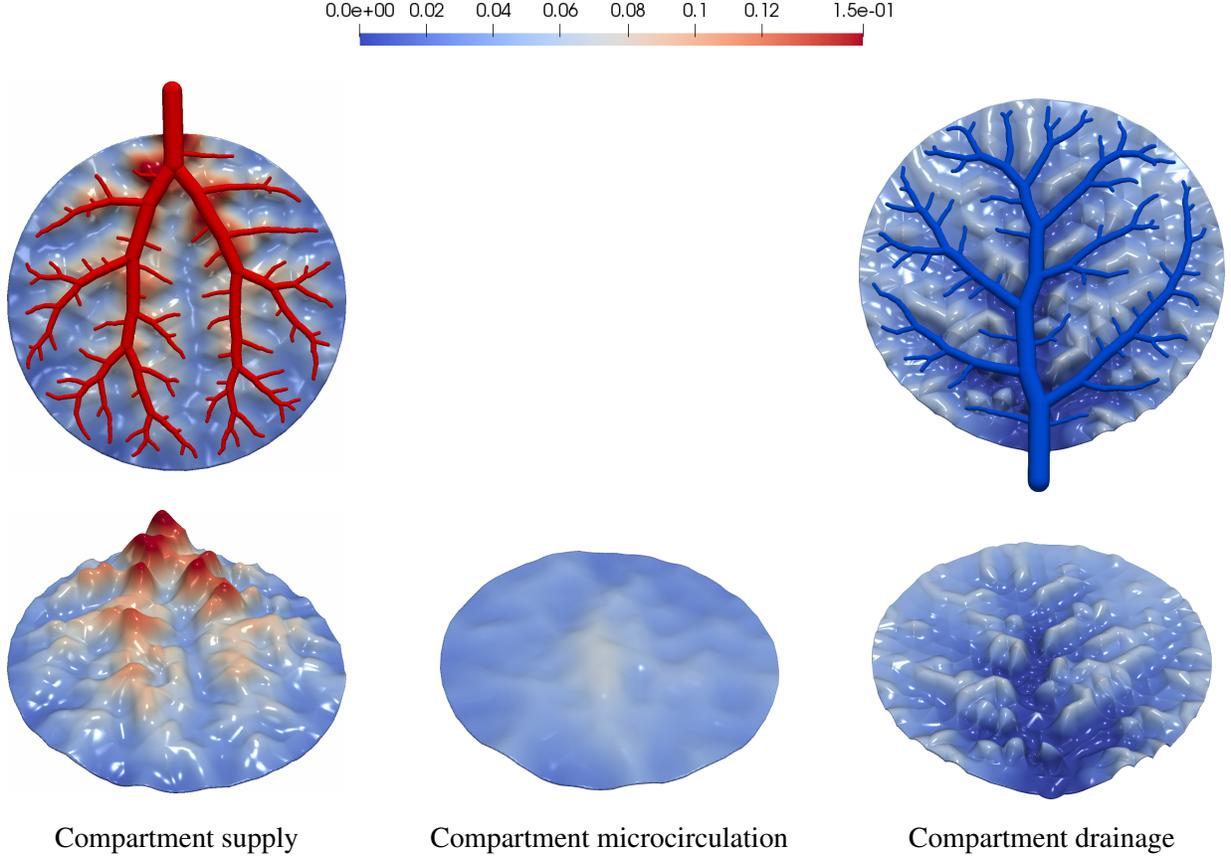
\begin{figure}[ht]
\centering
    \begin{tikzpicture}
       \node[] (pic) at (5,3.35) {\includegraphics[width=75mm, angle = 0]{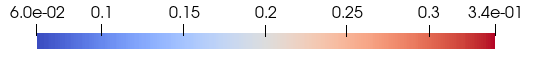}};
      \node[] (pic) at (-0.75,-0.25) {\includegraphics[width=50mm, angle = 0]{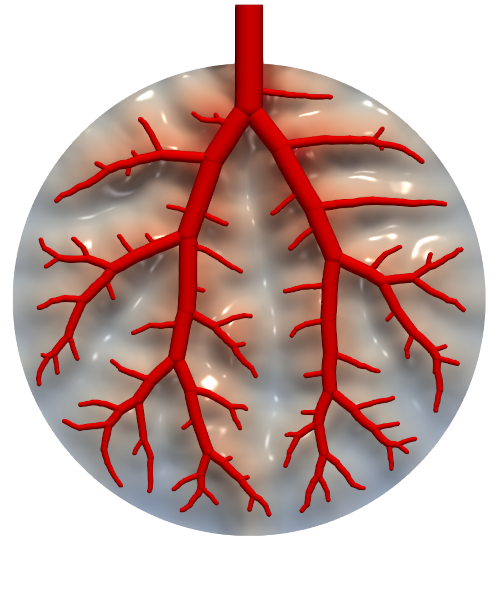}};
       \node[] (pic) at (10.75,0) {\includegraphics[width=50mm, angle = 0]{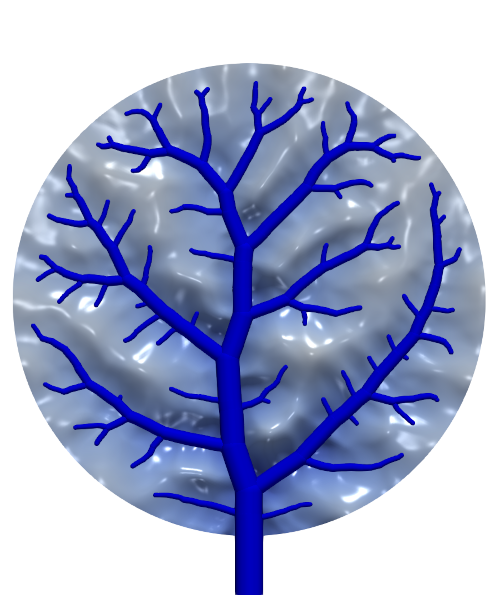}};
      \node[] (pic) at (-0.75,-5.25) {\includegraphics[width=50mm, angle = 0]{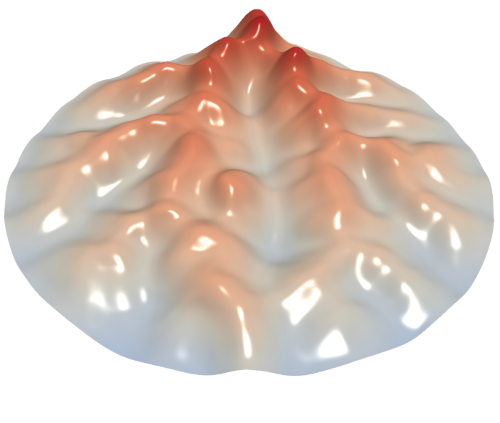}};  
      \node[] at (-0.75,-7.75) {Compartment supply};
      \node[] (pic) at (5,-5.25) {\includegraphics[width=50mm, angle = 0]{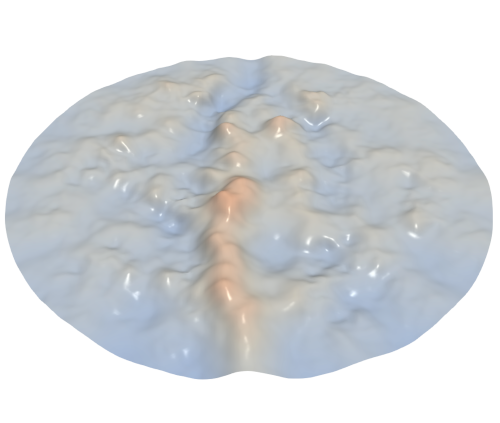}};
      \node[] at (5,-7.75) {Compartment microcirculation};
        \node[] (pic) at (10.75,-5.25) {\includegraphics[width=50mm, angle = 0]{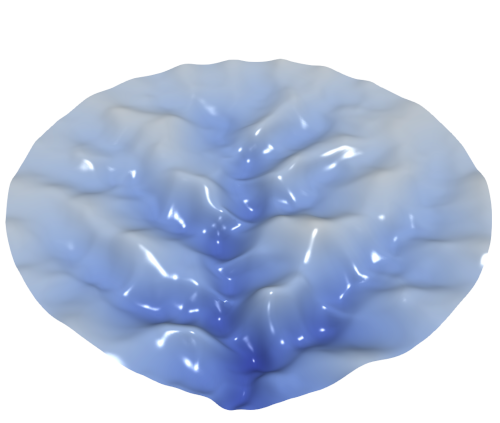}};  
      \node[] at (10.75,-7.75) {Compartment drainage};
     \end{tikzpicture}
    \caption{Solution of the pressure fields $p_i$ [$\text{kg} \; \text{mm}^{-1} \; \text{s}^{-2}$] for the model problem of the circular domain.}
    \label{fig:pressure_RVE6}
\end{figure} 

\begin{figure}[ht]
\centering
    \begin{tikzpicture}
        \node[] (pic) at (-0.7,1.5) {\includegraphics[height=6mm, angle = 0]{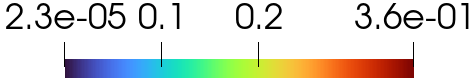}};  
        \node[] (pic) at (5,1.5) {\includegraphics[height=6mm, angle = 0]{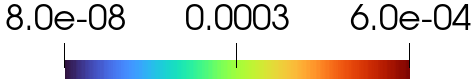}};
        \node[] (pic) at (10.7,1.5) {\includegraphics[height=6mm, angle = 0]{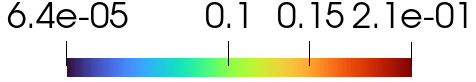}}; 
        \node[] (pic) at (-0.7,-2) {\includegraphics[width=50mm, angle = 0]{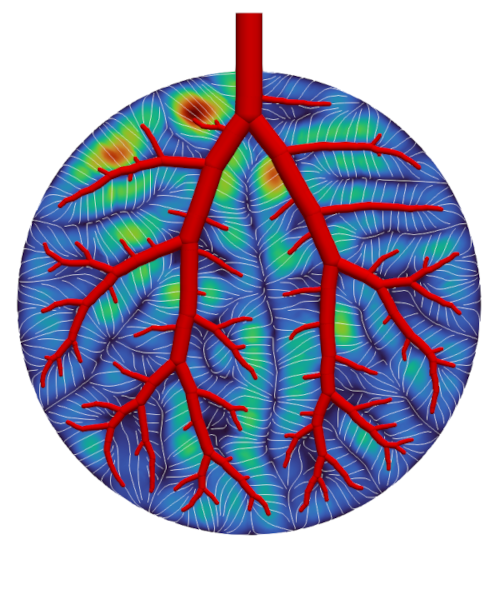}}; 
        \node[] (pic) at (5,-2) {\includegraphics[width=50mm, angle = 0]{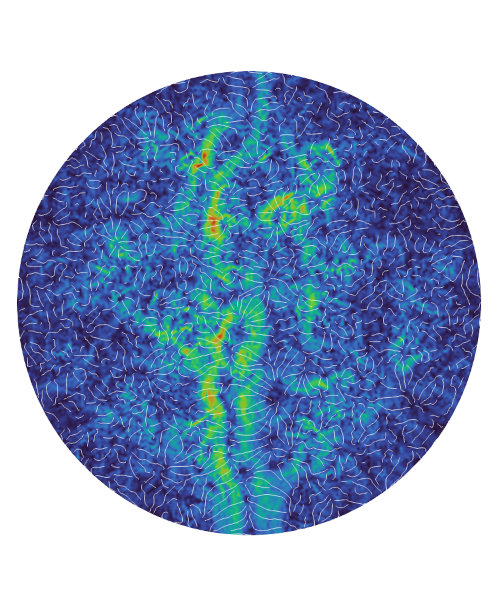}};
        \node[] (pic) at (10.7,-2) {\includegraphics[width=50mm, angle = 0]{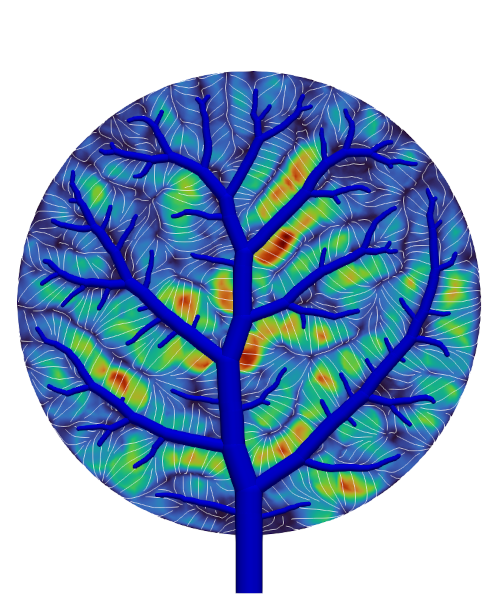}}; 
        \node[] at (-0.7,-5.6) {$w_{\text{\tiny{supply}}}$};
        \node[] at (5,-5.6) {$w_{\text{\tiny{micro}}}$};
        \node[] at (10.7,-5.6) {$w_{\text{\tiny{drain}}}$};
     \end{tikzpicture}
    \caption{Solution of the homogenized velocity magnitude fields $w_i$ [$ \text{mm}\; \text{s}^{-1}$] for the model problem of the circular domain with streamlines indicating the directions of flow.}
    \label{fig:circle_velocities}
\end{figure}

We %illustrate homogenized material parameters %permeability tensor by
plot the perfusion coefficient $\beta_{\text{\tiny{supply}}}$ component %$K_{yy}$ 
in Fig. \ref{fig:beta12_RVE1_5}. %{fig:k11_RVE1_5}. 
We can observe in Fig. \ref{a2} that for an AV radius of 1 mm, geometry and topology of the resolved larger vessels is reflected in the material parameters of the supply and drainage compartments. This result is plausible, as the presence of a (resolved) larger vessels implies the absence of smaller vessels (to be homogenized), which directly leads to a decrease in the corresponding permeability tensor and perfusion coefficient. In our case, we would like to maintain this mechanism to (a) implicitly represent the flow obstacle due to the resolved larger vessels in the homogenized compartments, and (b) mitigate intercompartmental flow from the compartment supply into the microcirculation and from the microcirculation into compartment drainage. % can not occur at locations where larger vessels of the upper hierarchies are present, as observed in the case of the smaller radius %Since compartment supply did not account for vessels of the upper hierarchy, which have relatively large spatial dimensions, fewer vessels of the lower hierarchy are consequently present at those locations. 
Hence, we conclude from Fig. \ref{a2} that the AV radius size must be appropriately chosen in the sense that macroscopic heterogeneities of the material parameters are not smoothed too heavily over the entire domain. An example is given in Fig. \ref{b2}, where we apply a larger AV radius of 5 mm. We note that unlike in the current 2D case, blood can flow around the obstacles in a 3D liver representation by using the third dimension. %This also prevents fluid in the the lower hierarchy from "flowing" unhindered through the larger vessels. This observation does not apply to a larger AV radius, such as 5 mm, as demonstrated in Fig. \ref{b2}.

%\begin{figure}[ht]
%\centering
%\subfigure[AV radius of 1 mm \label{a2}]{
% \begin{tikzpicture}
%    \node[] (pic) at (-3,-1.75) {\includegraphics[width=10mm, angle = 0]{"./Images/axes".png}};
%   \node[] (pic) at (0,0) {\includegraphics[width=45mm, angle = 0]{"./Images/K11_RVE1_new".png}};
%   \node[] (pic) at (3.25,0) {\includegraphics[width=10.0mm]{"./Images/K11_Colormap".png}};
%\end{tikzpicture}}
%\hfill
%\subfigure[AV radius of 5 mm \label{b2}]{
% \begin{tikzpicture}
%   \node[] (pic) at (0,0) {\includegraphics[width=45mm, angle = 0]{"./Images/K11_RVE5_new".png}};
%   \node[] (pic) at (3.25,0) {\includegraphics[width=10.0mm]{"./Images/K11_Colormap".png}};
%\end{tikzpicture}}
%  \caption{Permeability component $K_{yy}$ [$\text{mm}^3~\text{s}~\text{kg}^{-1}$] for compartment 1 (supply).}
%\label{fig:k11_RVE1_5}
%\end{figure}

%Fig. \ref{fig:beta12_RVE1_5} shows the perfusion coefficient $\beta_{12}$ for two different AV radii. Again, the basic tree structure is recognizable for the smaller radius and similar observations to those in the case of permeability can be made. Additionally, intercompartmental flow between compartments supply and microcirculation can not occur at locations where larger vessels of the upper hierarchies are present, as observed in the case of the smaller radius (Fig. \ref{aa2}).

% Reference a different paper for its implementation: Standard Lagrange basis functions with quadratic polynomials

We solve the associated coupled boundary value problems \eqref{eqn:reduced_multi_compartment1} and \eqref{eqn: bnd condition} of the compartmentalized model via the open-source finite element framework FEniCS \cite{RefFenics}. %, using standard quadratic nodal basis functions on the triangular mesh shown in Fig. \ref{a2}.
%The averaged outflow pressure $p_\text{out}$ is enforced by a sufficiently large penalty parameter $\beta_\text{out} = 10^5$. 
We use an isotropic and homogeneous permeability $K_{\text{\tiny{micro}}}$ = $1/180 \; \text{mm}^3~\text{s}~\text{kg}^{-1}$ \cite{RefDebbaut2,RefDebbaut3} in the compartment microcirculation.

\begin{figure}[ht]
\centering
    \begin{tikzpicture}
        \node[] (pic) at (-0.7,1) {\includegraphics[height=6mm, angle = 0]{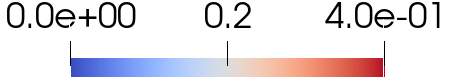}};  
        \node[] (pic) at (5,1) {\includegraphics[height=6mm, angle = 0]{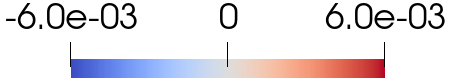}};
        \node[] (pic) at (10.7,1) {\includegraphics[height=6mm, angle = 0]{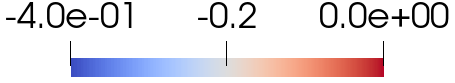}}; 
        \node[] (pic) at (-0.7,-2) {\includegraphics[width=50mm, angle = 0]{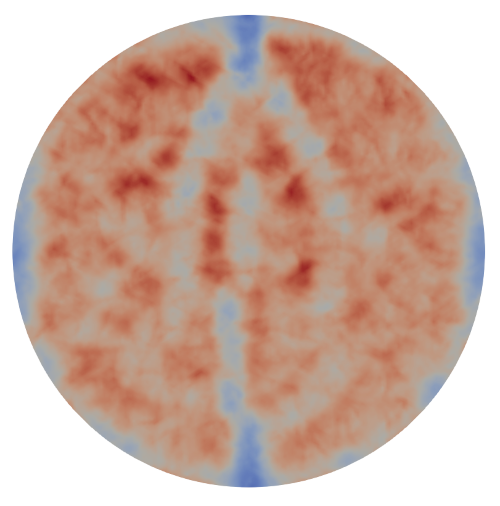}};  
        \node[] (pic) at (5,-2) {\includegraphics[width=50mm, angle = 0]{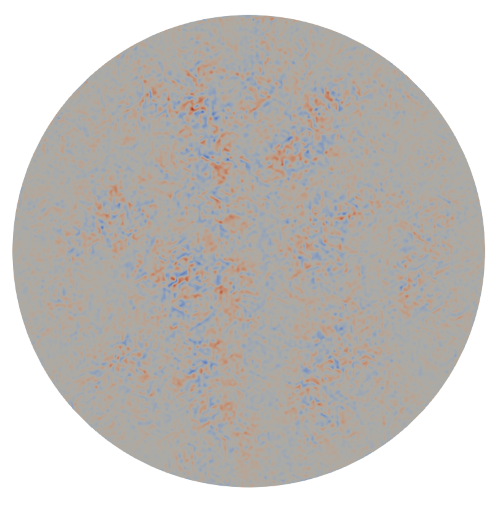}};
        \node[] (pic) at (10.7,-2) {\includegraphics[width=50mm, angle = 0]{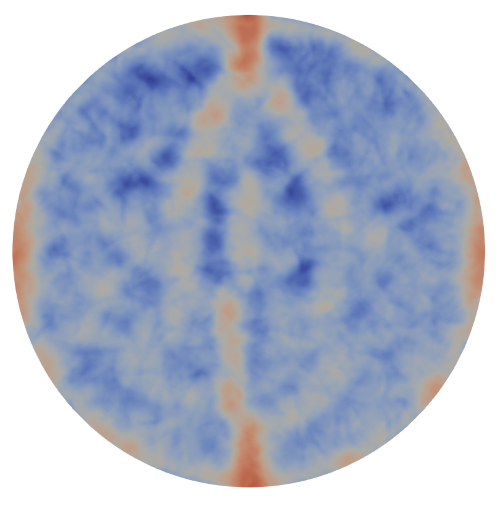}};  
        \node[] at (-0.7,-5) {$q_{\text{\tiny{supply}}}$};
        \node[] at (5,-5) {$q_{\text{\tiny{micro}}}$};
        \node[] at (10.7,-5) {-$q_{\text{\tiny{drain}}}$};
        \node[] at (-3,0) {(a)};
        \node[] at (2.7,0) {(b)};
        \node[] at (8.3,0) {(c)};        
     \end{tikzpicture}
    \caption{Solution of the homogenized intercompartmental flow rate densities $q_i$ [$ \text{s}^{-1}$] representing (a) the supply into and (c) the drainage from the compartment microcirculation. The difference (b) between supply and drainage shows that the redistribution of fluid within the compartment microcirculation is two orders of magnitude smaller and hence negligibly small.}
    \label{fig:circle_intercompartmental_flow}
\end{figure}
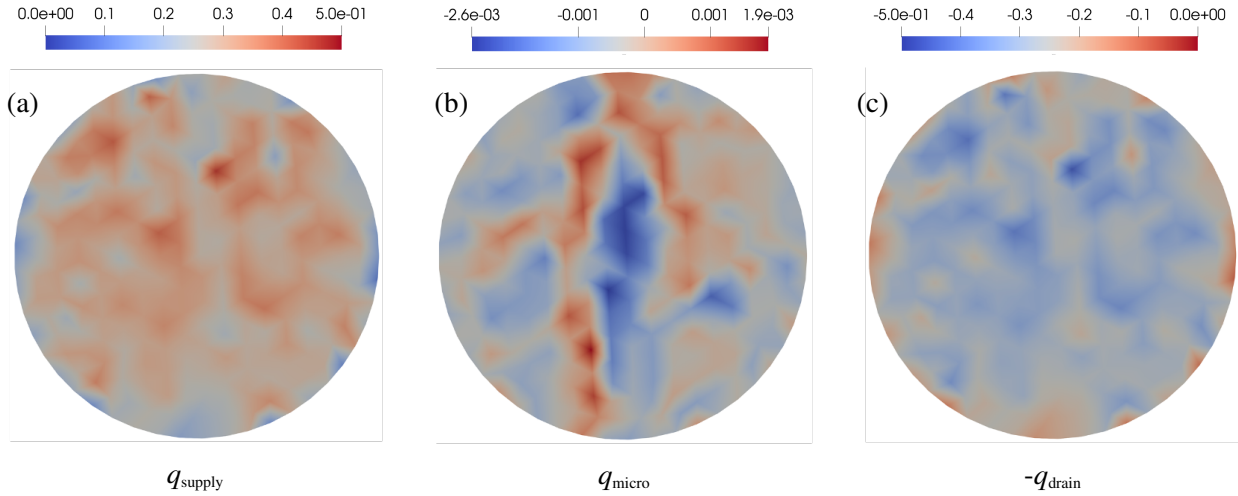 

The results of the pressure fields for all three compartments are shown in Fig. \ref{fig:pressure_RVE6}. We observe that the compartments supply and drainage exhibit pronounced pressure differences, following the structure of the resolved supplying and draining trees. In the compartment microcirculation, we can see an almost uniform pressure field. In particular, the pressure in the compartment supply decreases from the root to the terminals. In the compartment drainage, we observe that the pressure decreases from the terminals to the root. 
%For an in-depth validation, we refer the interested reader to \cite{Jannes}, where we compare the compartmental pressure field of the homogenized model with an averaged pressure field of the discrete Poiseuille network model. %demonstrating excellent agreement.

Figure \ref{fig:circle_velocities} presents the velocity magnitudes across all three compartments. The streamlines indicate that flow consistently moves away from upper hierarchy vessels within the compartment supply and towards them within the compartment drainage. The observed flow directions agree with the orientations of the vessels in the lower hierarchies of the corresponding vascular tree, as depicted in Fig. \ref{fig:tree_2D}. 

%In contrast, no organized flow pattern is observed 
In the compartment microcirculation, the velocity magnitudes are three orders of magnitude lower than those observed in the compartments supply and drainage. Therefore, a flow redistribution through the compartment microcirculation does not occur. This observation is in agreement with our interpretation that the compartment microcirculation represents the flow resistance of the microscale sinusoids, but cannot involve significant flow redistribution at the macroscale, as the lobular structure would not allow that.

%This aligns with expectations, as the model does not sufficiently capture microcirculatory flow, and no significant macro-scale fluid redistribution occurs. 

Figure \ref{fig:circle_intercompartmental_flow} shows the intercompartmental flow rate densities, computed from \eqref{eq:45a}, \eqref{eq:45b} and \eqref{eq:45c}. The patterns for supply into and drainage from the microcirculation do not show large variations across the domain. The difference between supply and drainage confirms that a significant fluid redistribution at the macroscale within the compartment microcirculation does practically not occur, as its magnitude is two orders lower than supply and drainage. In other words, the fluid flowing into the microcirculation at a particular location also flows out at the same location. % These findings suggest that the compartmentalized homogenized flow model provides an accurate representation of blood flow and pressure in the lower hierarchies of the vascular trees, as well as the perfusion characteristics of the microcirculation at the meso-scale.

% point out order of magnitude difference of velocity in microcirculation, reference values for velocities in microcirculation: 
% \begin{itemize}
%     \item Ricken: Darcy-velocities (?) range 0 - 400 $\mu m/s$, avg 152 $\mu$m/s, $\phi = 0.25$ ,  
%     \item Debbaut: real velocities range 0 - 2 $mm/s$, avg $10^{-4} m/s$, Darcy-velocity: $2.6*10^{-3} mm/s$, $\phi = 0.143$ ,
% \end{itemize}

\section{A hyperperfusion-driven poroelastic growth model}
\label{sec:Poroelasticity_growth}

In this section, we present our model for compensatory liver regrowth driven by hyperperfusion in the microcirculation. To this end, we first extend the multi-compartment perfusion model by a poroelastic growth model that is defined solely at the compartment microcirculation, but provides a growth map for the resolved vessel trees and the compartments supply and drainage. We then describe a flow-dependent evolution equation that connects hyperperfusion to volumetric growth. %Additionally, we provide an interpretation how our model represents lobular remodeling at the microcirculation as well as remodeling in the vascular tree. 
The growth evolution equation and the poroelastic growth model correspond to part B and part C, respectively, of our modeling framework illustrated in Fig. \ref{fig:Modelingframework}.

\subsection{Poroelasticity framework} \label{sec:Growth}

The coupling of tissue deformation and fluid flow can be achieved within the framework of poroelasticity. %, which can be rigorously based on continuum mixture theory.
For a comprehensive review, we refer the interested reader to \cite{RefBookCoussy,RefBookEhlers,RefBookBoer}, and particularly for large deformation formulations to \cite{RefAdnan,RefChapelle,RefPhysAppl,RefWallPorous}. Models to describe growing poroelastic media have also been proposed \cite{collis2017effective,penta2014effective}, including tissue growth \cite{ricken2010remodeling,ehlers2009computational}, biomass growth \cite{sacco2017poroelastic} or tumor growth \cite{xue2016biochemomechanical,ambrosi2017solid,fraldi2018cells}.

\subsubsection{Homogenization and porosity}

We start by focusing on the liver microcirculation, see Section \ref{sec:2.1} and Fig. \ref{fig:liveranatomy}. In line with our discussion on perfusion in Section \ref{sec:MultiCompartment}, we idealize the microcirculation as a porous material in the sense of a heterogeneous mixture. In our model, it consists of two constituents: a skeleton phase that includes all cells, the extracellular matrix and the interstitial fluid, and one perfusion phase that represents the blood moving through the sinusoids. In the following, corresponding quantities are indexed by \textit{skel} and \textit{perf}, respectively.  %Hence, the corresponding porous medium is fully saturated at all times.

\begin{figure}[t]
\centering
    \begin{tikzpicture}
   \node[] (pic) at (0.0,0) {\includegraphics[width=\textwidth]{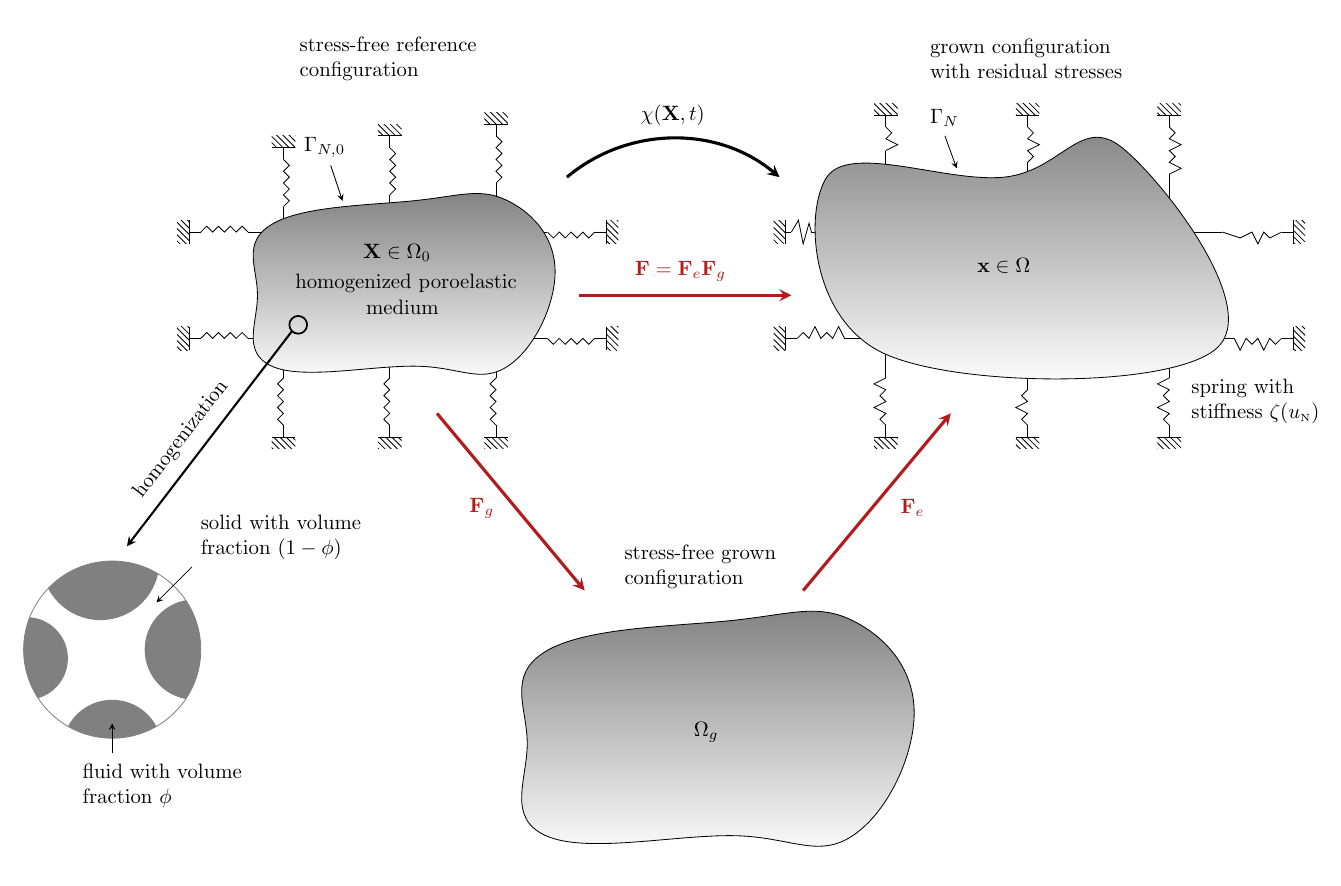}};
     \end{tikzpicture}
    \caption{Kinematics of finite growth of a poroelastic medium.}
    \label{fig:mapping}
\end{figure}

We recall the notion of an averaging volume (AV), discussed in Section \ref{sec:Modelparameters}. The two-constituent porous material is illustrated at the micro- and macroscale in Fig. \ref{fig:mapping}. Following averaging over a representative AV at the microscale, the volume fraction of the perfusion phase can be described by the porosity $\phi = \phi^{\textit{perf}}$. The porosity can be expressed at the macroscale as
\begin{align}
\phi(\mathbf{x},t) = \frac{dV^{\textit{perf}}}{dV}
\end{align}
where $dV^{\textit{perf}}$ is the incremental volume of the perfusion phase after homogenization and $dV$ is the incremental total volume. In a two-phase mixture under the assumption of absent void spaces (full saturation), one can express the volume fraction of the skeleton as $\phi^{\textit{skel}} = 1 - \phi$. We note that the corresponding total mass at each point of the macroscale domain is
\begin{align}
dm = \rho dV
\end{align}
where $\rho = \rho^{\textit{perf}} \phi + \rho^{\textit{skel}}(1 - \phi)$ is the total density, and $\rho^{\textit{perf}}$ and $\rho^{\textit{skel}}$ are the specific densities of the perfusing blood and the tissue skeleton, respectively. We recall that $\rho^{\textit{perf}}$ is constant based on our incompressibility assumption for blood. %, and in addition, we assume $\rho^{\textit{skel}}$ to be constant as well.

\subsubsection{Kinematics of growth}

Each of the two constituents of the porous material simultaneously occupies a common spatial region. We assume that the Lagrangian configuration of the constituents coincide, allowing us to work with a single Lagrangian description. Accordingly, we introduce the deformation map to describe the spatial position of a particle, illustrated in Fig. \ref{fig:mapping}:
\begin{equation}
    \mathbf{x} = \mathbf{\chi}(\mathbf{X},t),
\end{equation}
where $\mathbf{X} \in \Omega_0$ is the Lagrangian position of a particle of the liver microcirculation, $\mathbf{x}\in \Omega$ denotes the spatial position, and $t$ describes the time. Here, $\Omega_0$ and $\Omega$ refer to the reference and current domain of the porous material, respectively.

The displacement $\mathbf{u}$ %and Lagrangian velocity $\mathbf{v}$ 
of the porous material can be described by $\mathbf{u}=\mathbf{x}-\mathbf{X}$. %and $\mathbf{v}=\dot{\mathbf{u}}$ with the dot denoting the material derivative.
Accordingly, the standard kinematic quantities can be introduced:
\begin{align}
         \mathbf{F} & = \mathbf{I} + \nabla \mathbf{u} \\
         \mathbf{C} & = \mathbf{F}^T \mathbf{F}, \\
         \mathbf{E} & = \frac{1}{2}(\mathbf{C}-\mathbf{I}),
\end{align}
with $\mathbf{I}$ being the identity tensor, $\mathbf{F}$ the deformation gradient, $\mathbf{C}$ the right Cauchy-Green tensor, and $\mathbf{E}$ the Green-Lagrange strain tensor. The determinant of the deformation gradient associated with volumetric deformation is denoted as $J = \text{det}\; \mathbf{F}$.

To incorporate growth into the framework, the deformation map is decomposed into two components, illustrated in Fig. \ref{fig:mapping}. A material point is first mapped into an intermediate, incompatible growth configuration $\Omega_g$.
This state is considered stress-free, and only mass generation occurs between $\Omega_0$ and $\Omega_g$.
% This deformation is given by the 
 To ensure compatibility of the domain, elastic deformations are then applied to the intermediate configuration \cite{RefRodriguez}. 
The split of the deformation map into two consecutive mappings leads to a multiplicative decomposition of the deformation gradient
\begin{align}
		\mathbf{F} = \mathbf{F}_e \mathbf{F}_g,
\end{align}
into a purely elastic part $\mathbf{F}_{e}$ and a growth deformation gradient $\mathbf{F}_{g}$ \cite{RefRodriguez}, analogous to the decomposition of the deformation gradient in elastoplasticity \cite{RefLee}. 
Accordingly, we introduce the elastic part of the right Cauchy-Green strain tensor
\begin{align}
		 \mathbf{C}_{e} = \mathbf{F}_{e}^T \mathbf{F}_{e}.
\end{align}

\subsubsection{Quasi-static balance of linear momentum}

Liver regeneration is a dynamic and multiscale process that occurs over various scales in space and time. As our growth model is formulated at the macroscale, where the time scale consists of days or weeks, we can describe liver growth within this time window by a sequence of quasi-static growth processes, within which we can neglect any time-dependent effects. 

We then introduce the momentum balance, assuming a quasi-static setting and the absence of body forces, in both the current and reference configurations as
\begin{align}\label{eq: momentum equation}
    \nabla \cdot \mathbf{\sigma} &= \mathbf{0}  \; \;\; \, \text{in} \; \Omega, \\
    \nabla_0 \cdot \left(\mathbf{F} \mathbf{S} \right) &= \mathbf{0}  \; \;\; \, \text{in} \; \Omega_0.
\end{align}
where $\mathbf{\sigma}$ %  = J^{-1}\mathbf{F} \mathbf{S} \mathbf{F}^T$ 
describes the Cauchy stress tensor and $\mathbf{S}$ denotes the second Piola-Kirchhoff stress tensor.

In the scope of this work, we consider the contact of the liver 
with surrounding organs, which we model by nonlinear elastic springs at the outer boundary of the liver, as indicated in Fig. \ref{fig:mapping}. The effect of spring elements at the boundary are best included variationally in the weak formulation of the balance of linear momentum. For details, we refer the interested reader to \cite{RefAdnan,wriggers2006computational} and the references therein. 

To account for the elastic contact on the outer boundary of the liver, we choose the following interaction term
\begin{align}
    W_c(u_\text{N}) = \zeta u_\text{N},
\end{align}
with $\zeta$ representing the nonlinear stiffness of the spring element.
The contact contribution is applied in the normal direction, where $\mathbf{N}$ denotes the outward unit normal vector on the boundary $\Gamma_{N,0}$, and  $u_\text{N} = \mathbf{u} \cdot \mathbf{N}$ is the normal component of the displacement vector. We adopt the deformation-dependent stiffness:
%\begin{align}
%      \zeta(u_\text{N}) = \frac{2\alpha}{1+e^{-2 u_\text{N}/u_\text{0}}}-\alpha,
%    \label{eqn:stiffnesscontact}  
%\end{align}
\begin{align}
      \zeta(u_\text{N}) = \frac{2\alpha}{1+\exp{\left(\frac{-2 u_\text{N}}{u_\text{0}}\right)}}-\alpha,
    \label{eqn:stiffnesscontact}  
\end{align}
which saturates towards a maximum stiffness value $\alpha$ with increasing displacement $u_\text{N}$. {\color{blue}The parameter $u_0 = 1$ mm is a reference displacement to non-dimensionalize the exponent.}

\subsubsection{Mass balance: integration within the multi-compartment flow model}

We recall the pressure equation in \eqref{eqn:reduced_multi_compartment1},
which we obtained upon substituting Darcy's law \eqref{eqn:full_multi_compartment1} into the balance of fluid mass \eqref{eqn:full_multi_compartment2}. Due to coupling with compartments supply and drainage, the mass exchange term in terms of the net flow rate density $q_{\text{\tiny{micro}}}$ appears, see \eqref{eqn:reduced_multi_compartment1b}.
We use the pull-back operation $\nabla = \mathbf{F}^{-T}\nabla_0$ and the identity $\nabla\cdot\mathbf{w}\,d\Omega = \nabla_0\cdot(J\mathbf{F}^{-1}\mathbf{w})\,d\Omega_0$ for the mapping of \eqref{eqn:reduced_multi_compartment1} to the reference configuration $\Omega_0$
\begin{align}\label{eq: mass fluid}
    -\nabla_0 \cdot (\mathbf{K}_{i,0} \nabla_0 p_i) + \sum_{k=1}^N \beta_{i,k,0}(p_i-p_k) &= \theta_{i,0} \; \;\; \, \text{in} \; \Omega_0, \;\;\, i=\text{supply, micro, drain}
\end{align}
with $\nabla_0$ denoting the material gradient and $\mathbf{K}_{i,0},\, \beta_{i,k,0} = J\beta_{i,k},\, \theta_{i,0}=J\theta_{i}$ denoting the permeability tensor, the perfusion coefficient, and the source term in the reference configuration, respectively. The pull-back operation for the permeability tensor is defined by
\begin{align}
  \mathbf{K}_{i,0} = J \mathbf{F}^{-1}\mathbf{K}_{i}\mathbf{F}^{-T}. \label{eq:pullbackpermeability}
\end{align}

For the compartment microcirculation, we directly apply \eqref{eq:pullbackpermeability} for $K_{\text{\tiny{micro},0}}$, as it represents the resistance of the sinusoids, which is preserved due to lobular remodeling during growth, see also Section \ref{sec:4.3} below.

For the compartments supply and drainage, we can find simple update relation for the permeability tensors and perfusion coefficients based on the assumption that the perfusion properties of the vascular network are maintained during growth. Given the deformation field for each quasi-static iteration step, the update formulations are 
\begin{align}
  \mathbf{K}_i &= J^{-1}\mathbf{F}\,\mathbf{K}_{i}\big|_{t=0}\mathbf{F}^T, \quad i=\text{supply, drain}\label{eq:updatepermeability}\\
  \beta_{i,k} &= J^{-1} \beta_{i,k}\big|_{t=0}, \quad \quad \;\;\, i=\text{supply, drain}  \label{eq:updateperfusioncoeff}  
\end{align}
where $\mathbf{K}_{i}|_{t=0}$ and $\beta_{i,k}|_{t=0}$ denote the permeability and perfusion coefficient in the initial state, respectively.  Using these relations, we avoid the costly recomputation of the model parameters in each iteration step. Inserting \eqref{eq:updatepermeability} into \eqref{eq:pullbackpermeability} shows that the permeability tensor remains constant in the reference configuration, as does the perfusion coefficient.

\subsubsection{Constitutive equations}
%(Start with Dissipation inequality?)
{\color{teal}We introduce the Jacobian weighted by the volume fraction of the skeleton phase:
\begin{align}
J^{skel} = J(1-\phi),
\label{Jskel}
\end{align}
and utilize the following Helmholtz free energy density:
\begin{align}
\Psi^{skel}(\mathbf{C}_{e},J^{skel}) = \Psi^{{hyp}}(\mathbf{C}_{e}) +   \Psi^{{vol}}(J^{skel}), 
\label{PSI}
\end{align}
where $\Psi^{{hyp}}(\mathbf{C}_{e})$ denotes the hyperelastic potential of the liver skeleton and $\Psi^{{vol}}(J^{skel})$ accounts for the volume change of the structure caused by the fluid pressure \cite{RefWallPorous}.

In \eqref{PSI}, we choose a hyperelastic material model of Neo-Hookean type for the liver skeleton: 
\begin{equation}
\Psi^{hyp}(\mathbf{C}_{e}) = \frac{1}{8}\lambda \text{ln}^2(I_3)+\frac{1}{2}\mu[I_1 - 3 - \text{ln}(I_3)],
\end{equation}
where $\lambda$ and $\mu$ denote the Lamé parameters and $I_1 = \text{tr}\; \mathbf{C}_{e}$ and $I_3 = \text{det}\; \mathbf{C}_{e}$ are invariants of the elastic right Cauchy-Green tensor.
Additionally, we select
\begin{align}
    \Psi^{vol}(J^{skel}) = \kappa \left( \frac{J^{skel}}{1-\phi_0} - 1 - \text{ln}\left( \frac{J^{skel}}{1-\phi_0}\right) \right),
    \label{eqn:volumetric}
\end{align}
{\color{blue}where $\phi_0$ denotes the porosity in the stress-free grown configuration, which in our model is equivalent to the porosity in the reference configuration}, and $\kappa = E/(3(1-2\nu))$ is the bulk modulus of the skeleton \cite{RefWallPorous}.

We decompose the Cauchy stress tensor into the effective stress $\mathbf{\sigma}'$ and a fluid pressure component via:
\begin{align}
    \mathbf{\sigma} = \mathbf{\sigma}'- p_\text{\tiny{micro}}\mathbf{I},
\end{align}
where $p_\text{\tiny{micro}}$ is the fluid pressure of blood that moves through the sinusoids in the microcirculation. The effective stress $\mathbf{\sigma}'$ determines the deformation of the skeleton. 
The constitutive relations take the following form \cite{RefWallPorous,RefHimpel}:
\begin{subequations}
    \begin{align}
   \mathbf{S}&=\mathbf{S}'- J p_\text{\tiny{micro}}\mathbf{C}^{-1},\\
   \mathbf{S}'&= 2 \; \mathbf{F}_g^{-1} \frac{\partial \Psi^{hyp}}{\partial \mathbf{C}_{e}} \mathbf{F}_g^{-T},\\
    p_\text{\tiny{micro}} &=-\frac{\partial \Psi^{vol}}{\partial J^{skel}}, \label{eqn:pressureconsitituive}
\end{align}
\end{subequations}
where $\mathbf{S}'$ is associated with the effective stress via $\mathbf{\sigma}' = J^{-1}\mathbf{F} \mathbf{S}' \mathbf{F}^T$. Equation \eqref{eqn:pressureconsitituive} relates $J^{skel}$ to the fluid pressure $p_\text{\tiny{micro}}$, and consequently connects the porosity $\phi$ to $p_\text{\tiny{micro}}$.}

We note that these constitutive relations are motivated by macroscopic thermodynamic considerations. For a review of poroelasticity from a microscopic perspective and the derivation of constitutive equations by means of a micro-macro approach, we refer to \cite{RefDormieux2}.

{\color{blue}
\begin{remark}
The mass balance equations of the solid skeleton do not have to be explicitly considered here, since the skeleton partial density $\rho^{skel}(1-\phi)$ does not appear in the linear momentum equations \eqref{eq: momentum equation} and the fluid mass balance equations \eqref{eq: mass fluid}. Therefore, the only unknowns of the model are the pressure fields in the three compartments supply, microcirculation and drainage, and the displacement vector field. The final porosity in the grown configuration is not a variable, but can be computed as a postprocessing step via \eqref{eqn:pressureconsitituive}  ($p_\text{\tiny{micro}}\rightarrow J^{skel}$) and \eqref{Jskel} ($J^{skel}\rightarrow \phi$). %In case inertia or body forces are present, the skeleton density may be incorporated via: $\rho^{skel}(1-\phi) J^{skel} = \rho^{skel}(1-\phi_0) \Leftrightarrow (1-\phi) J^{skel} = (1-\phi_0)$. 
\end{remark}}

\subsection{Evolution equation for hyperperfusion-driven isotropic growth}
\label{sec:evolutionequation}

Growth evolution laws describe how living tissues alter their shape in response to external stimuli. These laws establish a relationship between the growth tensor $\mathbf{F}_{g}$ and mechanical fields, chemical fields, or biological signaling \cite{RefReviewGrowth,RefGoriely}.

\subsubsection{Isotropic compensatory growth}

Following our exposition in Section \ref{sec:2.2}, we recall that hepatocyte proliferation is the central mechanism by which the liver regains its mass. Growth factors and cytokines that drive hepatocyte proliferation at the cell level quickly spread within a lobule. %due to the thorough vascularization, ensuring that regenerative signals reach all liver parts. 
Additionally, lobular remodeling supports the even distribution of blood flow through the regenerating lobule tissue. 
Therefore, we assume that hepatocytes proliferate uniformly, %throughout the remaining liver tissue, 
leading to a consistent increase in liver mass in all directions. One might argue that at the microscale, growth might still occur orthotropically, as hepatocyte proliferation might differ along the sinusoid axis in flow direction and perpendicular to the sinusoid axis. We can assume, however, that even in this case, liver regrowth can be regarded as isotropic from a macroscopic viewpoint, since any potential direction dependence will be averaged out in the homogenization process. 

We hence consider compensatory volumetric growth to occur \textit{isotropically}, expressed in the following classical form:
\begin{align}
	\mathbf{F}_{g} = \vartheta^{1/3} \; \mathbf{I},
\end{align}
where the growth factor $\vartheta$ denotes the volumetric change due to growth \cite{RefLubarda}. It corresponds to the determinant of the growth deformation tensor
\begin{align}
	\vartheta = \text{det} \; \mathbf{F}_{g} = J_g = \frac{dV^g}{dV^0}
\end{align}
which denotes the ratio between the volume increase in the intermediate configuration and the initial volume in the reference configuration at each point of the macroscale domain.
In the context of our quasi-static poroelastic model, we assume that we can keep the same porosity when we map each AV from the initial to the intermediate configuration. Hence, we can write for the change in mass at the macroscale:
{\color{teal}
\begin{align} \label{eq62}
	dm^g  = \rho dV^g  = \frac{dV^g}{dV^0}\rho dV^0 = \vartheta \, dm^0. 
\end{align} }
We emphasize that in \eqref{eq62}, the growth factor is not only applied to the tissue skeleton, but also to the perfusing blood. In the context of our quasi-static model, the associated ``growth" of the fluid, $\vartheta \rho^{\textit{perf}} \phi dV^0$, represents the added mass of the blood that occupies additional sinusoid space. We also refer to the discussion in terms of lobular remodeling in Section \ref{sec:4.3} below.

Following \eqref{eq62}, the interpretation of the growth factor $\vartheta$ in terms of the change in homogenized mass under consistent microstructure is straightforward:
\begin{align}
	\vartheta = \frac{dm^g}{dm^0}.
 \label{eqn:theta_mass}
\end{align}

\subsubsection{Hyperperfusion stimulates regrowth}

{\color{red}Following the exposition in Section \ref{sec:2.2}, hyperperfusion in the microcirculation is the main stimulus for liver regrowth.}
We now establish a link between compensatory volumetric growth, described through an evolution equation, to (hyper-)perfusion through a data-based phenomenological relationship. In this work, we adopt the inflow rate density into the compartment microcirculation, $q_{\text{\tiny{supply}}}$ given in \eqref{eq:45a}, as an effective measure of (hyper-)perfusion in the microcirculation. The complex cell-scale mechanisms briefly touched upon in Section \ref{sec:2.2} and sketched in Fig. \ref{fig:drivingfactor} are not explicitly represented in this model, but implicitly incorporated through data calibration.

Motivated by existing growth models for other biological tissues \cite{RefKuhl2014}, we adopt the following form of the evolution equation for the growth factor $\vartheta$ that results from the multiplication of a growth scaling factor with a mechanism-specific growth criterion:
\begin{align}
\dot{\vartheta}(\mathbf{X}) &= \begin{cases}
k_{\vartheta}(\vartheta) \;\; \gamma_g(q_{\text{\tiny{supply}}}) & \text{for} ~ \gamma_g(q_{\text{\tiny{supply}}}) > 0,\\
0 & \text{for} ~ \gamma_g(q_{\text{\tiny{supply}}}) \leq 0.
\end{cases} 
\label{eq:theta_dot}
\end{align}
which is defined at each point of the macroscale domain.
The specific form \eqref{eq:theta_dot} enables the separation of the dependence on the size of the growth factor $\vartheta$ itself and the mechanism-specific field variable $q_{\text{\tiny{supply}}}$ into two separate variables $k_{\vartheta}$ and $\gamma_g$, which can then be considered one at a time.

We begin with the liver-specific growth criterion $\gamma_g$. Following our discussion on hyperperfusion-driven growth, we propose the following new growth criterion $\gamma_g$, 
\begin{align}
%\gamma_g = \lVert \mathbf{w}_{\text{\tiny{micro}} \rVert - \lVert \mathbf{\hat{w}}_\text{{equi}} \rVert
\gamma_g =  \frac{| q_\text{\tiny{supply}} | - | \tilde{q}_\text{\tiny{equi}} |}{ | \tilde{q}_\text{\tiny{equi}} |}.
\label{eq:growth_criterion}
\end{align}
which depends on the current homogenized blood flow rate density into the microcirculation (after resection). It is related to the homeostatic state $\tilde{q}_\text{\tiny{equi}}$, which we choose as the homogenized blood inflow rate density before resection. 
%At each point of the macroscale domain, we transfer a vector field into a scalar field by taking the Euclidean norm as follows 
%\begin{align}
%\|\mathbf{w}\| = \sqrt{\sum_{i=1}^{d} w_i^2}, \quad %\mathbf{X} \in \Omega
%\end{align}
%where $d=\{2,3\}$ is the spatial dimension. 
The growth criterion thus represents the relative increase of the current homogenized blood flow at each macroscale point with respect to the (supposedly healthy) homeostatic state before resection. We note that our choice of $\tilde{q}_\text{\tiny{equi}}$ automatically accounts for the potential special perfusion characteristics represented in a patient-specific simulation model that is set up by including patient-specific data such as the overall geometry of the liver or the location of large vessels. 

We observe that our growth criterion \eqref{eq:growth_criterion} is designed to mitigate hyperperfusion by driving growth only at the location, where a discrepancy between the current and the preferred flow exists. In \eqref{eq:theta_dot}, growth is activated only if the current local flow rate into the microcirculation exceeds a physiological equilibrium value. % In the scope of this work, we assume the physiological equilibrium $\tilde{q}_\text{{equi}}$ simply as the flow rate into the liver's microcirculation before resection. 

%Please note, that equation (\ref{eq:growth_criterion}) solely drives growth. 

The growth scaling coefficient $k_{\vartheta}$ is commonly used to prevent unbounded growth and allow the calibration of a specific growth curve. For the growth scaling coefficient, we choose the following form \cite{RefLubarda}:
\begin{align}
    k_\vartheta(\vartheta) = k_\vartheta^+\left[\frac{\vartheta^+ -  \vartheta}{\vartheta^+-1} \right]^{m_\vartheta^+}.
    \label{eq:scaling_coefficient}
\end{align}
It has three well-defined material parameters, which can be adjusted based on experimental data.
The parameter $\vartheta^+$ is the limiting value of growth. % and can be determined based on clinical observations with regard to the maximum size of lobules after growth. 
The parameter $k_\vartheta^+$ represents the growth speed, which allows calibration of the growth model with respect to time. Additionally, the parameter $m_\vartheta^+$ allows the adjustment of the nonlinearity of the growth process.

\subsubsection{Calibration based on experimental data}

Experimentally determined graphs of liver mass regeneration typically show a rapid initial increase in mass due to the high regeneration rate, followed by a slower increase as the resected liver approaches its original mass \cite{RefLiverSize,RefFurchtgott,RefKoniaris}. We can now fit the parameters of the growth scaling coefficient \eqref{eq:scaling_coefficient} to match the growth curves that are observed in experimental measurements conducted on rats.

Available data refer to growth of a complete liver, and do not feature local growth in a particular region. We therefore assume that the local growth process with respect to the AV sizes that we will choose for the simulation scenarios in the following does not differ from the growth process averaged over the complete organ. In addition, during calibration of the growth scaling coefficient, we set the growth criterion $\gamma_g = 1$. By doing so, we treat $\gamma_g$ as a growth driver that merely indicates whether growth is activated or not, without specifying its intensity in different regions of the liver. This allows us to focus on calibrating the free parameters of the growth scaling function \eqref{eq:scaling_coefficient}, so that the overall growth curves produced by the model match the global trends observed in the experimental data available.

\begin{figure}[ht]
\centering
\subfigure[Variation of nonlinearity parameters  $m_\vartheta^+$ \label{a252}]{
 \begin{tikzpicture}
   \node[] (pic) at (0,0) {\includegraphics[width=75mm, angle = 0]{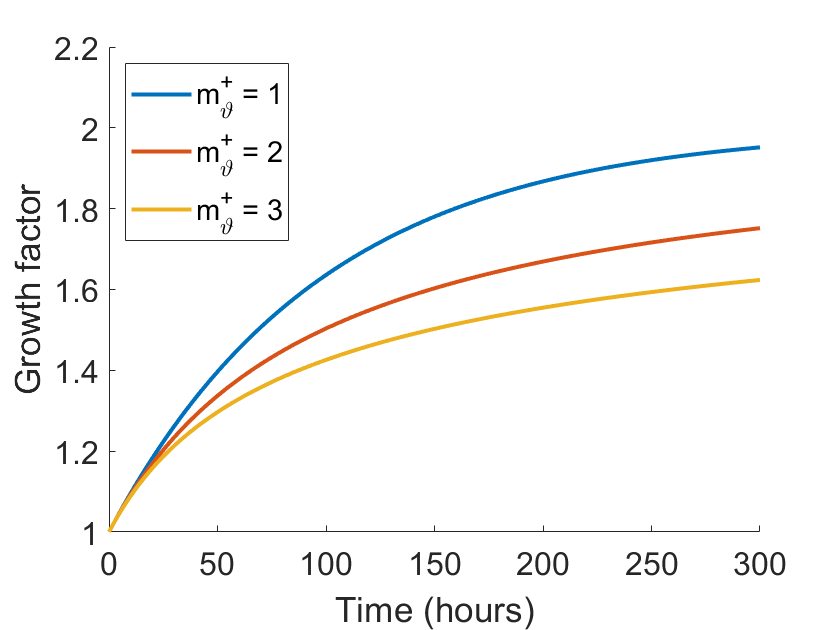}};
\end{tikzpicture}}
\hfill
\subfigure[Variation of growth speed parameters $k_\vartheta^+$ \label{b252}]{
 \begin{tikzpicture}
   \node[] (pic) at (3.25,0) {\includegraphics[width=75.0mm]{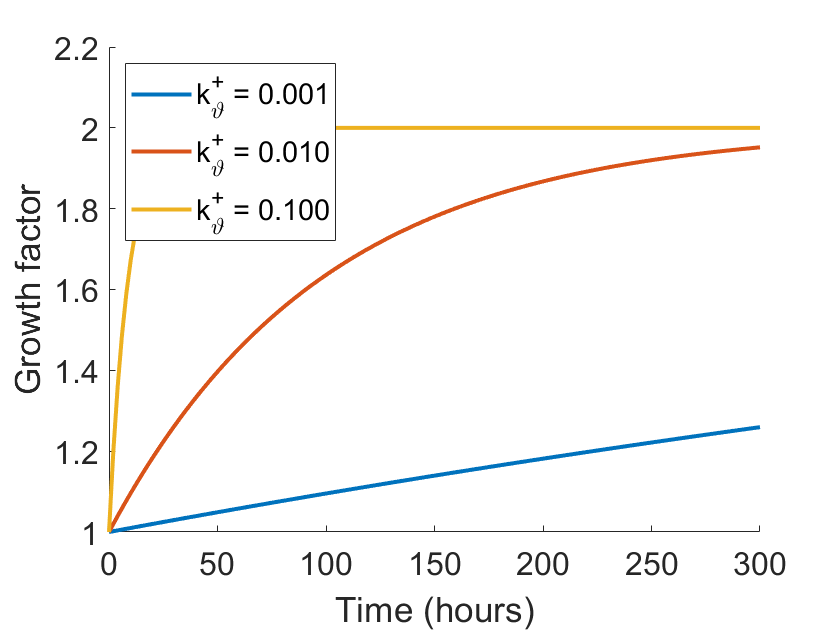}};
\end{tikzpicture}}
  \caption{Evolution of the growth factor $\vartheta$ over time.}
\label{fig:growthrfactorvariable}
\end{figure}

In the first step, we set the limiting value of growth $\vartheta^+ = 2.0$ for all examples as experimental measurements show that the grown lobules can increase to double their original size \cite{tsomaia2020liver}. In the next step, we conduct a sensitivity study for the remaining two parameters by numerically integrating the evolution equation \eqref{eq:theta_dot} with a simple forward Euler method and a time step $\Delta t = 2$ h. In Fig. \ref{a252}, we plot the resulting evolution of the growth factor $\vartheta$ versus time for different $m_\vartheta^+$ and a fixed $k_\vartheta^+ = 0.01$. We see that the nonlinearity of the growth process is influenced by the parameter $m_\vartheta^+$.
In Fig. \ref{b252}, we plot the evolution of the growth factor for different values of the growth speed $k_\vartheta^+$ and a fixed $m_\vartheta^+ = 1$.

\begin{figure}[t]
\centering
\subfigure[Remnant liver weight recovery reported in \cite{RefNishiyama} \label{a253}]{
 \begin{tikzpicture}
   \node[] (pic) at (0,0) {\includegraphics[width=75mm, angle = 0]{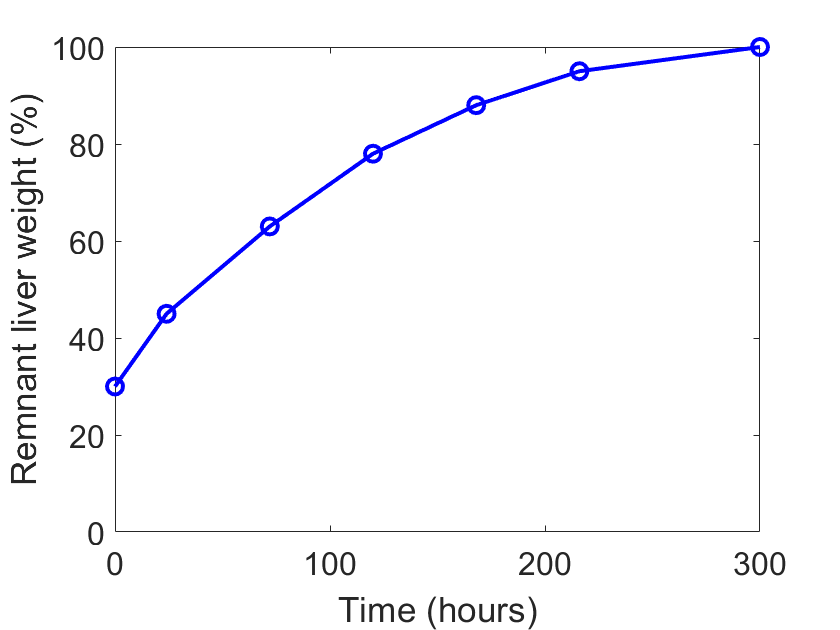}};
\end{tikzpicture}}
\hfill
\subfigure[Our model vs.\ experimental mass data \label{b253}]{
 \begin{tikzpicture}
   \node[] (pic) at (3.25,0) {\includegraphics[width=75.0mm]{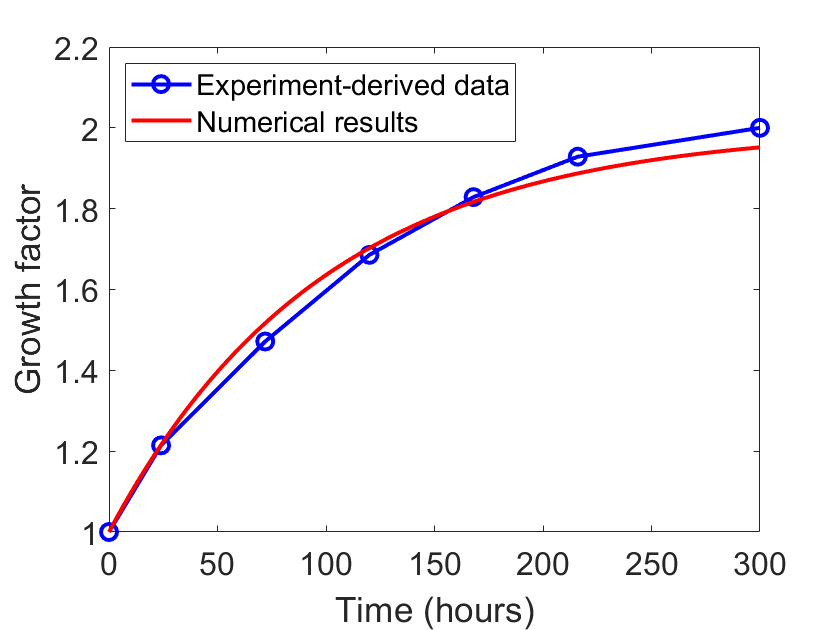}};
\end{tikzpicture}}
  \caption{Calibration of the growth factor $\vartheta$ with mass data reported in \cite{RefNishiyama}.}
\label{fig:growthrfactorexperimentalnumerical1}
\end{figure}

To calibrate the growth factor, we use the data reported in \cite{RefNishiyama} which come from experimental measurements conducted on mice. In these experiments, 70 \% of the liver was surgically removed. The recovery of liver mass over time was monitored and the remnant liver weight was measured at different time steps to assess the regeneration process (see Fig. \ref{a253}).

Next, we use the remnant liver weight (mass data) in Fig. \ref{a253} to determine the corresponding growth factor $\vartheta$, which is defined in equation \eqref{eqn:theta_mass} and is directly linked to the mass. The determined growth factor (see blue curve in Fig. \ref{b253}) ranges between 1.0 and 2.0, where $\vartheta = 1.0$ corresponds to the initial mass (30\%) and $\vartheta = 2.0$ corresponds to the fully regenerated liver mass (100\%) due to the previously determined limiting value of growth $\vartheta^+ = 2.0$.

Based on {\color{red}these} data and the sensitivity study above, we choose the nonlinearity parameter of the growth process to be $m_\vartheta^+ = 1.0$ and the growth speed to be $k_\vartheta^+ = 0.01$.
In Fig. \ref{b253}, we plot the resulting evolution of the modeled growth factor alongside the growth factor derived from the experimental mass data in \cite{RefNishiyama}. We observe a {\color{red} good agreement} for our choice of parameters. In addition, we observe that the duration of the growth process of 300 hours (12.5 days) also matches the available data  \cite{RefNishiyama,RefFurchtgott}.

\subsection{Volumetric growth vs.\ lobular remodeling}
\label{sec:4.3}

% Adnan: here we need reference for the remodeling phenomena I refer to. Please add 2-3 in each paragraph.

We recall from Section \ref{sec:2.2} that in addition to the restoration of lost mass via hepatocyte proliferation, liver regeneration also involves changes across the hierarchical vasculature \cite{lorenz2018mechanosensing,grosse2021role}. The compartmentalization of our model illustrated in Fig. \ref{fig:Multicompartment} allows the following interpretation with respect to lobular remodeling.

For the largest vessels of the resected supplying and draining tree that we keep resolved,
we assume that remodeling solely involves adaptations in length and position rather than angiogenic changes in their topology. In our model, the vessels remain straight cylindrical tubes during growth. Furthermore, we assume that they maintain their supply and drainage activity in the same way as before, but scaling up their throughput after resection via a corresponding change in diameter (vasodilation) \cite{grosse2021role}. In our model, we therefore update the positions of the nodes of these vessels according to the displacement field of the grown tissue represented in the compartment microcirculation, and then adjust the diameter according to the required blood flow \cite{Jannes}.

In our model, the medium- and smaller-sized vessels in the intrahepatic branches are homogenized into two compartments for supply and drainage. In a growing liver, these vessels 
also undergo angiogenesis and remodeling to connect new tissue areas with the existing vascular network, supporting uniform blood distribution within the liver \cite{RefMichalopoulos,grosse2021role}. In our model, we assume that these reorganization effects are reflected through geometric updates during homogenization. We therefore update the material parameters of the homogenized model as defined in equations \eqref{eq:updatepermeability} and \eqref{eq:updateperfusioncoeff}, using the geometry-related quantities $\mathbf{F}$ and $J$ \cite{Jannes}.

At the capillary level, the existing network of sinusoids must undergo angiogenesis to ensure efficient metabolic activity of the proliferated hepatocytes \cite{RefMichalopoulos,grosse2021role}. To this end, new microvessels sprout from existing ones to support the increased metabolic demand of the regenerating tissue. These angiogenic processes are dynamic and can therefore not be explicitly represented in our quasi-static model. In our approach, we therefore assume that they take place between the quasi-static states that we compute, and their effect is therefore available instantly. As a consequence, we assume that the resistance of the sinusoid network the regrown liver tissue always corresponds to the same (healthy) tissue and therefore use the same permeability in the compartment microcirculation at all times.

\subsection{Prototypical model problem in 2D}\label{sec:Numerical_example_disk_growth}

We utilize the same two-dimensional test problem of a circular disk with planar trees used in Section \ref{sec:Numerical_example_disk} to verify the basic behavior of the multi-compartment poroelastic growth model. We only add the poroelastic growth model presented in \ref{sec:Growth} and \ref{sec:evolutionequation}, while the rest of the setup (perfusion-related parameters, discretization, supplying and draining tree structures, compartmentalization and homogenization procedure) remains identical to what was shown in Section \ref{sec:Numerical_example_disk}. We adopt Young's modulus $E = 5 \; \text{kg}\; \text{mm}^{-1}\;\text{s}^{-2}$ and Poisson's ratio $\nu = 0.35$ reported for the liver in \cite{RefElasticPropertiesLiver}. For the contact boundary condition on the outer surface of the circular disk, we require a parameter in the nonlinear stiffness relation \eqref{eqn:stiffnesscontact}, which we choose as $\alpha = 5 \times 10^{-3}$. This choice is based on the elastic properties of the surrounding organs of the liver \cite{RefStiffnessOrgans}.

For simplicity, we set the equilibrium flow rate density $\tilde{q}_\text{{equi}}$ to the supply flow rate $q_{\text{\tiny{supply}}}$, plotted in Fig. \ref{fig:circle_intercompartmental_flow}. %The solution field $q_{\text{\tiny{supply}}}$ results from the parameters provided in Tab. \ref{tab_vesseltrees2D}, which define the baseline physiological state of the system. 
To induce growth within the model, we introduce a perturbation by doubling the flow rate through the supplying vascular tree. This increase in flow rate serves as a stimulus, pushing the system out of its equilibrium state and mimicking tissue growth driven by hyperperfusion.

\begin{figure}[ht]
\centering
\hspace{-1cm}
\begin{tikzpicture}[transform shape]
  \node[] at (0,4.0) {t = 0 h};
  \node[] at (8.0,4.0) {t = 300 h};
  \node[] (pic) at (4,5.0) {\includegraphics[width=60mm, angle = 0]{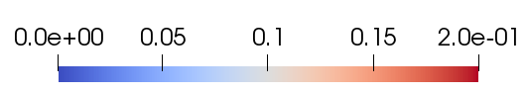}};
  \node[] (pic) at (0,0.5) {\includegraphics[width=65mm, angle = 0]{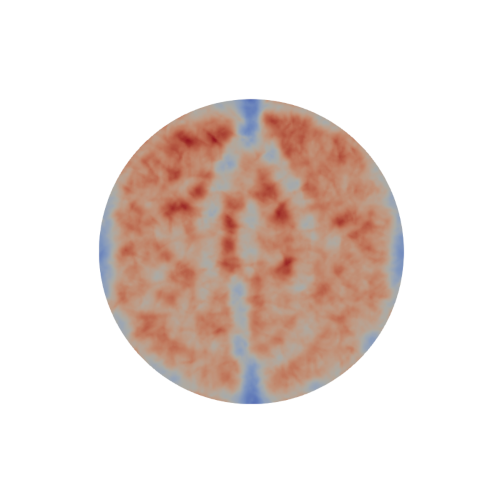}};
  \node[] at (0.0,-3.0) {$A_{C} = 314.16 \; \text{mm}^2$};
  \node[] (pic) at (8.0,0.5) {\includegraphics[width=65mm, angle = 0]{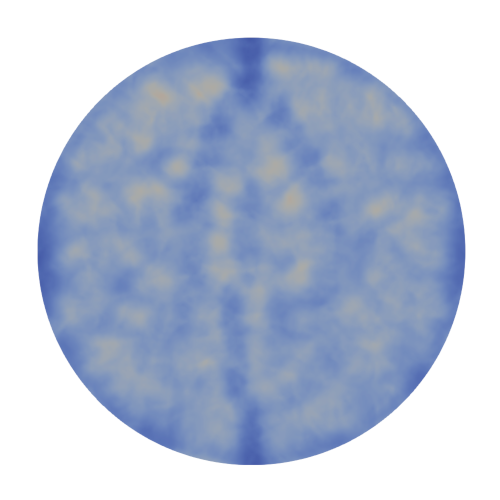}};
  \node[] at (8.0,-3.0) {$A_{C} = 627.08 \; \text{mm}^2$};
\end{tikzpicture}
\caption{Homogenized volumetric flow rate $q_{\text{\tiny{supply}}}$ [$ \text{s}^{-1}$] after doubling the inflow rate through the vascular tree. $A_{C}$ denotes the area of the disk.}
\label{fig:growth_flow_circle}
\end{figure}
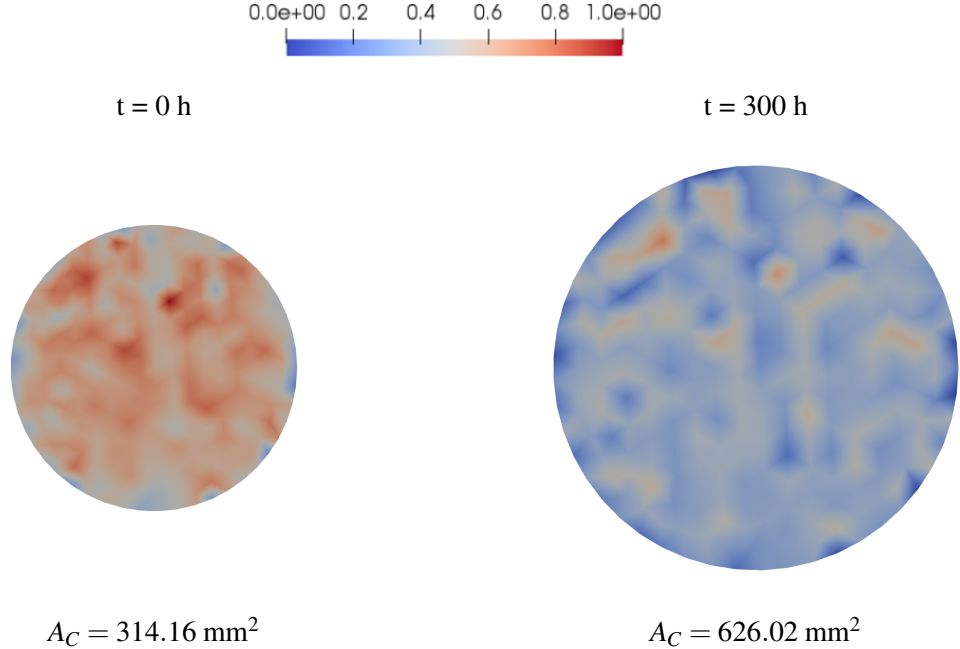

Figure \ref{fig:growth_flow_circle} illustrates the resulting growth of the circular disk. We observe that the disk grows nearly uniformly, with no noticeable areas exhibiting disproportionate growth. This aligns with our expectation as the full circular disk receives a relatively uniform blood supply from the vascular network. We also observe that the fully regrown circular disk has doubled in size after 300 hours, such that double the flow rate through the supplying vascular tree leads to a flow rate density close to our choice of $\tilde{q}_\text{{equi}}$.

%\section{Modeling liver regrowth}
%\section{Numerical examples}
\section{Towards simulation based assessment of liver regrowth}
\label{sec:Numerical_examples_Liver}

%\subsection{Describe liver regeneration (+schematic overview of modeling approach)}
%\subsection{Choose driving factors of growth model}
%\subsection{Growth direction based on reducing residual stresses (in lower scales of multi-compartment model) $\Rightarrow$ Anistropic growth}
%\subsection{Construct liver growth evolution equation based on driving factors and growth direction $\Rightarrow$ Connection of stress field with growth parameters}

%Liver resection triggers a complex regenerative process that is heavily influenced by the organ's vascular architecture and the redistribution of blood flow. The close relationship between liver regeneration and blood perfusion is well-known, as the liver’s highly vascularized nature plays a crucial role in maintaining essential functions like metabolism and detoxification \cite{RefMichalopoulos}.

%In the previous sections, we established a multiscale modeling framework for simulating liver perfusion and hyperperfusion-driven tissue growth. 
We now utilize our modeling framework to computationally investigate liver growth after surgical resection. We first describe the patient-specific liver model used in this study, along with the associated vascular geometry. We then focus on simulations of liver perfusion, which we need as an input to our growth evolution model. Finally, we focus on computationally analyzing liver regrowth as represented by our model.

\subsection{Patient-specific geometry and synthetic vascularization}
\begin{figure}[ht]
\centering
\subfigure[2D slice \label{a4}]{
 \begin{tikzpicture}
   \node[] (pic) at (0,0) {\includegraphics[width=60mm]{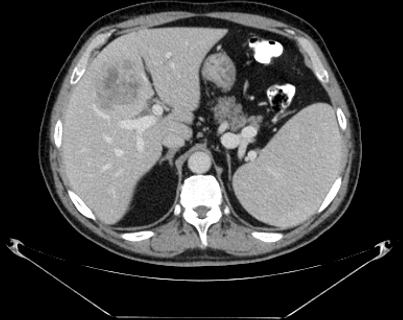}};
\end{tikzpicture}}
\hfill
\subfigure[Segmentation mask of liver (green) \label{b4}]{
 \begin{tikzpicture}
   \node[] (pic) at (0,0) {\includegraphics[width=60mm]{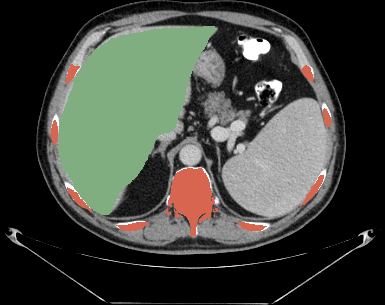}};
\end{tikzpicture}}
    \caption{Abdominal CT scan with a resolution of 0.977 x 0.977 mm and a slice thickness of 2.5 mm.} 
  \label{fig:segmentation}
\end{figure}

Our patient-specific liver geometry is obtained from CT scans provided in \cite{RefSegmentation}. %Typically, a 3D voxel model can be created by stacking 2D slices on top of each other. 
We make use of the open source software package 3D Slicer\footnote[1]{https://www.slicer.org/} and the free software Autodesk Meshmixer\footnote[2]{https://meshmixer.com/} to segment the liver.
Figure \ref{a4} shows a 2D slice of a CT scan, which has a resolution of 0.977 x 0.977 mm within each image, with a spacing of 2.5 mm between the slices. The segmentation mask of the liver domain is illustrated in Fig. \ref{b4}. 

\begin{figure}[ht]
\centering
\subfigure[Anterior view \label{a5}]{
 \begin{tikzpicture}
   \node[] (pic) at (0,0) {\includegraphics[width=78mm]{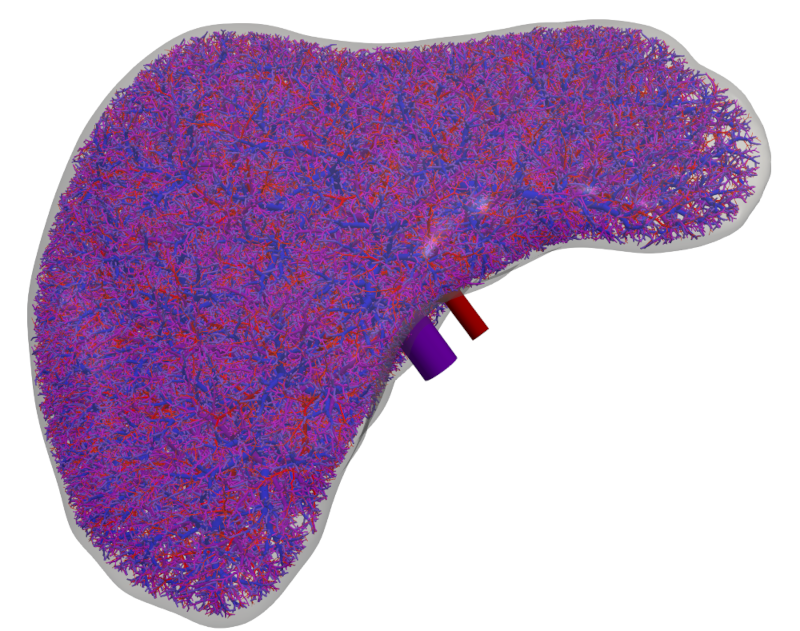}};
\end{tikzpicture}}
\hfill
\subfigure[Inferior view \label{b5}]{
 \begin{tikzpicture}
   \node[] (pic) at (0,0) {\includegraphics[width=78mm]{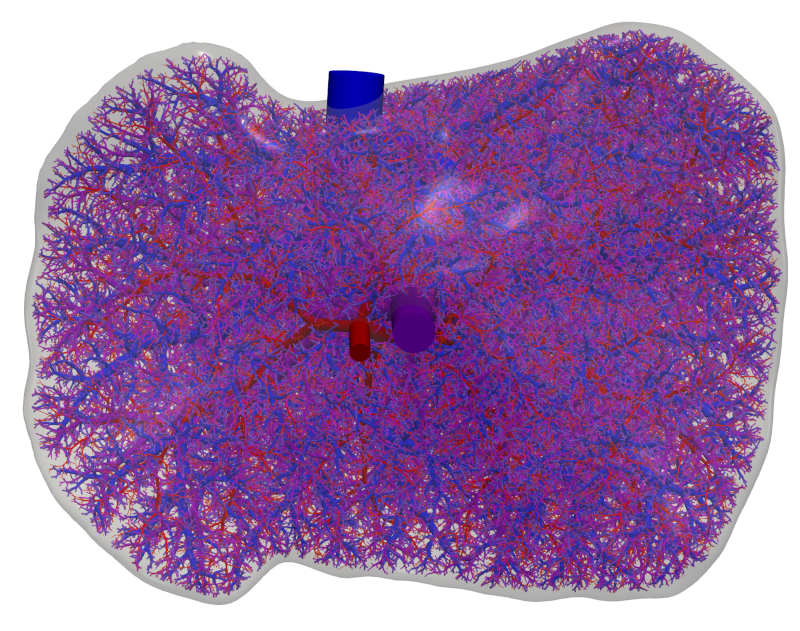}};
\end{tikzpicture}}
  \caption{Patient-specific liver domain with vascular trees: hepatic artery (red), portal vein (purple) and hepatic vein (blue). 
  Only 113,000 vessels per tree are visualized. {\color{teal}For a comprehensive visual impression of the synthetic tree structure, we refer to the supplementary video, which is part of this article.}}
  \label{fig:LiverModelWithVessels}
\end{figure}

We then generate the vasculature of the liver synthetically as described in Section \ref{sec:VascularTrees}, using the patient-specific liver domain as a boundary and locations of the patient-specific root locations. %We emphasize again that we lump the portal vein and the hepatic artery into one supplying tree, as their trees run largely in parallel.
Figure \ref{fig:LiverModelWithVessels} shows the patient-specific liver domain with the synthetically generated hepatic artery, portal vein and hepatic vein. %This allows for accurate modeling of blood flow into the poroelastic domain while maintaining computational efficiency. 
The underlying parameters for synthetic tree generation are summarized in Table \ref{tab_vesseltrees3D}.

\jh{
\begin{table}[t]
%\centering
\caption{Parameters for synthetic tree generation of the full liver example.}
\label{tab_vesseltrees3D}
\begin{center}
\begin{tabular}{l l l l l l l l} 
 \hline
  & $N_\text{vessel}$ & $N_\text{term}$ & $\Delta p$ [$\frac{\text{kg}}{\text{mm} \, \text{s}}$]  & $\hat{Q}_\text{perf}$ [$\frac{\text{mm}^3}{\text{s}}$] & $m_b$ [$\frac{\mu \text{W}^3}{\text{mm}^3}$] & $\eta$ [$\frac{\text{kg}}{\text{mm}\,\text{s}}$]\\ [0.5ex] 
 \hline
 Hepatic artery & 1,781,564 & 890,782 & 1.176  & 4,000 & 0.6 & $3.6 \times 10^{-6}$ \\ 
 Portal vein & 1,767,920 & 883,960 & 0.217  & 16,000 & 0.6 & $3.6 \times 10^{-6}$ \\ 
 %\hline
 %\hline
 Hepatic vein & 1,756,428 & 878,214 & -0.045  & 20,000 & 0.6 & $3.6 \times 10^{-6}$ \\ [1ex] 
 \hline
\end{tabular}
\end{center}
\end{table}
}

\begin{figure}[t]
    \centering
    \includegraphics[width=0.9\textwidth]{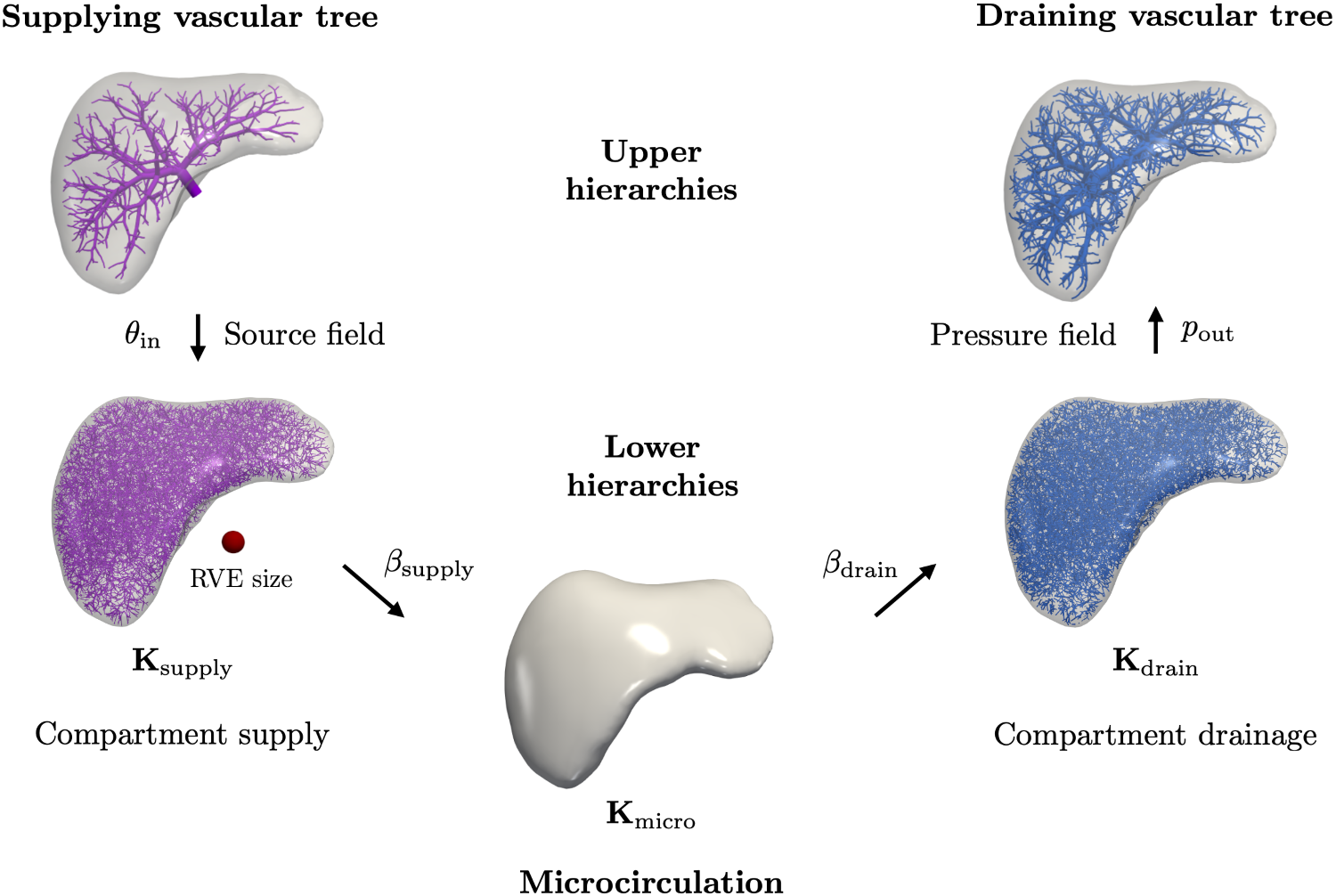}
    \caption{Compartmentalization of vascular trees for the liver.}
    \label{fig:MulticompartmentLiver}
\end{figure}

\subsection{Model results for liver (hyper-)perfusion}

\begin{figure}[ht]
\centering
\subfigure[Full liver \label{fig:Livermesh}]{
 \begin{tikzpicture}
   \node[] (pic) at (0,0) {\includegraphics[height=6.2cm]{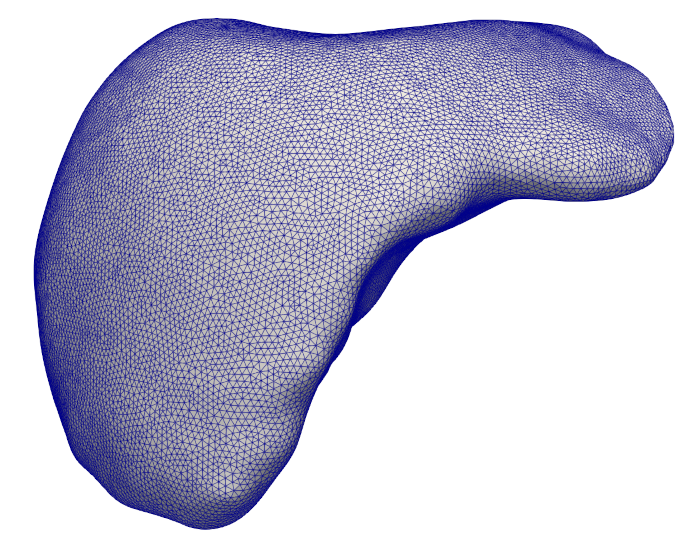}};
\end{tikzpicture}}
\hspace{1cm}
\subfigure[Resected liver \label{fig:LivermeshResected}]{
 \begin{tikzpicture}
   \node[] (pic) at (0,0) {\includegraphics[height=6.2cm]{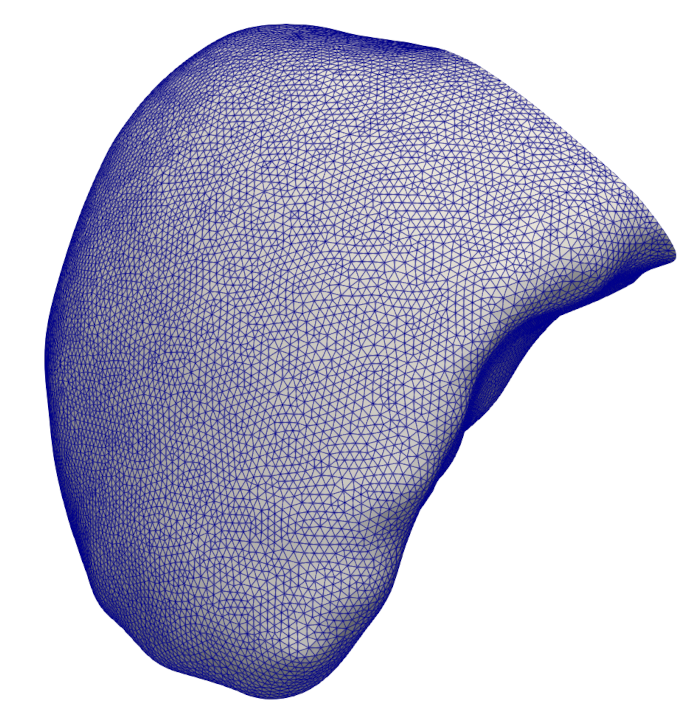}};
\end{tikzpicture}}
  \caption{Finite element meshes.} 
  \label{fig:MeshLiver}
\end{figure}

We first focus on simulating liver perfusion in the unresected liver, assuming a healthy state.  % These parameters provide the foundational data required for accurate modeling of liver perfusion.
We divide the synthetic vascular network generated in the previous section into spatially co-existing compartments. To this end, vessels with a radius exceeding $r_{\textit{thresh}} = 0.2\,mm$ belong to the upper hierarchies that are kept resolved, while those with a radius of $r_{\textit{thresh}} = 0.2\,mm$ or less are assigned to the lower hierarchies. The compartmentalization of the liver perfusion trees is illustrated in Fig. \ref{fig:MulticompartmentLiver}. We emphasize again that for the homogenization process we lump the portal vein and the hepatic artery into one supplying tree, as their trees run largely in parallel.
We then generate a finite element mesh illustrated in Fig. \ref{fig:Livermesh} that consists of 870,626 tetrahedral elements. %For our multi-compartment model, we use the the model parameters as in the test problem above. %, summarized in Table \ref{table_simulation_parameters_circle}. 
For the computation of the homogenized material parameters, we choose a radius of the spherical AV of 5 mm. 

{\color{teal}Given a characteristic length of the complete liver of approx.\ 12.5 cm, an AV radius of 5 mm, and a maximum radius $r_{\textit{thresh}} = 0.2\,mm$ of the vessels to be homogenized, we obtain a scale separation well above one order of magnitude. We note that choosing a larger value of $r_{\textit{thresh}}$ (for instance to 2 mm) requires in turn an AV size of at least 2 cm, making scale separation questionable. Conversely, refining $r_{\textit{thresh}}$ (for instance to 0.02 mm) falls below the smallest arterioles and venules represented in our synthetic tree. Given these limitations in both directions, the AV size and $r_{\textit{thresh}}$ can realistically be varied by a factor of two or three, but this has only a limited impact on the model's outcome.}

\subsubsection{Full liver before resection}

\begin{figure}[h!t]
\centering
    \begin{tikzpicture}
      \node[] (pic) at (-0.75,2.5) {\includegraphics[height=6mm, angle = 0]{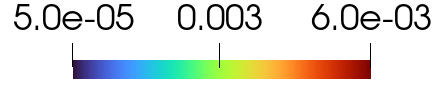}};
      \node[] (pic) at (-0.75,0) {\includegraphics[width=50mm, angle = 0]{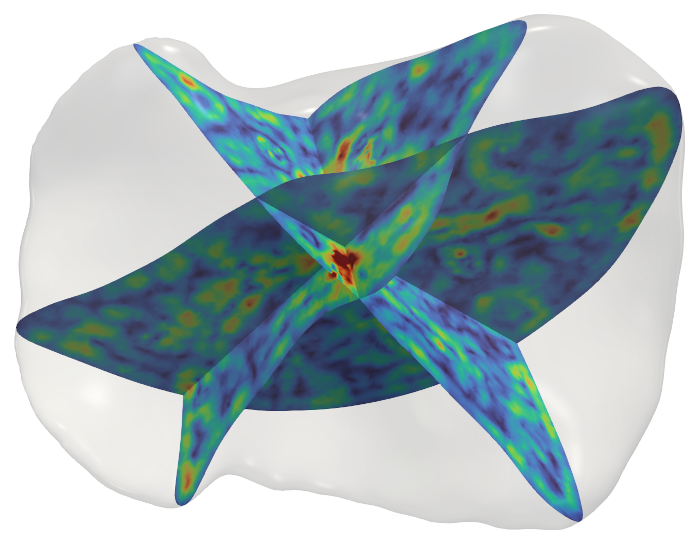}};
      \node[] at (-0.75,-2.75) {Compartment supply};
      \node[] (pic) at (5,2.5) {\includegraphics[height=6mm, angle = 0]{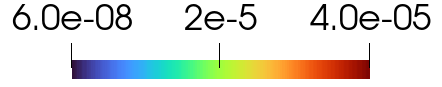}};
      \node[] (pic) at (5,0) {\includegraphics[width=50mm, angle = 0]{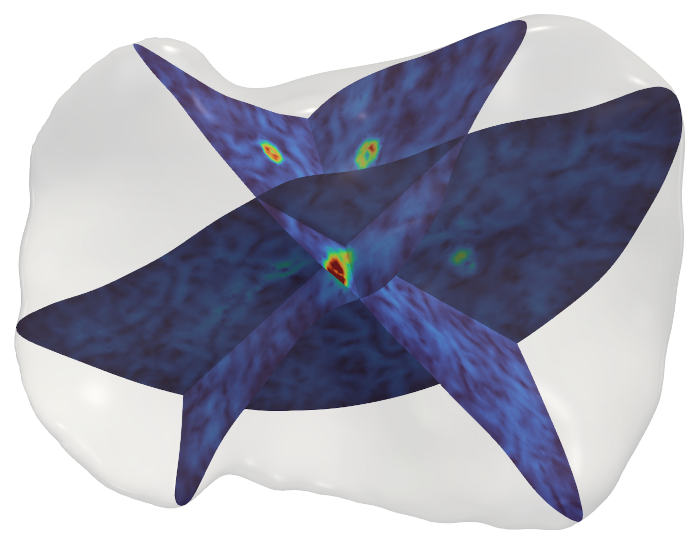}};
      \node[] at (5,-2.75) {Compartment microcirculation};
      \node[] (pic) at (10.75,2.5) {\includegraphics[height=6mm, angle = 0]{Images/v_colorbar_supply_drain.png}};
      \node[] (pic) at (10.75,0) {\includegraphics[width=50mm, angle = 0]{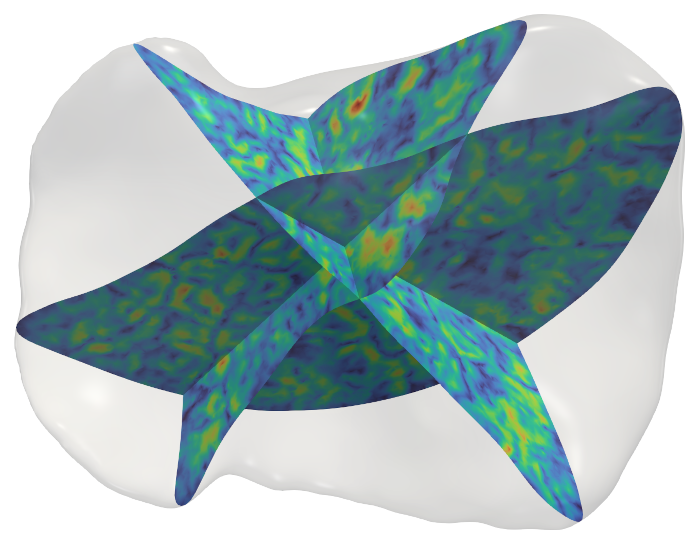}};
      \node[] at (10.75,-2.75) {Compartment drainage};
     \end{tikzpicture}
    \caption{Magnitude of the homogenized velocity fields $\lVert \mathbf{w}_i \rVert$ [$\text{mm} \; \text{s}^{-1}$] of the full liver before resection.}
    \label{fig:velocity_Full_Liver}

\vspace{1.cm}

\begin{tikzpicture}[transform shape, scale=0.875]
  \node[] (pic) at (,0) {\includegraphics[height = 7 mm, angle=0]{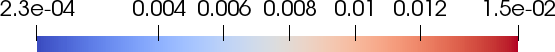}};
%  \node[] (pic) at (0.75,0.5) {\includegraphics[width=75mm, angle=0]{./Images/Full_Liver_Q.png}};
%  \node[] at (0.75,-3.) {Inferior view};
  \node[] (pic) at (,-5.1) {\includegraphics[width=120mm, angle=0]{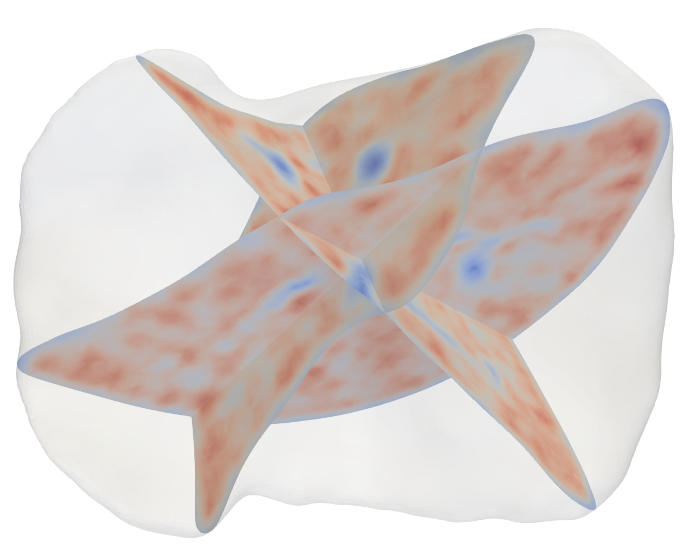}};
%  \node[] at (,-10) {View with slices};
\end{tikzpicture}
\caption{Homogenized volumetric flow rate density $q_{\text{\tiny{supply}}}$ [$ \text{s}^{-1}$] of the full liver before resection.}
\label{fig:flow_Full_Liver_same_scale}
\end{figure}

We first consider the perfusion results before resection.
In Fig. \ref{fig:velocity_Full_Liver}, we show the corresponding magnitudes of the homogenized velocities. In compartments supply and drainage, we can observe a larger flow along vessel segments of both supplying and draining trees. In the compartment microcirculation, we observe %a much more uniform flow distribution and significantly lower velocity magnitudes. %As previously in the 2D example (Section \ref{sec:Numerical_example_disk}) discussed, no organized flow pattern is also observed within the microcirculation of the liver. 
velocity magnitudes that are at three %to five 
orders of magnitude lower than those in the compartments supply and drainage. This confirms that the compartment microcirculation connects the compartments supply and drainage by representing the resistance of the sinusoids, and does not redistribute flow across the macroscale domain. %$  and no significant fluid redistribution occurs at the microscale level. 

Figure \ref{fig:flow_Full_Liver_same_scale} shows the homogenized volumetric flow rate density $q_{\text{\tiny{supply}}}$ of the full liver. We can observe an almost uniform blood supply to the liver which is to be expected in the healthy state.

\subsubsection{Liver after partial resection}

Partial anatomical resections involve removing portions of the liver along anatomical boundaries, e.g.\ to remove a liver tumor.
For planning a partial resection, surgeons often follow the Couinaud classification, which partitions the liver into eight functionally independent segments. Each segment is supplied by its own larger branch of the supplying tree, which splits into smaller vessels within the segment, and is connected to a larger branch of the draining tree. % (see Fig. \ref{fig:LiverSegments}). 
This enables segmental resections without affecting other segments \cite{RefVibert}. Examples include left hepatectomy, right hepatectomy, and segmental resections. In addition to removing the pathology, selecting corresponding cutting planes is also influenced by the objective of preserving as much liver tissue as possible. Surgeons have to carefully balance these two conflicting objectives to ensure sufficient liver function while achieving the best possible outcome in terms of removing the pathology \cite{RefChrist}. 

%A non-anatomical resection involves removing a part of the liver that does not strictly follow the anatomical boundaries, for instance, when the tumor is distributed over multiple segments.

In this context, assessing changes in perfusion, flow redistribution and the mechanical response of the resected liver has clinical relevance \cite{RefMichalopoulos}. An insufficient blood supply can result in ischemia, where the tissue does not receive enough oxygen and nutrients. Prolonged ischemia can cause tissue damage and death. Therefore, the liver's regenerative capacity can be impaired if a significant part of the tissue does not receive prolonged sufficient blood supply. The parts of the associated isolated vasculature, which are cut off and no longer supplied with blood, are often denoted as \textit{orphans}.

\begin{figure}[h!t]
\centering
    \begin{tikzpicture}
      \node[] (pic) at (-5.1,0) {\includegraphics[width=75mm, angle = 0]{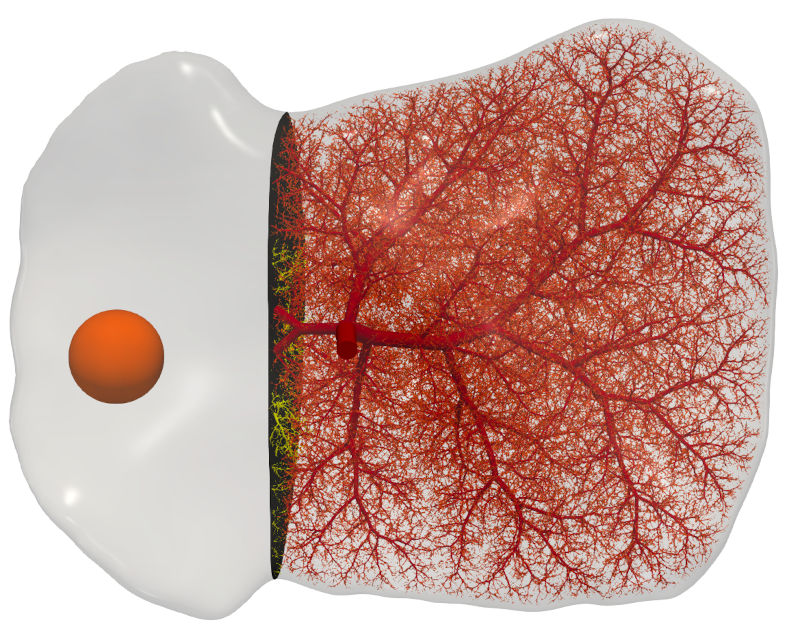}};
      \node[] at (-5.1,-3.5) {Hepatic artery};
      \node[] (pic) at (3.1,0) {\includegraphics[width=75mm, angle = 0]{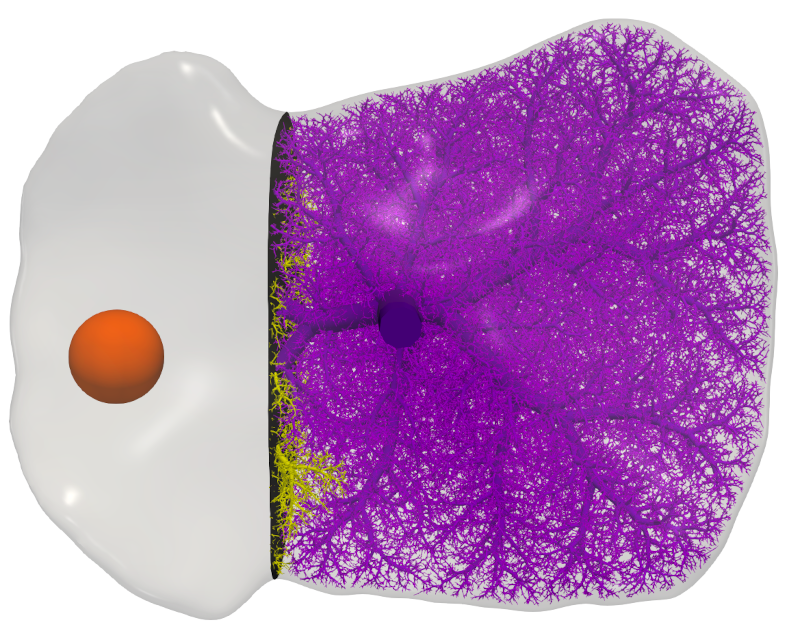}};
      \node[] at (3.1,-3.5) {Portal vein};  
      \node[] (pic) at (-1,-7) {\includegraphics[width=75mm, angle = 0]{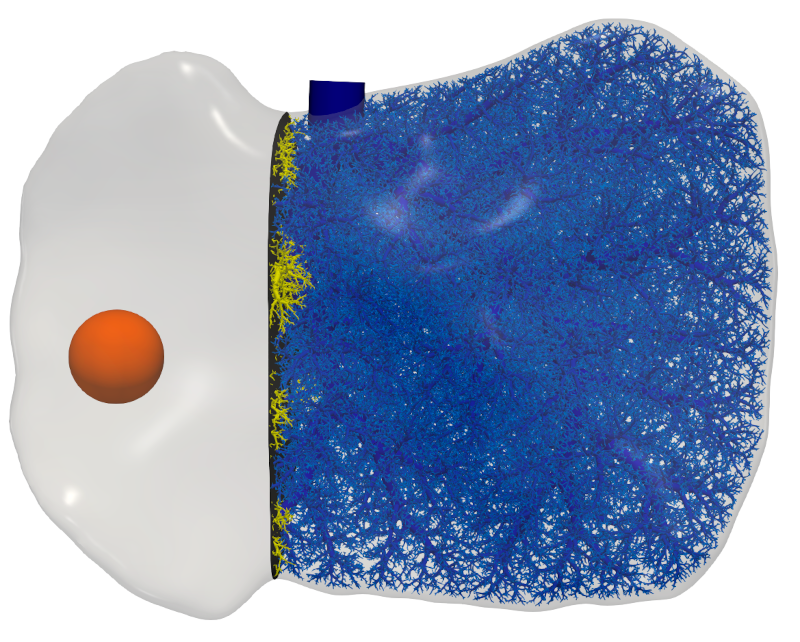}};
      \node[] at (-1,-10.5) {Hepatic vein};
      \node[] at (-6.6,-5.6) {Orphan vessels};
      \draw [->] (-5.2,-5.6) -- (-2.25,-6.5);
      \node[] at (-7,-7.3) {Tumor affected region};
      \draw [->] (-5.2,-7.3) -- (-4.2,-7.3);
      \node[] at (-6.4,-9) {Cutting plane};
      \draw [->] (-5.2,-9) -- (-2.4,-8.5);
     \end{tikzpicture}
    \caption{Synthetic vasculature after resection (inferior view). Orphan vessels are highlighted in yellow.}
    \label{fig:Synthetic vasculature after resection}
\end{figure}

We now consider a scenario where a tumor is located in the left lateral section of the liver as illustrated in Fig. \ref{fig:Synthetic vasculature after resection}. The partial resection required to remove the tumor-affected region involves a standard surgical cut that removes liver segments two and three according to the Couinaud classification. %(see Fig. \ref{fig:LiverSegments}). 
The cut also implies cutting the vasculature. We illustrate the ``active" vessels that are still connected to the root of the corresponding tree. All orphan vessels that lost this connection are highlighted in yellow. The number of remaining active vessel segments, terminal nodes and orphan vessel segments are tabulated in Table \ref{tab_vesseltrees3D_resected}.

\begin{table}[t]
%\centering
\caption{Vascular tree data for the resected liver.}
\label{tab_vesseltrees3D_resected}
\begin{center}
\begin{tabular}{l l l l} 
 \hline
  & $N_\text{vessel}$ & $N_\text{term}$ & $N_\text{orphan}$\\ [0.5ex] 
 \hline
 Hepatic artery & 1,384,866 & 692,058 & 12,245\\ 
 Portal vein & 1,355,034 & 677,256 & 30,817 \\ 
 Hepatic vein & 1,361,280 & 680,257 & 16,097 \\ [1ex] 
 \hline
\end{tabular}
\end{center}
\end{table}

Vessels that are cut are sealed during surgery to prevent blood loss. In our simulations, we thus assume that blood flow through any cut segment does not occur. The blood flow associated with cut segments is then redistributed evenly across the remaining portion of the active tree model, in line with our assumptions discussed in Section \ref{sec:4.3}. We note that after partial resection, the cardiovascular system tries to reduce the overall blood flow rate in the liver by reducing the flow in the hepatic artery. In the scope of the present study, this effect is not considered. The resected liver domain is discretized by a finite element mesh of 674,389 tetrahedral elements, which is illustrated in Fig. \ref{fig:LivermeshResected}. The homogenization procedure described above is repeated based on the cut synthetic vasculature and the perfusion simulation is re-run.

Figure \ref{resectedliver} plots the resulting homogenized volumetric flow rate density of the resected liver before regrowth. On the one hand, we can clearly see a much higher overall flow rate as compared to the results for the full liver in Fig. \ref{fig:flow_Full_Liver_same_scale}. The increased flow rate density represents the state of hyperperfusion, as in our model, the same amount of blood needs to pass through a smaller domain after partial liver resection. On the other hand, we see a significantly lower flow rate density along the cut plane. This is due to the existence of orphans in this area that do not receive blood supply. In Fig. \ref{fig:Synthetic vasculature after resection}, we can see that in the corresponding region, there are no active vessels of the portal vein.

\subsection{Model results for liver regrowth}
\label{subsec:Liver_regrowth}

In the next step, we consider the modeling of liver regrowth, where our focus is on how well our modeling framework represents characteristic phenomena associated with liver regeneration at the organ scale. %Our approach involves detailed simulation and examination of the regrowth process using our multiscale-multiphysics model presented in Fig. \ref{fig:Modelingframework}.
Our growth criterion \eqref{eq:growth_criterion} requires a measure for the homeostatic perfusion state, which we have chosen to express via the homogenized flow rate density $\tilde{q}_\text{\tiny{equi}}$ of the full liver before resection. Otherwise, we use the same parameters as specified in Section \ref{sec:Numerical_example_disk_growth}. We integrate with respect to time using a standard explicit forward Euler method with a time step of $\Delta t = 15$ hours, up to a final time of t = 300 hours after resection.

% and Table \ref{table_simulation_parameters_circle}.

\begin{figure}[t]
\centering
\begin{tikzpicture}
  % Colorbar at the top
  \node[] (pic) at (5,3.0) {\includegraphics[width=75mm, angle=0]{./Images/Q_supply_colorbar.png}};
  \end{tikzpicture}

\subfigure[Resected liver at t = 0, {$V_{L} = 1,918,501 \; \text{mm}^3$} \label{resectedliver}]
{\includegraphics[width=80mm]{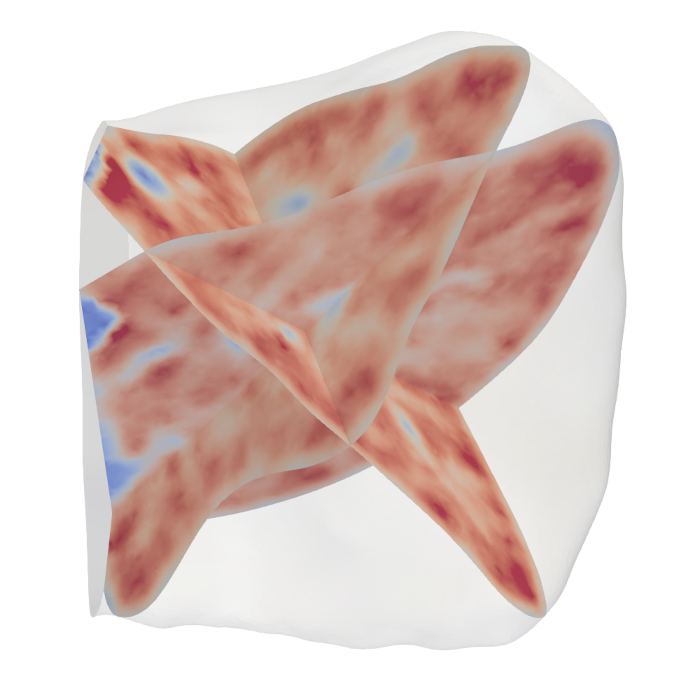}} \hfill
\subfigure[Regrown liver at t = 300 h, {$V_{L} = 2,370,351 \; \text{mm}^3$}\label{grownliver}] {\includegraphics[width=80mm]{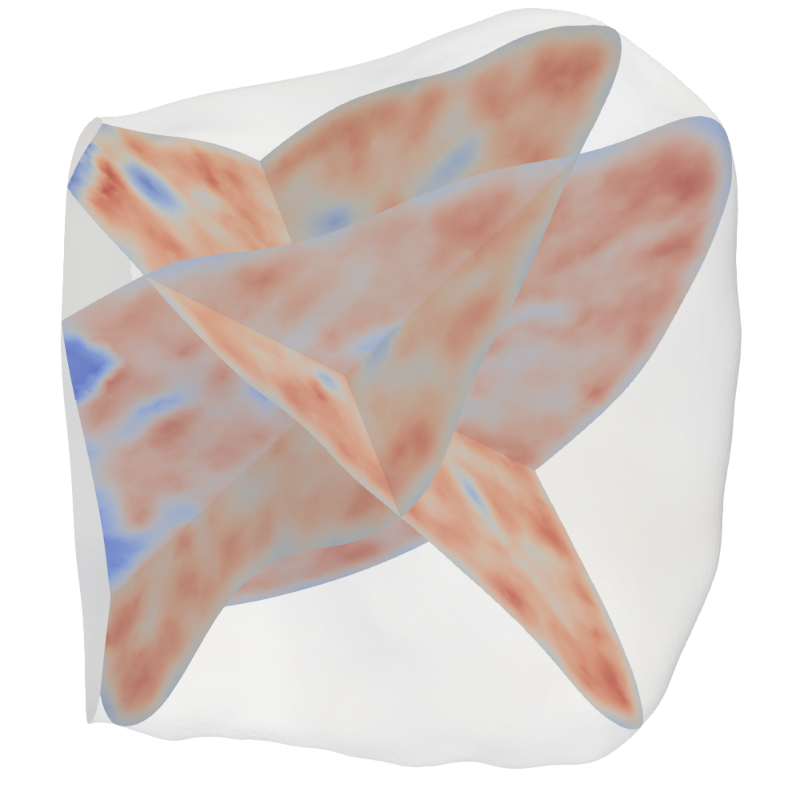}}

\caption{Homogenized volumetric flow rate density $q_{\text{\tiny{supply}}}$ [$ \text{s}^{-1}$] plotted on the deformed configuration at the beginning and after 300 hours of the regrowth process. $V_{L}$ denotes the current volume of the liver domain.}
\label{fig:grownliverflow}
\end{figure}

Figure \ref{fig:grownliverflow} compares the resected liver at the beginning and after 300 hours of the simulated regeneration process. We also report the associated volume $V_{L}$ of the current liver domain. Referring these volumes to the volume of the full liver before resection ($V_{L} = 2,403,749 \; \text{mm}^3$), we see that directly after resection, the liver is reduced to 79.8\% of its original volume, but after 300 hours has recovered 98.6\% of its original volume. 

Additionally, we plot the homogenized blood flow rate density of the regrown state after 300 hours in Fig.~\ref{grownliver}. Comparing these results to the results of the full liver in Fig. \ref{fig:flow_Full_Liver_same_scale}, we observe that the flow rate density reduces over time to the level of the homeostatic state in the unresected full liver. 
On the one hand, the intensity of the flow rate decreases in areas that initially experienced the highest increase. On the other hand, the flow rate does not recover in the area, where the flow rate dropped due to orphans.

\subsection{Flow rate variability across the liver domain}

We now assess the effect of partial liver resection on the homogenized flow rate density $q_{\text{\tiny{supply}}}$ into the compartment microcirculation, which we see as a suitable measure of the quality of perfusion at the meso- and microscale. To this end, we construct histograms to compare the variability and distribution of the homogenized flow rate $q_{\text{\tiny{supply}}}$ across the liver domain before and after partial resection.

{\color{red}
For constructing histograms for the full liver and the resected liver, we require sample data at points that are covering each domain sufficiently close to an equal distribution. For simplicity, we collect the flow rate density at the element midpoints of the two tetrahedral meshes shown in Figs. \ref{fig:Livermesh} and \ref{fig:LivermeshResected} and weight the frequency by the element volume. To mitigate the effect of a potential boundary layer, we leave away all boundary elements. For both cases, we group the values in the same 150 flow rate bins, which we found suitable to clearly visualize the underlying distribution patterns.  
To eliminate the effect of the difference in the number of elements in the two cases, the histograms report the relative frequency in each bin.}

\begin{figure}[ht]
\begin{subfigure}
    \centering
    \begin{tikzpicture}
      % Colorbar at the top
      \node[] (pic) at (-7.5,0) {\includegraphics[height=55mm]{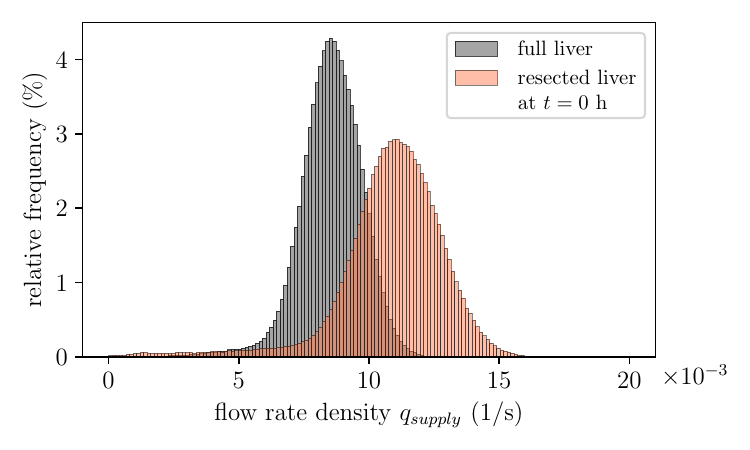}};
      \node[] (pic) at (0.5,0) {\includegraphics[height=55mm]{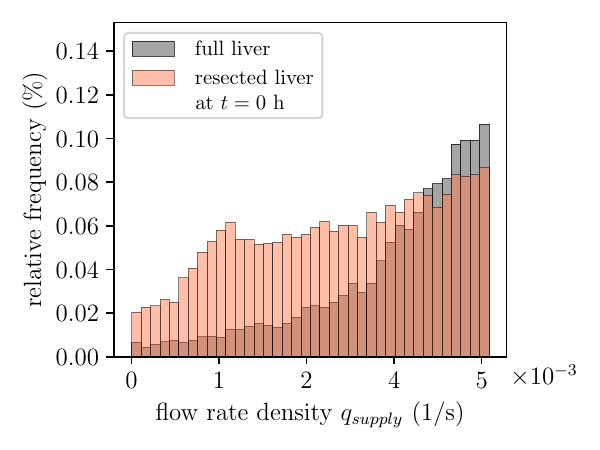}};
    \end{tikzpicture}
    % \centering
    % \includegraphics[height = 60mm]{./Images/flow_rate_densities_histogramm.png}
    \caption{\color{red}Variability of the homogenized flow rate density $q_{\text{\tiny{supply}}}$ [$ \text{s}^{-1}$] in the full and resected liver. The left pictures zooms in to show the left-hand tails.}
\label{fig:histogramm_flow_full_resected}
\end{subfigure}
\vspace{0.5cm}
\begin{subfigure}
\centering
    \begin{tikzpicture}
      % Colorbar at the top
      \node[] (pic) at (-7.5,0) {\includegraphics[height=55mm]{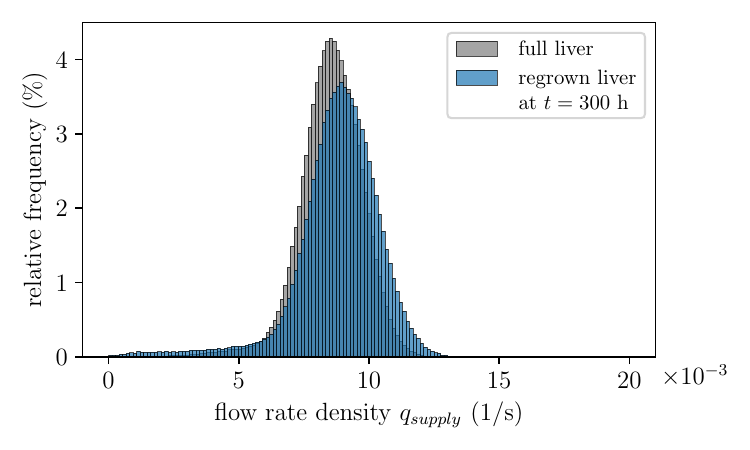}};
      \node[] (pic) at (0.5,0) {\includegraphics[height=55mm]{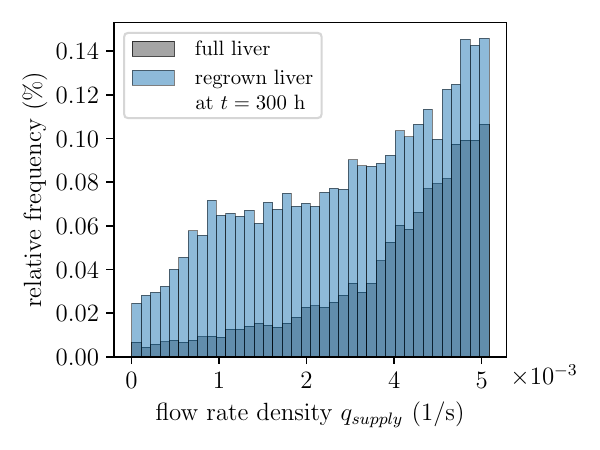}};
    \end{tikzpicture}
  \caption{\color{red}Variability of the homogenized flow rate density $q_{\text{\tiny{supply}}}$ [$ \text{s}^{-1}$] at the beginning of the regrowth process and after 300 hours. As a homeostatic reference, we include the results for the liver before resection.} 
\label{fig:histogramm_flow_grown_liver}
\end{subfigure}
\end{figure}

\subsubsection{Before and after partial resection}

Figure \ref{fig:histogramm_flow_full_resected} compares the change in variability and distribution of the homogenized volumetric flow rate density $q_{\text{\tiny{supply}}}$, when we move from the full liver (in blue) to the resected liver without regrowth (in orange). Table \ref{tab_flow1} reports the corresponding maximum values, the mean, and the standard deviation.%, and the skewness.
%Using our simulation results for the full and resected liver sampled at the nodes of the corresponding finite element meshes, we arrive at the histograms plotted in Fig. \ref{fig:histogramm_flow_full_resected} for the homogenized volumetric flow rate $q_{\text{\tiny{supply}}}$. 
We observe a distinct increase in the mean flow rate density in the case of the resected liver, demonstrating the state of hyperperfusion compared to the normal homeostatic conditions in the case of the full liver. We can observe that there is also a significant rise in the variability and dispersion of the blood supply $q_{\text{\tiny{supply}}}$ in the case of the resected liver. This reflects the state of hyperperfusion in most regions, but also the effect of orphans in the vasculature that practically lead to the inhibition of perfusion.
%The larger standard deviations observed in the histograms of the resected liver suggest a much more heterogeneous distribution of homogenized flow rate. This variability is likely due to the altered flow dynamics and uneven blood supply following the partial resection.

\begin{table}[t]
%\small % Adjust font size to match the text
\centering
\caption{Variability measures for the homogenized volumetric flow rate $q_{\text{\tiny{supply}}}$ at different liver states.}
\label{tab_flow1}
\begin{tabular}{l r r c } 
 \hline
  & max. [$s^{-1}$] & mean [$s^{-1}$] %& min. [$s^{-1}$] 
  & stand.\ dev.\ [$s^{-1}$] \\ [0.5ex] %& sk [-] \\ [0.5ex] 
 \hline
 Full liver (homeostatic reference) & $14.6\cdot 10^{-3}$ & $7.8\cdot 10^{-3}$ %& $0.0$ 
 & $2.0\cdot 10^{-3}$  \\%& $0.941$\\ 
 %\hline
 Resected liver at t = 0 & $21.5\cdot 10^{-3}$ & $10.0\cdot 10^{-3}$ & %$0.0$ & 
 $2.8\cdot 10^{-3}$  \\%& $0.2765$\\
 %\hline
 Regrown liver at t = 45 h & $18.9\cdot 10^{-3}$ & $8.9\cdot 10^{-3}$ & %$0.0$ &
 $2.5\cdot 10^{-3}$  \\%& $0.496$ \\ 
 Regrown liver at t = 150 h & $16.5\cdot 10^{-3}$ & $8.3\cdot 10^{-3}$ & %$0.0$ &
 $2.3\cdot 10^{-3}$   \\%& $0.468$ \\
 Regrown liver at t = 300 h & $15.1\cdot 10^{-3}$ & $8.0\cdot 10^{-3}$ & %$0.0$ &
 $2.2\cdot 10^{-3}$   \\[1ex] %& $0.457$ \\[1ex] 
 \hline
\end{tabular}
\end{table}

These results demonstrate that our model is able to represent hyperperfusion as a consequence of partial resection. They also provide a clear indication how the surgical procedure disrupts normal perfusion patterns, leading to increased flow variability.

\subsubsection{During the regrowth process}

The histograms plotted in Fig. \ref{fig:histogramm_flow_grown_liver} compare the variability of the homogenized volumetric flow rate $q_{\text{\tiny{supply}}}$ in the resected liver at the beginning of the regrowth process and after 300 hours (in orange). As the homeostatic reference, we also show again the variability in the full unresected liver (in blue). For some intermediate states computed, Table \ref{tab_flow1} also reports the maximum values, the mean and the standard deviation at the beginning and after 45, 150 and 300 hours.

We observe that the mean and the standard deviation of the flow rate density reduces during regeneration. When we compare the histogram of the fully regrown state after 300 hours to the initial state of the full liver, we observe that the regrown liver exhibits a flow rate variability and distribution close to the initial homeostatic state. We also see that the lower end tail of the flow rate towards zero flow remains and does not recover. The slightly increased mean and the slight shift towards larger flow rates in the histogram are likely due to the presence of regions with inhibited perfusion, which the remaining regions need to compensate for. 

In general, our results demonstrate again that our model is able to account for the reduction of hyperperfusion towards a homeostatic perfusion state in the regrown liver.

\subsubsection{Local hyperperfusion vs.\ local hypoperfusion}

The results reported in the histograms above suggest a pronounced variability in the local perfusion state across the growing liver domain. In particular, they indicate that after resection, we do not only encounter hyperperfusion, but also a reduction of blood flow (hypoperfusion) in some regions of the compartment microcirculation. We would like to further investigate this variation in local perfusion behavior in our liver model, as it underscores the need for a localized growth criterion. Capturing variations in perfusion during regrowth is likely a key component for potential applications in simulation-based diagnosis and prediction, e.g., for hypoperfusion-driven ischemia.

\begin{figure}[ht]
    \centering
    \includegraphics[width=100mm]{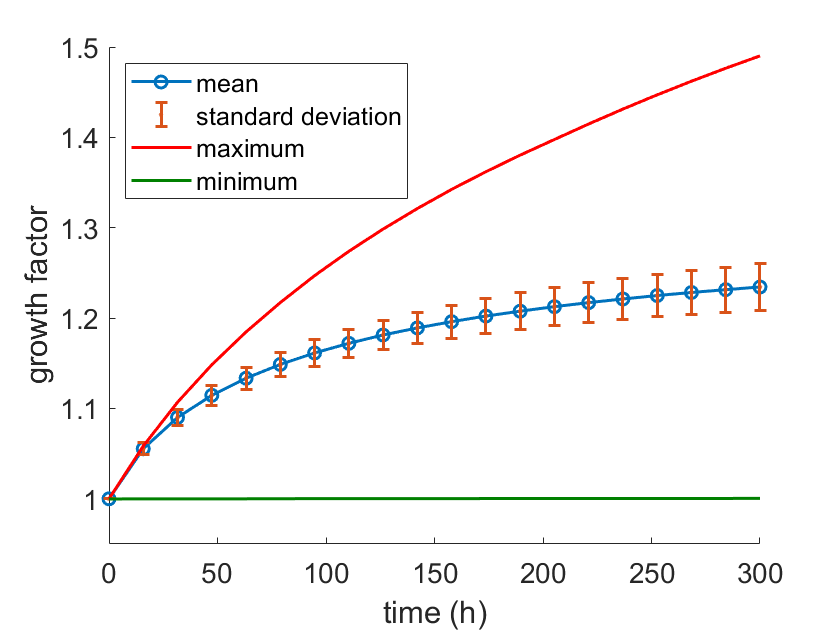}
    \caption{Variability of the growth factor $\vartheta$ across the growing liver domain over time.}
\label{fig:growthfactordistributionovertime}
\end{figure}

In Fig. \ref{fig:growthfactordistributionovertime}, we first plot the minimum and maximum values, the mean and the standard deviation of the local growth factor, monitored at the nodal points of the finite element mesh during regrowth. We observe that on average, each volume element of the resected liver eventually increases its volume by a factor of approx.\ 1.24, which agrees well with the global volume increase from 79.8\% to 98.6\% of the original volume of the unresected liver. The growth factor, however, ranges from 1.48 (twice the average growth factor) to 1.0 (no growth at all). 

\begin{figure}[t]
\centering
\begin{tikzpicture}[transform shape, scale=1.2]
  %\node[] (pic) at (0,2) {\includegraphics[width=25mm, angle=0]{./Images/Selected_areas_flow1.pdf}};
 %\node[] (pic) at (0,-2) {\includegraphics[width=25mm, angle=0]{./Images/Selected_areas_flow1_diagonal.pdf}};
 \node[] (pic) at (0,0) {\includegraphics[width=40mm, angle=0]{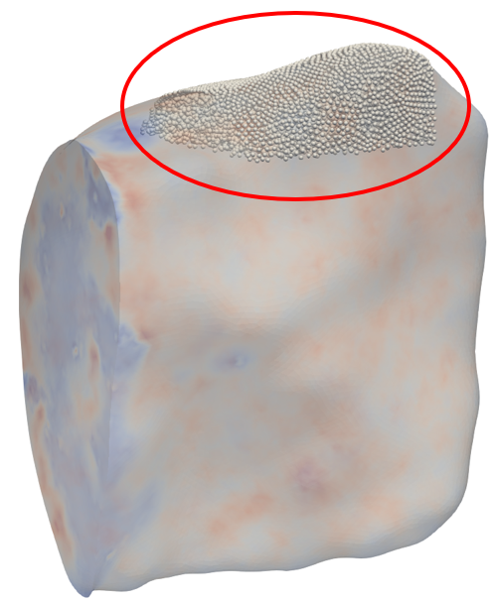}};
  \node[] (pic) at (7,0) {\includegraphics[width=70mm, angle=0]{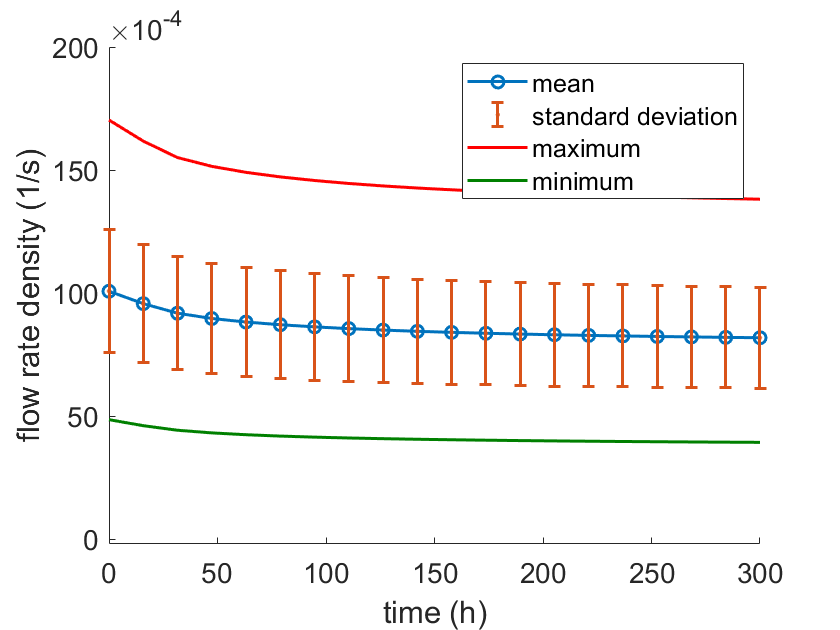}};
\end{tikzpicture}
\caption{Local hyperperfusion: variability of the homogenized volumetric flow rate $q_{\text{\tiny{supply}}}$ [$ \text{s}^{-1}$] across the selected region (red box) over time.}
\label{fig:FlowDistributionOverTime}
\end{figure}

We then focus on a specific region shown in Fig.~\ref{fig:FlowDistributionOverTime}, where the simulated homogenized flow rate density indicates hyperperfusion. To quantify local hyperperfusion, we restrict our selection of nodal points to this region and plot the local variability of the homogenized flow rate density $q_{\text{\tiny{supply}}}$ during regrowth in Fig.~\ref{fig:FlowDistributionOverTime}. 

We observe that on average, the flow rate density decreases from around 0.01 to 0.008 $\text{s}^{-1}$, which corresponds inversely to the average growth factor of 1.24 reported in Fig. \ref{fig:growthfactordistributionovertime} for the complete liver. We also observe a significant variability in the flow rate density in this region, with the maximum value at $t = 0$ as large as 0.017 $\text{s}^{-1}$ and the minimum value as low as 0.0049 $\text{s}^{-1}$. The latter can be attributed to the presence of very large vessels in the synthetic tree, which occupy a certain space with no smaller vessels to be homogenized. Hence, there are always regions with low permeability, effectively modeling the flow obstruction caused by these large vessels.

%This variability results from the variation in permeability, in particular due to the presence of larger vessel, and the inhomogeneous distribution of inflow source terms in the multi-compartment model, which both directly reflect properties of the synthetic vascular tree.

We observe that all curves - average, minimum and maximum - still exhibit the same relative decrease in flow rate density over time. It is straightforward to infer from our definitions of the evolution equation \eqref{eq:theta_dot} and the growth criterion \eqref{eq:growth_criterion} that at the locations of low flow rate density, the homeostatic reference $\tilde{q}_\text{\tiny{equi}}$ in \eqref{eq:growth_criterion}, taken from the full unresected liver, must have also been low - for instance, due to the existence of a larger resolved vessel at that location. We can therefore conclude that Figure \ref{fig:FlowDistributionOverTime}, including the curve of the minimum value, does not reflect a state of hypoperfusion.

\begin{figure}[t]
\centering
\begin{tikzpicture}[transform shape, scale=1.2]
  %\node[] (pic1) at (0,2) {\includegraphics[width=25mm, angle=0]{./Images/Selected_areas_flow_orphan1.pdf}};
   % \node[] (pic1) at (0,-2) {\includegraphics[width=25mm, angle=0]{./Images/Selected_areas_flow_orphan1_diagonal.pdf}};
   \node[] (pic1) at (0,0) {\includegraphics[width=40mm, angle=0]{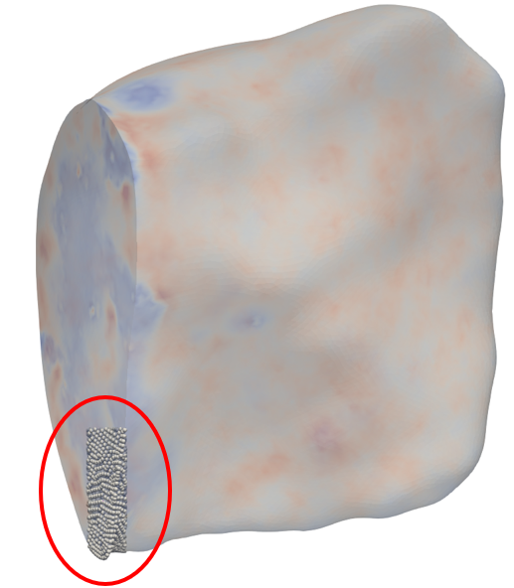}};
  \node[] (pic2) at (7,0) {\includegraphics[width=70mm, angle=0]{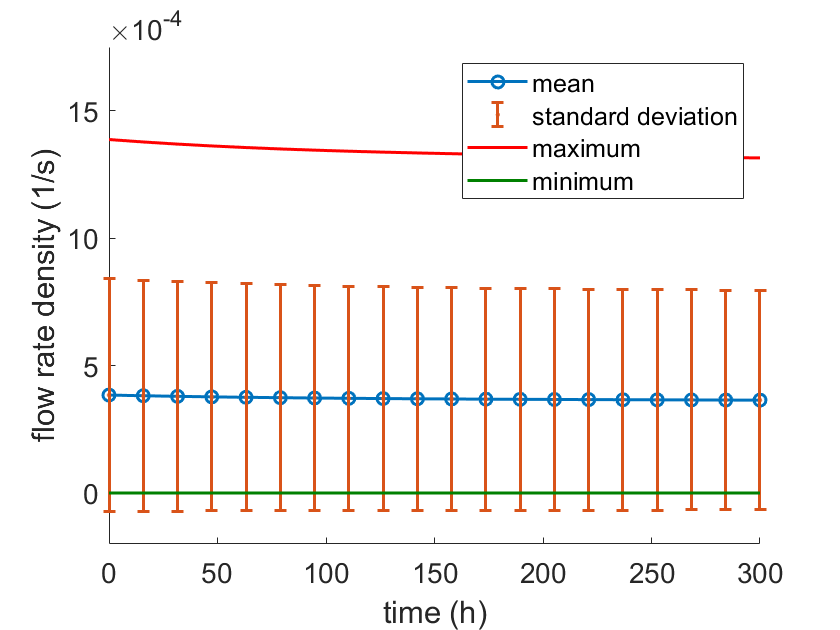}};
\end{tikzpicture}
\caption{Local hypoperfusion: variability of the homogenized volumetric flow rate $q_{\text{\tiny{supply}}}$ [$ \text{s}^{-1}$] across the selected region (red box) over time.}
\label{fig:FlowDistributionOrphanOverTime}
\end{figure}

We then focus on a specific region at the cut plane shown in Fig. \ref{fig:FlowDistributionOrphanOverTime}. We repeat the same procedure and plot the variability of the local homogenized flow rate density during regrowth in Fig. \ref{fig:FlowDistributionOrphanOverTime}. We now observe that only the curve for the maximum value shows a slight decrease over time, while the average and minimum curves remain constant. We therefore infer from \eqref{eq:theta_dot} and \eqref{eq:growth_criterion} that the homeostatic reference $\tilde{q}_\text{\tiny{equi}}$ in the unresected liver must have been larger in this region, such that an increase in the growth factor \eqref{eq:theta_dot} is excluded. We can therefore conclude that Figure \ref{fig:FlowDistributionOrphanOverTime} does reflect a state of hypoperfusion, which indicates the existence of orphans in the supplying tree. The plot of the active vasculature of the portal vein in Fig. \ref{fig:Synthetic vasculature after resection} confirms the existence of orphans in this region.

\section{Summary and conclusions}
\label{sec:Conclusion}
In this paper, we presented a novel framework for modeling liver regrowth on the organ scale. It combines the following three main model components: 
\begin{enumerate}[label=(\Alph*),leftmargin=*]
    \item \textit{a multiscale perfusion model} that combines synthetic vascular tree generation with a multi-compartment homogenized flow model, including a homogenization procedure to obtain the effective permeabilities and intercompartmental perfusion coefficients from the lower hierarchies of the synthetic tree structure.
\item \textit{a poroelastic finite growth model} that is defined in the compartment microcirculation, but acts also on the other compartments and the synthetic vascular tree structure by transfer of the resulting finite kinematics.

\item \textit{an evolution equation} for the local volumetric growth factor, driven by the homogenized flow rate density into the compartment microcirculation that we identified as a measure for local hyperperfusion.
\end{enumerate}

Our framework is based on a series of modeling assumptions and interpretations. The most important are:
\begin{enumerate}[label={(\theenumi)},leftmargin=*]
    \item We assumed that the hierarchical vascular structures of the liver can be approximated by straight vessel segments, where blood is a Newtonian fluid and Poiseuille's law holds. We then synthesized the vascular tree structure of the liver via non-intersecting directed graphs that minimize a combination of the metabolic demand and the hydrodynamic resistance \cite{Jessen3}. We combined the vascular trees of the portal vein and the hepatic artery into one supplying structure.
\item 
We divided the synthetic supplying and draining trees in larger vessel segments that are kept resolved, and smaller vessel segments that are replaced by two homogenized flow compartments, one for the lower hierarchies of the supplying tree and one for the lower hierarchies of the draining tree. We used the smaller vessel segments to find compartmental permeability and intercompartmental perfusion coefficients via a homogenization procedure, based on a scale separation of one order of magnitude between macroscale (full organ), representative averaging volume, and radii of the vessels to be homogenized. %This separation of scales is assumed to be sufficient for the homogenization of permeability and perfusion coefficients.

\item 
We assumed that the compartments supply and drainage are connected via a compartment microcirculation, whose permeability represents the resistance of the capillary network of sinusoids at the microscale. The compartments supply and drainage are coupled to the outlets of the resolved vessel segments via suitably {\color{red}estimated} source and sink terms in the continuity equation. %, but does not account for the flow patterns in the lobular structures at the mesoscale.

\item The characteristic time scale of our regrowth model formulated in the compartment microcirculation 
is days. We therefore describe liver regrowth by a sequence of quasi-static growth processes. We also assumed that liver regrowth can be regarded as isotropic from a macroscopic viewpoint.

\item  %The corresponding growth factor defined at each point of the macroscale domain denotes the volumetric change due to growth. 
We assumed that we can keep the same porosity during regrowth. Hence, the growth factor is not only applied to the tissue skeleton, but also to the perfusing blood, representing the added mass of the blood that occupies additional sinusoid space.

\item %Hyperperfusion in the microcirculation is the main stimulus for liver regrowth. 
We postulated a phenomenological link between the evolution equation for the growth factor, representing volumetric growth, and the inflow rate density into the compartment microcirculation, measuring \mbox{(hyper-)}perfusion. %The evolution equation multiplies a growth scaling factor with a mechanism-specific growth criterion. 
We then proposed a growth criterion that represents the relative increase of the current homogenized blood flow at each macroscale point with respect to the (supposedly healthy) homeostatic state before resection in the same liver. This homeostatic reference automatically accounts for potential individual perfusion characteristics in a patient-specific simulation model.

\item For the largest vessels of the resected vascular tree structure, we assumed that remodeling solely involves adaptations in length and position. Furthermore, their supply and drainage activity is scaled up after resection via a corresponding change in diameter (vasodilation). For medium- and smaller-sized vessels homogenized in the compartments supply and drainage, we assumed that remodeling is reflected through geometric updates of the homogenized permeability and perfusion coefficients. At the capillary level, we
assumed that remodeling is effective instantly, so that the resistance of the sinusoid network in terms of the permeability in the compartment microcirculation always corresponds to the same (healthy) tissue.

 \end{enumerate}

We calibrated our regrowth model with experimental liver data, adjusting parameters for the growth speed, nonlinearity, and maximum growth factor, to match observed liver volume regeneration curves. We then applied the resulting modeling framework to a full-scale patient-specific liver example, for which we assumed a common surgical resection cut. The cut also involved orphan vessels that loose connection to the root of the supplying tree and hence induce local insufficient blood supply of the microcirculation (hypoperfusion). We conducted finite element simulations of the perfusion behavior of the unresected full liver and the regeneration of the resected liver, with a focus on how well our modeling framework represents characteristic phenomena at the organ scale. 

% Recovers well in 300 Hours.
We observed that the resection
reduced our example liver to 79.8\% of its original volume, but after 300 hours it recovered to 98.6\%. {\color{red}The overall regrowth dynamics of our model thus corresponds well with common clinical observations} \cite{yamanaka1993dynamics,RefYamamoto}. 
We furthermore observed that the homogenized flow rate density significantly increased after resection and reduced over time to the level of the the homeostatic state in the unresected full liver. These results demonstrate that our model is able to represent hyperperfusion as a consequence of partial
resection and to account for the reduction of hyperperfusion towards a homeostatic perfusion state in the regrown liver.

%amd a driver of regrowth. They also provide a clear indication how the surgical procedure disrupts normal perfusion patterns, leading to increased flow variability.

Furthermore, the simulation results suggest a pronounced variability in the local perfusion state
across the growing liver domain, demonstrated via histograms and distribution parameters. The growth factor observed ranges from 1.48 (twice the average) to 1.0 (no growth at all), the homogenized flow rate density from 0.017 to 0.0 $s^{-1}$. Hence, we do not only capture hyperperfusion, but also the expected local hypoperfusion in the vicinity of the orphan vessels. Capturing variations in perfusion during regrowth is likely a key component for potential applications in simulation-based diagnosis and prediction, e.g., for hypoperfusion-driven ischemia or for the preoperative identification of suitable cut patterns for partial liver resection. These observations emphasize the need for a localized growth criterion such as the one proposed in this work.

The current model provides a basis for further research and refinement with respect to a number of aspects, some of which are:
\begin{enumerate}[label={(\theenumi)},leftmargin=*]
\item While our model provides insights into perfusion and regrowth dynamics, it does not yet account for liver functions, metabolism, or pathological preconditions, which can be additional important factors in a more comprehensive analysis of liver regeneration processes. %One approach to achieve this could involve the integration of ODEs that incorporate various risk factors associated with postoperative liver failure.
%Incorporating these aspects is essential in advancing the proposed model towards a more evidence-based physiological simulation tool suitable in clinical practice. For example, including metabolic activity could enhance predictions of liver regeneration in the context of different metabolic conditions or diseases.

\item The current compartment microcirculation with one global permeability parameter that represents the flow resistance of the capillary network of sinusoids could be extended or replaced by a lobule-level perfusion approach that can represent the variation of specific properties relevant for perfusion and regrowth across the microcirculation, e.g., based on an efficient reduced-order model of a lobule \cite{siddiqui2024reduced}. 

\item The current approach assumes isotropic growth, but liver regrowth may exhibit anisotropic behavior. Future research could explore anisotropic growth models to provide a possibly more realistic representation of liver regeneration. This involves defining preferred directions for growth based on local mechanical or biological factors.

{\color{red}
\item To better understand the impact of parameter variations on model outcomes, a detailed sensitivity analysis (e.g., for the multi-compartment model) is of interest. This may involve systematically varying model parameters (e.g., size of representative averaging volume, number of compartments, separation criterion for macroscopic and microscopic vasculature) to identify how changes affect predictions of liver perfusion and regrowth. %Such analyses could also guide more accurate model calibration.

\item At the current (early) stage of development of our modeling framework, comprehensive validation studies against experimental and clinical data have not been conducted yet, but constitute a crucial next step. For the current model, our plans include the monitoring of tissue growth rates and changes in the vascular system, based on available experimental and clinical CT data for humans \cite{forbes2016liver} or in comparison with liver ischemia-reperfusion experiments in mice. We also plan to assess our results in comparison with existing simpler ODE-based regrowth models \cite{RefFurchtgott,RefChrist}.

\item Along the same lines, the current model benefits from further personalization to reflect individual patient characteristics. This includes refining model parameters based on patient-specific data.}

\end{enumerate}

%\newpage
\section*{Acknowledgments}
The results presented in this work were achieved as part of the ERC Starting grant project $\mathrm{'}$ImageToSim$\mathrm{'}$ that has received funding from the European Research Council (ERC) under the European Union’s Horizon 2020 research and innovation programme (grant agreement no. 759001). The authors gratefully acknowledge this support. 
%The authors also gratefully acknowledge the computing time provided to them on the high-performance computer Lichtenberg at the NHR Centers NHR4CES at TU Darmstadt. This is funded by the Federal Ministry of Education and Research and the State of Hesse.

%\bibliographystyle{plain}
%\bibliographystyle{unsrt}
%\bibliography{bibliography_new.bib}

%%\bibliographystyle{plain}
%\bibliographystyle{unsrt}
%\bibliography{bibliography_new.bib}

\printbibliography % This command replaces the traditional \bibliography command

\end{document}